\documentclass[12pt, draftclsnofoot, onecolumn]{IEEEtran}
\usepackage{amssymb}
\usepackage[cmex10]{amsmath}
\usepackage{stfloats}
\usepackage{graphicx}
\usepackage{subfigure}
\usepackage{tabularx}
\usepackage{epsfig,epsf,color,balance,cite}
\usepackage{verbatim}
\usepackage{url}
\usepackage{bm}
\usepackage{mathrsfs}
\usepackage{amsthm}
\usepackage{multirow}
\DeclareMathAlphabet\mathbfcal{OMS}{cmsy}{b}{n}

\newcommand{\code}[1]{\colorbox[RGB]{245,245,245}{\texttt{\detokenize{#1}}}}
\newtheorem{theorem}{Theorem}
\newtheorem{definition}{Definition}

\newtheorem{lemma}{Lemma}
\newtheorem{remark}{Remark}

\usepackage{algorithm}
\usepackage{algpseudocode}
\DeclareFontFamily{OT1}{pzc}{}
\DeclareFontShape{OT1}{pzc}{m}{it}{<-> s * [1.40] pzcmi7t}{}
\DeclareMathAlphabet{\mathpzc}{OT1}{pzc}{m}{it}

\usepackage{graphicx}
\usepackage{epstopdf}
\IEEEoverridecommandlockouts

\newfont{\bbb}{msbm10 scaled 700}

\newfont{\bb}{msbm10 scaled 1100}

\usepackage{stmaryrd} 

\usepackage{hyperref}
\hypersetup{
    bookmarks=true,         
    unicode=false,          
    pdftoolbar=true,        
    pdfmenubar=true,        
    pdffitwindow=false,     
    pdfstartview={FitH},    
    pdfnewwindow=true,      
    colorlinks=true,       
    linkcolor=red,          
    citecolor=green,        
    filecolor=blue,      
    urlcolor=blue           
}

\begin{document}

\title{A New Achievable Region of the $K$-User MAC Wiretap Channel with Confidential and Open Messages Under Strong Secrecy}

\author{
	\IEEEauthorblockN{Hao Xu, \emph{Member, IEEE}\IEEEauthorrefmark{0},
		Kai-Kit Wong, \emph{Fellow, IEEE}\IEEEauthorrefmark{0},
		and
		Giuseppe Caire, \emph{Fellow, IEEE}\IEEEauthorrefmark{0}
	}
	\thanks{
		This work was supported by the European Union's Horizon 2020 Research and Innovation Programme under Marie Skłodowska-Curie Grant No. 101024636 and the Alexander von Humboldt Foundation.
		Partial of this work has been presented at the 2023 IEEE International Symposium on Information Theory (ISIT) \cite{xu2023achievable}.
		
		H. Xu and K.-K. Wong are with the Department of Electronic and Electrical Engineering, University College London, London WC1E 7JE, UK (e-mail: hao.xu@ucl.ac.uk; kai-kit.wong@ucl.ac.uk).
		
		G. Caire is with the Faculty of Electrical Engineering and Computer Science at the Technical University of Berlin, 10587 Berlin, Germany (e-mail: caire@tu-berlin.de).
	}
}

\maketitle

\begin{abstract}
This paper investigates the achievable region of a $K$-user discrete memoryless (DM) multiple access wiretap (MAC-WT) channel, where each user transmits both secret and open (i.e., non-confidential) messages.
All these messages are intended for the legitimate receiver (Bob), while the eavesdropper (Eve) is only interested in the secret messages.
In the achievable coding strategy, the confidential information is protected by open messages and also by 
the introduction of auxiliary messages.
When introducing an auxiliary message, one has to ensure that, on one hand, its rate is large enough for protecting the secret message from Eve and, on the other hand, the resulting sum rate (together with the secret and open message rate) does not exceed Bob's decoding capability.
This yields an inequality structure involving the rates of all users' secret, open, and auxiliary messages.
To obtain the rate region, the auxiliary message rates must be eliminated from the system of inequalities.
A direct application of the Fourier-Motzkin elimination procedure is elusive since a) it requires that the number of users $K$ is explicitly given, and b) even for small $K = 3, 4, \ldots$, the number of inequalities becomes extremely large.
We prove the result for general $K$ through the combined use of Fourier-Motzkin elimination procedure and mathematical induction.
This paper adopts the {\em strong secrecy} metric, characterized by {\em information leakage}. 
To prove the achievability under this criterion, we analyze the resolvability region of a $K$-user DM-MAC channel (not necessarily a wiretap channel).
In addition, we show that users with zero secrecy rate can play different roles and use different strategies 
in encoding their messages. 
These strategies yield non-redundant (i.e., not mutually dominating) rate inequalities.
By considering all possible coding strategies, we provide a new achievable region for the considered channel, and show that it strictly improves those already known in the existing literature by considering a specific example.
\end{abstract}

\begin{IEEEkeywords}
	Discrete memoryless (DM) multiple access wiretap (MAC-WT) channel, strong secrecy, secret and open messages, achievable rate region.
\end{IEEEkeywords}

\IEEEpeerreviewmaketitle

\section{Introduction}
\label{section1}

Information theoretic security, also known as {\em physical layer security}, is an alternative to cryptographic security that may be attractive in cases where generating, distributing, and managing cryptographic keys is difficult, or if advances in quantum computing will make systems based on classical computational cryptography intrinsically insecure. 
Physical layer security techniques exploit the randomness of the transmission channel and particular code constructions to prevent the eavesdroppers (Eves) from wiretapping, and do not rely on Eves' limited computational capability assumptions.
After the first seminal works on the single-user wiretap channel \cite{shannon1949communication, wyner1975wire, leung1978gaussian, csiszar1978broadcast}, the research on physical layer security considered various network topologies such as multiple access (MAC) wiretap channels \cite{ekrem2008secrecy, 4036106, 4455769, 5961828, chen2017joint, chen2018collective, tekin2008gaussian, tekin2008general, 9174164, xu2022achievable, yassaee2010multiple, nafea2019generalizing, Hayashi2019secrecy}, broadcast channels \cite{5730586, 5605348, 6584931, 9133130}, interference channels \cite{4529283, 5752448, 6006610, 7313047}, and relay-aided channels \cite{4608977, 5352243, 7105936, 7551149}.

This paper focuses on the MAC wiretap (MAC-WT) channel. 
To put our work in context, we review the main related literature below. 
In \cite{ekrem2008secrecy, 4036106, 4455769}, two-user discrete memoryless (DM) MAC-WT systems were studied, where \cite{ekrem2008secrecy} developed inner and outer bounds for a channel with a weaker Eve, and \cite{4036106} and \cite{4455769} studied a channel where two users communicate with a common receiver and see each other as an Eve.
In \cite{chen2017joint, chen2018collective, tekin2008gaussian, tekin2008general, 9174164, xu2022achievable}, the more general scenario with an arbitrary number $K$ of users was investigated.
Specifically, \cite{chen2017joint} and \cite{chen2018collective} developed achievable regions for DM MAC-WT channels, and \cite{tekin2008gaussian} studied a Gaussian MAC-WT system with a weaker Eve seeing a degraded channel.
Reference \cite{tekin2008general} extended the work of \cite{tekin2008gaussian} to the non-degraded Gaussian MAC and Gaussian two-way wiretap channels, where each user has, in addition to confidential information, also an open (i.e., non-confidential) message for the legitimate receiver. 
Achievable regions were derived, the sum secrecy rate was maximized by power control, and cooperative jamming was also proposed to enhance the secrecy.
Note that in some papers, ``open message'' is also called ``private message'' \cite{watanabe2014optimal}.
In \cite{9174164} and \cite{xu2022achievable}, the achievable regions of MAC-WT systems with open and secret messages were further studied, and it was shown that by introducing open messages to wiretap channels, the system spectral efficiency can be significantly increased while the secrecy performance remains unchanged.

The literature mentioned above considered the {\em weak secrecy} criterion, characterized by the {\em {information leakage rate}}.
It should be noted that a vanishing information leakage rate does not imply that a vanishing number of information bits of the secret message are leaked, because the length of the message in bits grows linearly with the block length $n$.  
To address this issue, {\em strong secrecy} was introduced in \cite{maurer2000information, csiszar2000common}, by considering directly the {\em information leakage} in terms of the multi-letter mutual information between messages and Eve's received signal, without normalization by $n$.
A comprehensive discussion on different secrecy metrics can be found in \cite{bloch2013strong}.
Under strong secrecy, \cite{yassaee2010multiple, nafea2019generalizing, Hayashi2019secrecy} considered the standard DM MAC-WT channel with only secret messages. 
Specifically, by analyzing the output statistics in terms of average variational distance and applying random coding, an achievable region was provided in \cite{yassaee2010multiple}.
In \cite{nafea2019generalizing}, the MAC-WT system with a DM main channel and different wiretapping scenarios was studied.
Both \cite{yassaee2010multiple} and \cite{nafea2019generalizing} considered the two-user case.
In \cite{Hayashi2019secrecy}, a $K$-user DM MAC-WT channel was investigated and the results in \cite{chen2017joint} and \cite{chen2018collective} were strengthened subject to the strong secrecy metric.
However, by checking the two-user case and comparing \cite[(14)]{Hayashi2019secrecy} with \cite[Theorem~$1$]{yassaee2010multiple}, it can be seen that the achievable region given in \cite{Hayashi2019secrecy} includes only 
the region ${\cal R}_1$ in \cite{yassaee2010multiple} but not ${\cal R}_2$ and ${\cal R}_3$, indicating that even for the MAC-WT channel with only secret messages, there is still space for improvement of the achievable region.


In this work, we study the information-theoretic secrecy problem for a $K$-user DM MAC-WT channel where users have both
secret and open messages for the intended receiver (Bob) while preserving
the confidentiality of the secret messages with respect to Eve.
Eve aims to wiretap the confidential information of all users and the users do not care if their open messages may be decoded by Eve. 
Our contributions in this work include new proof techniques, a new insight in the role
of open messages, and a thorough numerical analysis of a special case, showing
that the new achievable region can be strictly larger than what was previously known. 
These main points are summarized below.


\begin{itemize}
\item {\em Proof techniques:} In wiretap channels, typical achievability strategies are based on introducing auxiliary messages to protect the confidential information. 
When introducing an auxiliary message, one has to ensure that, on one hand, its rate is large enough for protecting the secret message from Eve and, on the other hand, the resulting sum rate (together with the secret and open message rate) does not exceed Bob's decoding capability.
This yields an inequality structure involving the rates of all users' secret, open, and auxiliary messages.
In Theorem~\ref{lemma_FM_gene_K2} we provide such structure and also give the conditions under which this structure can be satisfied. 
An essential step in the proof of Theorem~\ref{lemma_FM_gene_K2} is provided by Lemma~\ref{theorem_FM}.  
A two-user case of Lemma~\ref{theorem_FM} was proven in \cite[Lemma~$7$]{xu2022achievable} by direct application of the Fourier-Motzkin elimination procedure \cite[Appendix D]{el2011network}.
However, as $K$ grows, this direct proof becomes unmanageable due to the excessively large number of inequalities.
To exemplify the complexity of the direct Fourier-Motzkin application, we provided the case $K = 3$ in the unpublished research note \cite{xu2022note}, from which we see that even in this relatively simple case, over $130$ inequalities are generated in the direct elimination procedure.
Besides the extremely high complexity, another problem of the direct elimination strategy is that it can be applied in principle only if the number of users $K$ is explicitly given (e.g., $K = 1, 2, 3, \ldots$) 
so that all inequalities can be explicitly listed. This makes the direct proof approach 
inappropriate for Lemma~\ref{theorem_FM}, since it is stated for a generic $K$.
In Appendix~\ref{Prove_theorem_FM}, we circumvent this problem 
through the combined use of Fourier-Motzkin elimination procedure and mathematical induction.
Then, using this lemma, the general proof of Theorem~\ref{lemma_FM_gene_K2} follows.

We prove the achievability under the {\em strong secrecy} metric.
To this end, we analyze the resolvability region for a $K$-user standard DM-MAC channel (not necessarily a wiretap channel) in Theorem~\ref{theo_varia_dist}.\footnote{One may refer to \cite{bloch2013strong}, \cite{bloch2011achieving}, and \cite{frey2018mac} for the notion of resolvability and resolvability region, and their relationship with secrecy.}

\item {\em Role of open messages:}
In the achievability strategy of this paper, confidential information is protected by open messages and also the 
introduced auxiliary messages.
In general, the sum rate of these messages should exceed a certain amount such that the confidential information can be perfectly protected.
However, a novel observation of this paper is that if a user has no secret message, such 
user can play different roles and has two options: 1) introducing an additional auxiliary message such that the sum rate of its messages is beyond Eve's decoding capability; 
2) simply transmitting its open message as in a standard MAC channel with no wiretapping.
We show that these two options yield non-redundant (i.e., not mutually dominating) rate inequalities. 
Therefore, by considering the union over all options, i.e., all coding strategies for the users with zero secret rates, we obtain a larger achievable region (Theorem~\ref{lemma_DM_exten}) than previously known.
Interestingly, by simply letting the open message rate of all users be zero, we obtain an achievable region for the standard DM MAC-WT channel with only secret messages directly from Theorem~\ref{lemma_DM_exten}, and this region generalizes those provided in \cite{yassaee2010multiple} and \cite{Hayashi2019secrecy}.

\item 
{\em Numerical analysis:} To compare the proposed new achievable region with that provided in \cite{xu2022achievable} (largest known region before the present paper), 
we consider a two-user binary-input real adder channel with Bernoulli-distributed noise.
For such a channel, all rate bounds in the derived achievable region can be computed for all possible input distributions. Nevertheless, comparing the resulting rate regions (in general, 4-dimensional convex non-polytopes) is not a trivial task. 
In our numerical analysis, we use a support vector machine (SVM) to  show the existence of points in the new region that can be  linearly separated from the convex hull of regions in \cite{xu2022achievable}.
This suffices to show that, the results of this paper strictly enlarge 
the known achievability region for the DM MAC-WT with secret and open messages.
\end{itemize}

The rest of this paper is organized as follows. In Section~\ref{DM_model} we introduce the $K$-user DM MAC-WT channel model and give the definition of ``achievability''.
In Section~\ref{DM_region} we give the main results. The achievability proof of the new region is provided in Section~\ref{achie_proof}. In Section~\ref{BAC}, the proposed new region is compared with the region in the existing literature in a two-user binary-input real adder channel.
Section~\ref{conclusion} points out some concluding remarks. 
Auxiliary technical results are given in the appendices.

{\bf Notations:} We use upper and lower case letters to denote random variables and their realizations, e.g., $X$, $x$.
$P_X (\cdot)$ denotes the probability mass function (pmf)~\footnote{Loosely referred to as ``distributions'' in the following.} of $X$ in the sense that $P_X(\cdot) \triangleq \{ P_X(x) : x\in {\cal X}\}$ where $P_X (x) = {\text {Pr}} \{ X = x \}$.
For two different distributions ${\hat P}_X (\cdot)$ and ${\tilde P}_X (\cdot)$ on the same alphabet 
${\cal X}$, their total variational distance is  defined as the $\ell_1$ norm of the difference of the corresponding pmfs, i.e., 
$\left\| {\hat P}_X (\cdot) - {\tilde P}_X (\cdot) \right\|_1 = \sum_{x \in {\cal X}} \left| {\hat P}_X (x) - {\tilde P}_X (x) \right|$.
We use calligraphic capital letters to denote sets, $|{\cal X}|$ indicates the cardinality of a set ${\cal X}$, 
and ${\cal X}_1 \setminus {\cal X}_2$ denotes set subtraction,  and ${\cal X}_1 \times {\cal X}_2$ is the Cartesian product set. 
We use line over a calligraphic letter to indicate it is the complement of a set, e.g., ${\overline {{\cal K}'}} = {\cal K} \setminus {\cal K}'$ if ${\cal K}' \subseteq {\cal K}$, and calligraphic subscript to denote the set of elements whose indexes take 
values from the subscript set, e.g., $X_{\cal K} = \{X_k :  k \in {\cal K}\}$. Finally, we use $[\cdot]^+ \triangleq \max (\cdot,0)$.


\section{Channel Model and Problem Definition}
\label{DM_model}

\begin{figure}
	\centering
	\includegraphics[scale=0.80]{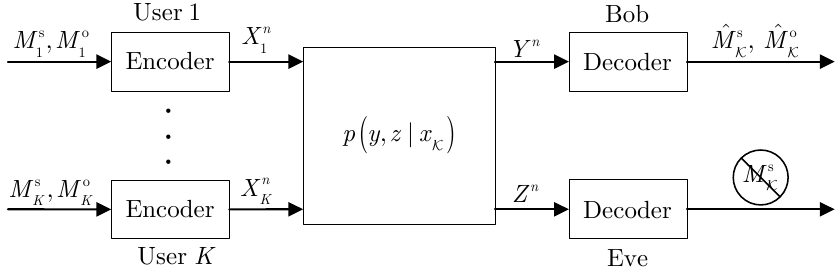}
	\caption{Block diagram of a $K$-user DM MAC-WT channel with secret (i.e., confidential) and open (i.e., non-confidential) messages. Both secret and open messages are intended for Bob, while Eve is interested only in the secret messages.}
	\label{Fig1}
\end{figure}

Fig.~\ref{Fig1} shows the considered  DM MAC-WT channel with $K$ transmitters, 
a legitimate receiver (Bob), and an eavesdropper (Eve).
Let ${\cal K} = \{1, \cdots, K\}$ denote the set of all users.
The DM MAC-WT channel is defined by $\left({\cal X}_{\cal K}, P_{Y, Z| X_{\cal K}}, {\cal Y}, {\cal Z}\right)$ (in short $P_{Y, Z| X_{\cal K}}$), where ${\cal X}_k$, $\cal Y$, and $\cal Z$ are finite alphabets, $x_k \in {\cal X}_k$ is the channel input of user 
$k$, and $y \in {\cal Y}$ and $z \in {\cal Z}$ are respectively channel outputs at Bob and Eve. 
Each user $k \in {\cal K}$ has a secret (i.e., confidential) message $M_k^{\text s}$ and also an open (i.e., non-confidential) message $M_k^{\text o}$ intended for Bob. 
We require that Bob decodes all messages with vanishing probability of error in the limit of large coding block length.  
Eve's goal is to decode the secret messages, while there is no decoding requirement for the open messages, although Eve 
may decode (some of) them as part of an eavesdropping strategy.
User $k$ encodes its information into a codeword $X_k^n$ and then transmits it over the channel.
Upon receiving the sequence $Y^n$, Bob decodes the messages of all users.
To avoid leakage of confidential information to Eve, the secret messages of all users, i.e., $M_{\cal K}^{\text s}$, should be protected. 

Let $R_k^{\text s}$ and $R_k^{\text o}$ denote the rate of user $k$'s secret and open messages. 
Then, a secrecy code for the considered DM MAC-WT channel consists of
\begin{itemize}
	\item Secret and open message sets: ${\cal M}_k^{\text s} = \left[1:2^{n R_k^{\text s}}\right]$ and ${\cal M}_k^{\text o} \!=\! \left[1:2^{n R_k^{\text o}}\right], \forall k \!\in\! {\cal K}$.
	Messages $M_k^{\text s}$ and $M_k^{\text o}$ are uniformly distributed over the corresponding sets ${\cal M}_k^{\text s}$ and ${\cal M}_k^{\text o}$. 
	\item $K$ randomized encoders: the encoder of user $k$ maps the message pair $(M_k^{\text s}, M_k^{\text o}) \in {\cal M}_k^{\text s} \times {\cal M}_k^{\text o}$ to a (possibly random) codeword $X_k^n$.
	\item A decoder at Bob which maps the received sequence $Y^n$ to the message estimate $\left( {\hat M}_k^{\text s}, {\hat M}_k^{\text o} \right) \in {\cal M}_k^{\text s} \times {\cal M}_k^{\text o}, \forall k \in {\cal K}$.
\end{itemize}

Next, we define the criteria for achievability. 
These criteria are defined in an {\em operational sense}, i.e., in terms of the behavior of the error probability at Bob and information leakage at Eve.  
\begin{definition}\label{defi_achi}
	A rate-tuple $(R_1^{\text s}, R_1^{\text o},\cdots, R_K^{\text s}, R_K^{\text o})$ is said to be achievable if there exists a sequence of $\left( 2^{n R_1^{\text s}}, 2^{n R_1^{\text o}}, \cdots, 2^{n R_K^{\text s}}, 2^{n R_K^{\text o}}, n \right)$ codes such that 
	\begin{align}
		& \lim_{n \rightarrow \infty} {\text {Pr}} \left\{ \left( {\hat M}_{\cal K}^{\text s}, {\hat M}_{\cal K}^{\text o} \right) \neq \left( M_{\cal K}^{\text s}, M_{\cal K}^{\text o} \right) \right\} = 0, \label{Pe}\\
		& \lim_{n \rightarrow \infty} I (M_{\cal K}^{\text s}; Z^n) = 0. \label{secrecy_metric}
	\end{align}	
$\phantom{x}$	 \hfill $\lozenge$
\end{definition}

Given an ensemble of coding strategies,  the corresponding achievable region  ${\mathscr R}$ for the DM MAC-WT  is the convex closure of all rate-tuples $(R_1^{\text s}, R_1^{\text o},\cdots, R_K^{\text s}, R_K^{\text o}) \in \mathbb{R}_+^{2K}$ satisfying Definition \ref{defi_achi} under such coding strategies. The determination of the largest possible achievable region (i.e., the capacity region) 
requires also the proof of a converse result. Converses (i.e., matching outer bounds) for the DM MAC-WT are generally yet an open problem even in the case of zero open message rates, while of course they are well-known in the case of zero secret message rates \cite{cover1999elements}. 
Focusing on achievability, in the following we provide a new achievable region which is generally 
larger than what was known in the existing literature.

\section{Main Results}
\label{DM_region}

In this section, we provide an achievable region for the considered DM MAC-WT channel under strong secrecy and show that it improves the one in \cite[Lemma~$1$]{xu2022achievable}. 
Using the result, a new achievable secrecy rate region can be directly obtained for the standard DM MAC-WT channel with only secret messages. Before giving the regions, we first provide some auxiliary results that are important for the achievability proof.

\subsection{Auxiliary Results}
\label{Motivations}

We shall consider achievability strategies where users may introduce auxiliary messages, denoted by the superscript ``a'', 
to protect the confidential information. In particular,  
let $M_k^{\text a}$ denote the auxiliary message introduced for user~$k$ and $R_k^{\text a}$ denote its rate.
Note that in this paper, there are three types of messages, i.e., secret, open, and auxiliary messages.
As we have explained at the beginning of Section~\ref{DM_model}, both the secret and open messages are intended for Bob, while Eve is only interested in the secret messages.
Note that the auxiliary messages are not intended for either Bob or Eve.
They are introduced only to add dummy randomness to protect or ``hide'' the secret messages.
As explained in the introduction, there should be a relationship or structure for $R_k^{\text s}$, $R_k^{\text o}$, and $R_k^{\text a}, \forall k \in {\cal K}$ such that the sum rate of the messages does not exceed Bob's decoding capability and the confidential information can be protected.
In the following theorem, we construct such a structure (see (\ref{region_FM2})) and also give the conditions (see (\ref{cond1}) and (\ref{region_DM0})) under which this structure can be satisfied.
\begin{theorem}\label{lemma_FM_gene_K2}
	Let $(X_{\cal K}, Y, Z) \sim \prod_{k=1}^K P_{X_k} P_{Y, Z| X_{\cal K}}$.
	For a given subset ${\cal K}' \subseteq {\cal K}$, if 
	\begin{equation}\label{cond1}
		I(X_{\cal S}; Y| X_{\overline {\cal S}}, X_{\overline {{\cal K}'}}) - I(X_{\cal S}; Z| X_{\overline {{\cal K}'}}) \geq 0, \forall {\cal S} \subseteq {\cal K}',
	\end{equation}
	then, for any rate-tuple
	$(R_1^{\text s}, R_1^{\text o},\cdots, R_K^{\text s}, R_K^{\text o})$ satisfying
	\begin{equation}\label{region_DM0}
		\left\{\!\!\!
		\begin{array}{ll}
			R_k^{\text s} = 0, \forall k \in {\overline {{\cal K}'}}, \\
			\sum\limits_{k \in \cal S} R_k^{\text s} + \sum\limits_{k \in {\cal S} \setminus {\cal S}'} R_k^{\text o} + \sum\limits_{k \in {\cal T}} R_k^{\text o} \leq I(X_{\cal S}, X_{\cal T}; Y| X_{\overline {\cal S}}, X_{\overline {\cal T}}) - I(X_{{\cal S}'}; Z| X_{\overline {{\cal K}'}}), \\
			\quad\quad\quad\quad\quad\quad\quad\quad\quad\quad\quad\quad \forall {\cal S} \subseteq {\cal K}', {\cal S}' \subseteq {\cal S}, {\cal T} \subseteq {\overline {{\cal K}'}},
		\end{array} \right.
	\end{equation}
	there exist $R_k^{\text a}, \forall k \in {\cal K}'$ such that
	\begin{equation}\label{region_FM2}
		\left\{\!\!\!
		\begin{array}{ll}
			R_k^{\text a} \geq 0, \forall k \in {\cal K}', \\
			\sum\limits_{k \in {\cal S}} (R_k^{\text s} + R_k^{\text o} + R_k^{\text a}) + \sum\limits_{k \in {\cal T}} R_k^{\text o} \leq I(X_{\cal S}, X_{\cal T}; Y| X_{\overline {\cal S}}, X_{\overline {\cal T}}), \forall {\cal S} \subseteq {\cal K}', {\cal T} \subseteq {\overline {{\cal K}'}}, \\
			\sum\limits_{k \in {\cal S}} (R_k^{\text o} + R_k^{\text a}) \geq I(X_{\cal S}; Z| X_{\overline {{\cal K}'}}), \forall {\cal S} \subseteq {\cal K}',
		\end{array} \right.
	\end{equation}
	where ${\overline {{\cal K}'}} = {\cal K} \setminus {\cal K}'$, ${\overline {\cal S}} = {\cal K}' \setminus {\cal S}$, and ${\overline {\cal T}} = {\overline {{\cal K}'}} \setminus {\cal T}$.
\end{theorem}
\itshape \textbf{Proof:} \upshape
In the following Lemma~\ref{theorem_FM} and Appendix~\ref{Prove_theorem_FM}, we give a special case of Theorem~\ref{lemma_FM_gene_K2} with ${\cal K}' = {\cal K}$ and its proof. 
Then, we prove Theorem~\ref{lemma_FM_gene_K2} in Appendix~\ref{Prove_lemma_FM_gene_K2} by using Lemma~\ref{theorem_FM}.
\hfill $\Box$

For any ${\cal K}' \subseteq {\cal K}$, using the chain rule and non-negativity of mutual information, we have
\begin{align}\label{relax12}
	I(X_{\cal S}, X_{\cal T}; Y| X_{\overline {\cal S}}, X_{\overline {\cal T}}) & = I(X_{\cal S}; Y| X_{\overline {\cal S}}, X_{\overline {{\cal K}'}}) + I(X_{\cal T}; Y| X_{\overline {\cal S}}, X_{\overline {\cal T}}) \nonumber\\
	& \geq I(X_{\cal S}; Y| X_{\overline {\cal S}}, X_{\overline {{\cal K}'}}), \forall {\cal S} \subseteq {\cal K}', {\cal T} \subseteq {\overline {{\cal K}'}}, \nonumber\\
	I(X_{\cal S}; Z| X_{\overline {{\cal K}'}}) & = I(X_{{\cal S}'}, X_{{\cal S} \setminus {\cal S}'}; Z| X_{\overline {{\cal K}'}}) \nonumber\\
	& = I(X_{{\cal S}'}; Z| X_{\overline {{\cal K}'}}) + I(X_{{\cal S} \setminus {\cal S}'}; Z| X_{{\cal S}'}, X_{\overline {{\cal K}'}}) \nonumber\\
	& \geq I(X_{{\cal S}'}; Z| X_{\overline {{\cal K}'}}), \forall {\cal S} \subseteq {\cal K}', {\cal S}' \subseteq {\cal S}.
\end{align}
If (\ref{cond1}) holds, due to (\ref{relax12}), we have
\begin{equation}\label{cond2}
	I(X_{\cal S}, X_{\cal T}; Y| X_{\overline {\cal S}}, X_{\overline {\cal T}}) - I(X_{{\cal S}'}; Z| X_{\overline {{\cal K}'}}) \geq 0, \forall {\cal S} \subseteq {\cal K}', {\cal S}' \subseteq {\cal S}, {\cal T} \subseteq {\overline {{\cal K}'}},
\end{equation}
i.e., each upper bound in (\ref{region_DM0}) is non-negative.
On the other hand, by setting ${\cal S}' = {\cal S}$ and ${\cal T} = \emptyset$ in (\ref{cond2}), we can obtain (\ref{cond1}) directly from (\ref{cond2}).
Hence, (\ref{cond1}) and (\ref{cond2}) are actually equivalent.

For each ${\cal K}' \subseteq {\cal K}$, Theorem~\ref{lemma_FM_gene_K2} gives the existence conditions of $R_k^{\text a}, \forall k \in {\cal K}'$ in (\ref{cond1}) and (\ref{region_DM0}).
To apply the coding scheme provided in the next section, it is also necessary to know how to obtain $R_k^{\text a}, \forall k \in {\cal K}'$.
This can be done efficiently as follows. For a given rate point $(R_1^{\text s}, R_1^{\text o},\cdots, R_K^{\text s}, R_K^{\text o})$, if (\ref{cond1}) and (\ref{region_DM0}) can be satisfied, the linear inequalities in (\ref{region_FM2}) define a polytope as a feasible region of $R_k^{\text a}, \forall k \in {\cal K}'$.
Then, we may apply Dantzig's simplex algorithm to obtain $R_k^{\text a}, \forall k \in {\cal K}'$ \cite{gass2003linear}.

Now we consider the special case of Theorem~\ref{lemma_FM_gene_K2} with ${\cal K}' = {\cal K}$.
The basic result is collected in the following lemma.
\begin{lemma}\label{theorem_FM}
	Let $(X_{\cal K}, Y, Z) \sim \prod_{k=1}^K P_{X_k} P_{Y, Z| X_{\cal K}}$.
	If 
	\begin{equation}\label{cond3}
		I(X_{\cal S}; Y| X_{\overline {\cal S}}) - I(X_{\cal S}; Z) \geq 0, \forall {\cal S} \subseteq {\cal K},
	\end{equation}
	then, for any rate-tuple $(R_1^{\text s}, R_1^{\text o},\cdots, R_K^{\text s}, R_K^{\text o})$ satisfying 
	\begin{align}\label{rate_region0}
		\sum_{k \in \cal S} R_k^{\text s} + \sum_{k \in {\cal S} \setminus {\cal S}'} R_k^{\text o} \leq I(X_{\cal S}; Y| X_{\overline {\cal S}}) - I(X_{{\cal S}'}; Z), \forall {\cal S} \subseteq {\cal K}, {\cal S}' \subseteq {\cal S},
	\end{align}
	there exist $R_k^{\text a}, \forall k \in {\cal K}$ such that
	\begin{equation}\label{region_FM1}
		\left\{\!\!\!
		\begin{array}{ll}
			R_k^{\text a} \geq 0, \forall k \in {\cal K}, \\
			\sum\limits_{k \in {\cal S}} (R_k^{\text s} + R_k^{\text o} + R_k^{\text a}) \leq I(X_{\cal S}; Y| X_{\overline {\cal S}}), \forall {\cal S} \subseteq {\cal K}, \\
			\sum\limits_{k \in {\cal S}} (R_k^{\text o} + R_k^{\text a}) \geq I(X_{\cal S}; Z), \forall {\cal S} \subseteq {\cal K},
		\end{array} \right.
	\end{equation}
	where ${\overline {\cal S}} = {\cal K} \setminus {\cal S}$.
\end{lemma}
\itshape \textbf{Proof:} \upshape
See Appendix~\ref{Prove_theorem_FM}.
\hfill $\Box$

\begin{remark}\label{remark_FM}
	Lemma~\ref{theorem_FM} extends \cite[Lemma~$7$]{xu2022achievable} from a two-user case to the general $K$-user case.
	For the system with a small $K$, e.g., $K = 1$ or $K = 2$, Lemma~\ref{theorem_FM} can be proven by eliminating $R_k^{\text a}$ in (\ref{region_FM1}) using directly the Fourier-Motzkin procedure \cite[Appendix D]{el2011network} and showing that (\ref{rate_region0}) is the projection of (\ref{region_FM1}) onto the hyperplane $\{ R_k^{\text a} = 0, \forall k \in {\cal K}\}$.
	However, when $K$ increases, the number of inequalities resulted in the elimination procedure grows very quickly (doubly exponentially), making it quite difficult or even impractical to prove this lemma by following this brute-force way.
Besides the extremely high complexity, another problem of the direct elimination strategy is that it can be applied in principle only if the number of users $K$ is explicitly given (e.g., $K = 1, 2, 3, \ldots$) so that all inequalities can be explicitly listed. 
This makes the direct proof approach inappropriate for Lemma~\ref{theorem_FM}, since it is stated for a generic $K$.

In the research note \cite{xu2022note}, we considered a system with $K = 3$ and proved Lemma~\ref{theorem_FM} by eliminating $R_1^{\text a}$, $R_2^{\text a}$, and $R_3^{\text a}$ one by one.
From (\ref{region_FM1}) we first got $4$ upper bounds and $5$ lower bounds on $R_1^{\text a}$.
By pairing up these lower and upper bounds, we got $20$ inequalities, based on which $8$ upper bounds and $7$ lower bounds on $R_2^{\text a}$ were obtained.
We then eliminated $R_2^{\text a}$ and got $56$ inequalities, most of which are redundant.
Neglecting the redundant terms, we further got $9$ upper bounds and $7$ lower bounds on $R_3^{\text a}$, and $63$ inequalities by pairing them up.
Neglecting the redundant terms, we showed that the remaining inequalities construct (\ref{rate_region0}).
This unpublished note is mentioned here and made public in \cite{xu2022note} to illustrate how difficult the brute-force Fourier-Motzkin elimination is, even in the simple case of $3$ users. \hfill $\lozenge$
\end{remark}

This paper employs the strong secrecy metric and proves the achievability based on the resolvability theory of the MAC channel.
The notion of channel resolvability was introduced in \cite{han1993approximation} to approximate the output distribution of single-user channels by simulating an input process.
Then, in \cite{steinberg1998resolvability} and \cite{oohama2013converse}, it was further developed to study the randomness needed for approximating the output distribution and the identification capacity for two-user MAC channels, respectively.
The resolvability theory has also been shown to be a powerful tool in proving strong secrecy in wiretap channels \cite{bloch2011achieving, bloch2013strong, yassaee2010multiple, helal2020resolvability}.
Specifically, in \cite{bloch2013strong} and \cite{bloch2011achieving}, channel resolvability was leveraged to establish the secrecy-capacity region for single-user wiretap channels under the strong secrecy metric.
In \cite{yassaee2010multiple} and \cite{helal2020resolvability}, achievable regions of two-user wiretap channels under different system settings and the strong secrecy metric were studied based on channel resolvability.
In this paper, we show that the resolvability theory can also be applied to prove the strong secrecy for the considered system, where the users transmit both secret and open messages.
To this end, we first extend the resolvability result in \cite{yassaee2010multiple} from a two-user case to the $K$-user DM-MAC channel (not necessarily a wiretap channel) in Theorem~\ref{theo_varia_dist}.

In particular, we consider a standard DM-MAC channel (not a wiretap channel) with $K$ users and a receiver.
Each user $k$ has a message $M_k$ at rate $Q_k$, and $M_k$ is uniformly distributed over ${\cal M}_k = \left[1:2^{n Q_k}\right]$.
User $k$ generates a codebook ${\mathpzc c}_k$ by randomly and independently generating $2^{n Q_k}$ sequences $x_k^n(m_k), \forall m_k \in {\cal M}_k$, each according to $\prod_{i=1}^n P_{X_k} (x_{ki})$.
Then, for a given message $m_k \in {\cal M}_k$, user $k$ transmits codeword $x_k^n(m_k)$ over the channel $P_{Z |X_{\cal K}}$.\footnote{Note that the coding scheme considered here is instrumental to prove Theorem~\ref{theo_varia_dist}, 
and it is different from that in Section~\ref{achie_proof}.}
The observation of the receiver is $Z^n$.
Let ${\cal C}_k$ denote the random choice of codebook ${\mathpzc c}_k$ and define the following conditional output distribution
\begin{align}\label{p1}
	P_{Z^n} (\cdot| {\cal C}_{\cal K}) & = \sum_{m_{\cal K} \in \prod_{k \in {\cal K}} {\cal M}_k} P_{M_{\cal K}} (m_{\cal K}) P_{Z^n} (\cdot| {\cal C}_{\cal K}, m_{\cal K}) \nonumber\\
	& = 2^{- n \sum_{k \in {\cal K}} Q_k } \sum_{m_{\cal K} \in \prod_{k \in {\cal K}} {\cal M}_k} P_{Z^n} \left(\cdot| \left\{ X_k^n (m_k) \right\}_{k \in {\cal K}} \right).
\end{align}
In addition, for a given ${\cal K}' \subseteq {\cal K}$, define
\begin{equation}\label{p2}
	P_{Z^n} (\cdot| {\cal C}_{\overline {{\cal K}'}}) = 2^{- n \sum_{k \in {\overline {{\cal K}'}}} Q_k } \sum_{m_{\overline {{\cal K}'}} \in \prod_{k \in {\overline {{\cal K}'}}} {\cal M}_k} P_{Z^n} \left(\cdot| \left\{ X_k^n (m_k) \right\}_{k \in {\overline {{\cal K}'}}} \right),
\end{equation}
where
\begin{equation}
	P_{Z^n} \left(\cdot| \left\{ X_k^n (m_k) \right\}_{k \in {\overline {{\cal K}'}}} \right) = \sum_{x_{{\cal K}'}^n \in \prod_{k \in {\cal K}'} {\cal X}_k^n} P_{X_{{\cal K}'}^n} (x_{{\cal K}'}^n) P_{Z^n} \left(\cdot| x_{{\cal K}'}^n, \left\{ X_k^n (m_k) \right\}_{k \in {\overline {{\cal K}'}}} \right).
\end{equation}
The following theorem shows that when certain conditions are satisfied, using the coding scheme provided above, the average variational distance between the output statistics $P_{Z^n} (\cdot| {\cal C}_{\cal K})$ and $P_{Z^n} (\cdot| {\cal C}_{\overline {{\cal K}'}})$ vanishes exponentially in $n$.
\begin{theorem}\label{theo_varia_dist}
	For given distribution $\prod_{k=1}^K P_{X_k} P_{Z| X_{\cal K}}$ and subset ${\cal K}' \subseteq {\cal K}$, if
	\begin{align}\label{condi}
		Q_k & = 0, \forall k \in {\overline {{\cal K}'}}, \nonumber\\
		\sum_{k \in {\cal S}} Q_k & > I(X_{\cal S}; Z| X_{\overline {{\cal K}'}}), \forall {\cal S} \subseteq {\cal K}', {\cal S} \neq \emptyset,
	\end{align}
	using the above coding scheme, there exists $\varepsilon > 0$ such that
	\begin{equation}\label{exp_vari_dist}
		{\mathbb E} \left\| P_{Z^n} (\cdot| {\cal C}_{\cal K}) - P_{Z^n} (\cdot| {\cal C}_{\overline {{\cal K}'}}) \right\|_1 \leq e^{-n \varepsilon},
	\end{equation}
	where the expectation is taken over the random codebooks.
\end{theorem}
\itshape \textbf{Proof:} \upshape
See Appendix \ref{prove_theo_varia_dist}.
\hfill $\Box$

By a standard random coding argument,  (\ref{exp_vari_dist}) implies that, if a rate-tuple $(Q_1, \cdots, Q_K)$ satisfies (\ref{condi}), there must exist codebooks ${\mathpzc c}_{\cal K}$ such that
\begin{equation}\label{resol}
	\left\| P_{Z^n} (\cdot| {\mathpzc c}_{\cal K}) - P_{Z^n} (\cdot| {\mathpzc c}_{\overline {{\cal K}'}}) \right\|_1 \leq e^{-n \varepsilon}.
\end{equation}
Therefore, the region defined by (\ref{condi}) is the resolvability region of the DM-MAC channel with input process $X_{\cal K} \sim \prod_k P_{X_k}$ \cite{frey2018mac}.
Theorem~\ref{theo_varia_dist} shows that if a rate point is in the resolvability region, as the block length $n$ goes to infinity, the induced output distribution $P_{Z^n} (\cdot| {\cal C}_{\cal K})$ conditioned on the codebooks approaches $P_{Z^n} (\cdot| {\cal C}_{\overline {{\cal K}'}})$, which is induced by the knowledge of some codebooks and randomly guessing (with the given distribution $\prod_k P_{X_k}$) the other input $ x_{{\cal K}'}^n$ over the whole alphabets $\prod_{k \in {\cal K}'} {\cal X}_k^n$.
Theorem~\ref{theo_varia_dist} applied to the case $K = 2$ and ${\cal K}' = \{1, 2\}$ coincides with  \cite[Theorem~$2$]{yassaee2010multiple}.

\subsection{Achievable Regions}
\label{achi_region_DM}

In the following theorem we provide the new achievable region for the considered DM MAC-WT channel with both secret and open messages.
\begin{theorem}\label{lemma_DM_exten}
	For given distribution $\prod_{k=1}^K P_{X_k} P_{Z| X_{\cal K}}$ and subset $ {\cal K}' \subseteq {\cal K}$, any rate-tuple $(R_1^{\text s}, R_1^{\text o},$ $\cdots, R_K^{\text s}, R_K^{\text o})$ satisfying
	\begin{equation}\label{region_DM_exten}
	\left\{
	\begin{array}{ll}
	R_k^{\text s} = 0, \forall k \in {\overline {{\cal K}'}}, \\
	\sum\limits_{k \in \cal S} R_k^{\text s} + \sum\limits_{k \in {\cal S} \setminus {\cal S}'} R_k^{\text o} + \sum\limits_{k \in {\cal T}} R_k^{\text o}	\leq \left[ I(X_{\cal S}, X_{\cal T}; Y| X_{\overline {\cal S}}, X_{\overline {\cal T}}) - I(X_{{\cal S}'}; Z| X_{\overline {{\cal K}'}}) \right]^+, \\
	\quad\quad\quad\quad\quad\quad\quad\quad\quad\quad\quad\quad \forall {\cal S} \subseteq {\cal K}', {\cal S}' \subseteq {\cal S}, {\cal T} \subseteq {\overline {{\cal K}'}},
	\end{array} \right.
	\end{equation}
	is achievable, where $\overline {\cal S}$, $\overline {\cal T}$, and $\overline {{\cal K}'}$ are defined in (\ref{region_FM2}).
	Let ${\mathscr R} (X_{\cal K}, {\cal K}')$ denote the set of rate-tuples satisfying (\ref{region_DM_exten}).
	Then, the convex hull of the union of ${\mathscr R} (X_{\cal K}, {\cal K}')$ over all $\prod_{k=1}^K P_{X_k}$ and $ {\cal K}' \subseteq {\cal K}$ is an achievable rate region of the DM MAC-WT channel.
\end{theorem}
\itshape \textbf{Proof:} \upshape
The proof is provided in Section~\ref{achie_proof}.
\hfill $\Box$

It can be found that \cite[Lemma~$1$]{xu2022achievable} is a special case of Theorem~\ref{lemma_DM_exten} with ${\cal K}' = {\cal K}$.
In the following remark, we explain why Theorem~\ref{lemma_DM_exten} can improve \cite[Lemma~$1$]{xu2022achievable}.

\begin{remark}\label{remark_lemma_DM_exten}
	The partitioning of ${\cal K}$ into ${\cal K}'$ and ${\overline {{\cal K}'}}$ is very important in determining the achievable region, where users in ${\cal K}'$ have zero secrecy rate.
	We observe that if a user has no secret message, it can play different roles and has two options: 1) introducing an additional auxiliary message such that the sum rate of its messages is beyond Eve's decoding capability; 
	2) simply encoding and transmitting its open message as in a standard MAC channel with no wiretapping. 
	The achievability proof in the next section shows that both these two options have advantages and disadvantages in determining the achievable regions. 
	Using the first option, the signal of the user plays as noise to Eve and can thus weaken its wiretapping capability.
	However, more constraints are imposed on the message rate, which is detrimental to determining the achievable region.
	The advantage and disadvantage of the second option are exactly the opposite, i.e., fewer constraints are imposed, but Eve has a stronger wiretapping capability.
    Therefore, to obtain a larger achievable region, all $2^K$ possible partitions ${\cal K} = {\cal K}' \cup {\overline {{\cal K}'}}$ should be taken into account.
    In \cite{xu2022achievable}, only the ${\cal K}' = {\cal K}$ case was considered.
    Hence, Theorem~\ref{lemma_DM_exten} improves the region in \cite[Lemma~$1$]{xu2022achievable}.
    We will further verify the improvement in Section~\ref{BAC} by considering a specific example. 
    $\phantom{x}$	 \hfill $\lozenge$
\end{remark}

Now we consider two special cases of Theorem~\ref{lemma_DM_exten} with respectively $R_k^{\text s} = 0, \forall k \in {\cal K}$ and $R_k^{\text o} = 0, \forall k \in {\cal K}$.
First, if $R_k^{\text s} = 0, \forall k \in {\cal K}$, each user has only one open message, which does not interest Eve.
The system can then be seen as a normal DM-MAC channel with no wiretapping.
As expected, in this case the region given in Theorem~\ref{lemma_DM_exten} reduces to the capacity region of the standard $K$-user DM-MAC channel with no evesropping \cite[Chapter~4]{el2011network} (the proof is omitted for brevity).

Next, we consider the case with $R_k^{\text o} = 0, \forall k \in {\cal K}$.
In this case, the system reduces to the conventional DM MAC-WT channel with only secret messages.
As anticipated in Section \ref{section1}, our results extend those in \cite{yassaee2010multiple} and \cite{Hayashi2019secrecy}. If $R_k^{\text o} = 0, \forall k \in {\cal K}$, (\ref{region_DM_exten}) becomes
\begin{equation}\label{region_DM_exten_1}
\left\{
\begin{array}{ll}
R_k^{\text s} = 0, \forall k \in {\overline {{\cal K}'}}, \\
\sum\limits_{k \in \cal S} R_k^{\text s} \leq \left[I(X_{\cal S}, X_{\cal T}; Y| X_{\overline {\cal S}}, X_{\overline {\cal T}}) - I(X_{{\cal S}'}; Z| X_{\overline {{\cal K}'}}) \right]^+, \forall {\cal S} \subseteq {\cal K}', {\cal S}' \subseteq {\cal S}, {\cal T} \subseteq {\overline {{\cal K}'}},
\end{array} \right.
\end{equation}
which can be simplified as (\ref{region_secrecy_only2}) due to (\ref{relax12}).
Then, we give the following lemma.
\begin{lemma}\label{achi_secrecy_only2}
	Let $R_k^{\text o} = 0, \forall k \in {\cal K}$. For any ${\cal K}' \subseteq {\cal K}$, any rate-tuple $(R_1^{\text s}, \cdots, R_K^{\text s})$ satisfying 
	\begin{equation}\label{region_secrecy_only2}
	\left\{\!\!\!
	\begin{array}{ll}
	R_k^{\text s} = 0, \forall k \in {\overline {{\cal K}'}}, \\
	\sum\limits_{k \in \cal S} R_k^{\text s} \leq \left[I(X_{\cal S}; Y| X_{\overline {\cal S}}, X_{\overline {{\cal K}'}}) - I(X_{\cal S}; Z| X_{\overline {{\cal K}'}}) \right]^+, \forall {\cal S} \subseteq {\cal K}',
	\end{array} \right.
	\end{equation}
	is achievable.
	Let ${\mathscr R}^{\text s} (X_{\cal K}, {\cal K}')$ denote the set of rate-tuples satisfying (\ref{region_secrecy_only2}).
	Then, the convex hull of the union of ${\mathscr R}^{\text s} (X_{\cal K}, {\cal K}')$ over all $\prod_{k=1}^K P_{X_k}$ and $ {\cal K}' \subseteq {\cal K}$ is an achievable secrecy rate region of the DM MAC-WT channel with only secret messages. \hfill $\Box$
\end{lemma}

\begin{remark}\label{remark3}
	It can be seen that \cite[Theorem~$1$]{yassaee2010multiple} and \cite[Theorem~$1$]{Hayashi2019secrecy} are special cases of Lemma~\ref{achi_secrecy_only2} by respectively setting $K = 2$ and considering only ${\cal K}' = {\cal K}$.
	As shown in the next section, for a given ${\cal K}'$, we introduce auxiliary messages, whose rates satisfy (\ref{region_FM2}), and then prove the achievability of ${\mathscr R} (X_{\cal K}, {\cal K}')$ by providing a coding scheme.
	However, (\ref{region_FM2}) can only be satisfied if (\ref{cond1}) is true.
	If it is not, the proof is no longer valid.
	As a matter of fact, this problem also exists in \cite{yassaee2010multiple} and \cite{Hayashi2019secrecy} (by respectively checking \cite[(7), (20)]{yassaee2010multiple} and \cite[(11), (13)]{Hayashi2019secrecy}), but was not considered, making the proof incomplete.
	In the next section we prove that if (\ref{cond1}) is not true, there always exists ${\cal K}'' \subsetneqq {\cal K}'$ such that (\ref{cond1}) becomes true for the reduced set ${\cal K}''$ and ${\mathscr R} (X_{\cal K}, {\cal K}')$ is included in ${\mathscr R} (X_{\cal K}, {\cal K}'')$, which can then be proved to be achievable.
	Therefore, this paper not only generalizes the results given by \cite{yassaee2010multiple} and \cite{Hayashi2019secrecy}, but also ``completes'' the proofs in these works. 
	\hfill $\lozenge$
\end{remark}

This paper studies wiretap channels.
Hence, we are especially concerned about the maximum achievable sum secrecy rate of the system.
In addition, apart from the secret message, each user also has an open message intended for Bob.
Then, an interesting question is if all users transmit their confidential information at the maximum sum secrecy rate, what is the maximum sum rate at which they could encode their open messages.
We give the answer in the following Theorem.
\begin{theorem}\label{max_R_s_joint}
	For the considered DM MAC-WT channel and given $\prod_{k=1}^K P_{X_k} P_{Z| X_{\cal K}}$, the maximum achievable sum secrecy rate $\sum_{k \in {\cal K}} R_k^{\text s}$ is
	\begin{equation}\label{R_s_joint_DM}
	R^{\text s} (X_{\cal K}) = \mathop {\max }\limits_{{\cal K}' \subseteq {\cal K}} \left\{ \left[I(X_{{\cal K}'}; Y| X_{\overline {{\cal K}'}}) - I(X_{{\cal K}'}; Z| X_{\overline {{\cal K}'}})\right]^+ \right\}.
	\end{equation}
	Let ${\cal K}'^*$ denote the subset in ${\cal K}$ which achieves (\ref{R_s_joint_DM}) and $ {\overline {{\cal K}'^*}} = {\cal K} \setminus {\cal K}'^*$.
	If $R^{\text s} (X_{\cal K}) > 0$ and all users transmit their confidential messages at sum rate $R^{\text s} (X_{\cal K})$, the maximum achievable sum rate at which users in ${\cal K}$ could send their open messages is given by\footnote{Here we assume $R^{\text s} (X_{\cal K}) > 0$ since otherwise we have $R_k^{\text s} = 0, \forall k \in {\cal K}$, i.e., the system reduces to a standard DM-MAC channel with no wiretapping.}
	\begin{equation} \label{R_o_K1_DM}
	R^{\text o} (X_{\cal K}) = I(X_{\overline {{\cal K}'^*}}; Y) + I(X_{{\cal K}'^*}; Z| X_{\overline {{\cal K}'^*}}).
	\end{equation}
\end{theorem}
\itshape \textbf{Proof:} \upshape
See Appendix \ref{Prove_max_R_s_joint}.
\hfill $\Box$

Theorem~\ref{max_R_s_joint} shows that the channel can support a non-trivial additional open sum rate even if the coding scheme is designed to maximize the sum secrecy rate.

\section{Achievability Proof}
\label{achie_proof}

In this section, we prove Theorem~\ref{lemma_DM_exten}.
We start from a special case with ${\cal K}' = \emptyset$ and ${\overline {{\cal K}'}} = {\cal K}$.
In this case, (\ref{region_DM_exten}) becomes
\begin{equation}
	\left\{
	\begin{array}{ll}
		R_k^{\text s} = 0, \forall k \in {\cal K}, \\
		\sum_{k \in {\cal T}} R_k^{\text o} \leq I (X_{\cal T}; Y| X_{\overline {\cal T}}), \forall {\cal T} \subseteq {\cal K}.
	\end{array} \right.
\end{equation}
The region ${\mathscr R} (X_{\cal K}, \emptyset)$ defined above is included in the capacity region of a standard DM-MAC channel with no wiretapping and its achievability proof is well known.
One may refer to \cite[Chapter~4]{el2011network} for the detailed proof.

Next, we prove the achievability of ${\mathscr R} (X_{\cal K}, {\cal K}')$ for any non-empty subset ${\cal K}' \subseteq {\cal K}$. 
Without loss of generality (w.l.o.g.), we always assume
\begin{equation}\label{assumption}
	I(X_{\cal S}; Y| X_{\overline {\cal S}}) > 0, \forall {\cal S} \subseteq {\cal K}, {\cal S} \neq \emptyset,
\end{equation}
since otherwise users in ${\cal S}$ cannot communicate with Bob.
Moreover, we assume
\begin{equation}\label{assumption_0}
	I(X_{\cal S}; Y| X_{\overline {\cal S}}, X_{\overline {{\cal K}'}}) - I(X_{\cal S}; Z| X_{\overline {{\cal K}'}}) > 0, \forall {\cal S} \subseteq {\cal K}', {\cal S} \neq \emptyset,
\end{equation}
which is (\ref{cond1}) with strict ``$>$''.
If (\ref{assumption_0}) can be satisfied, Theorem~\ref{lemma_FM_gene_K2} can be applied for the achievability proof.
Otherwise, we show later that the achievability could be proven by modifying the proof steps.

\subsection{Achievability Proof When (\ref{assumption_0}) Holds}
\label{assump_true}

In this subsection, we show that with (\ref{assumption_0}), any rate-tuple inside ${\mathscr R} (X_{\cal K}, {\cal K}')$ is achievable.
This, together with the standard time-sharing over coding strategies, suffices to prove the achievability of ${\mathscr R} (X_{\cal K}, {\cal K}')$. 
If (\ref{assumption_0}) can be satisfied, due to (\ref{relax12}), we have
\begin{align}\label{assumption_1}
	I(X_{\cal S}, X_{\cal T}; Y| X_{\overline {\cal S}}, X_{\overline {\cal T}}) - I(X_{{\cal S}'}; Z| X_{\overline {{\cal K}'}}) > 0, \forall {\cal S} \subseteq {\cal K}', {\cal S}' \subseteq {\cal S}, {\cal T} \subseteq {\overline {{\cal K}'}}, {\cal S} \neq \emptyset.
\end{align}
Then, the rate-tuples inside region ${\mathscr R} (X_{\cal K}, {\cal K}')$ satisfy
\begin{equation}\label{rate_region1}
	\left\{
	\begin{array}{ll}
		R_k^{\text s} = 0, \forall k \in {\overline {{\cal K}'}}, \\
		\sum\limits_{k \in \cal S} R_k^{\text s} + \sum\limits_{k \in {\cal S} \setminus {\cal S}'} R_k^{\text o} + \sum\limits_{k \in {\cal T}} R_k^{\text o} & < I(X_{\cal S}, X_{\cal T}; Y| X_{\overline {\cal S}}, X_{\overline {\cal T}}) - I(X_{{\cal S}'}; Z| X_{\overline {{\cal K}'}}) - \epsilon,\\
		& \forall {\cal S} \subseteq {\cal K}', {\cal S}' \subseteq {\cal S}, {\cal T} \subseteq {\overline {{\cal K}'}}, {\cal S} \cup {\cal T} \neq \emptyset,
	\end{array} \right.
\end{equation}
where $\epsilon$ is an arbitrarily small positive number. 
For a given rate-tuple $\left(R_1^{\text s}, R_1^{\text o},\cdots, R_K^{\text s}, R_K^{\text o}\right)$ satisfying (\ref{rate_region1}), it is known from Theorem~\ref{lemma_FM_gene_K2} that there exist $R_k^{\text a}, \forall k \in {\cal K}'$ such that
\begin{equation}\label{region_FM3}
	\left\{\!\!
	\begin{array}{ll}
		R_k^{\text a} \geq 0, \forall k \in {\cal K}', \\
		\sum\limits_{k \in {\cal S}} (R_k^{\text s} \!+\! R_k^{\text o} \!+\! R_k^{\text a}) \!+\! \sum\limits_{k \in {\cal T}} R_k^{\text o} \!<\! I(X_{\cal S}, X_{\cal T}; Y| X_{\overline {\cal S}}, X_{\overline {\cal T}}) \!-\! \epsilon, \forall {\cal S} \!\subseteq\! {\cal K}', {\cal T} \!\subseteq\! {\overline {{\cal K}'}}, {\cal S} \cup {\cal T} \!\neq\! \emptyset, \\
		\sum\limits_{k \in {\cal S}} (R_k^{\text o} + R_k^{\text a}) > I(X_{\cal S}; Z| X_{\overline {{\cal K}'}}), \forall {\cal S} \subseteq {\cal K}', {\cal S} \neq \emptyset,
	\end{array} \right.
\end{equation}
and $R_k^{\text a}, \forall k \in {\cal K}'$ can be found by applying Dantzig's simplex algorithm \cite{gass2003linear}.
Now we provide a coding scheme and show that the metrics in Definition~\ref{defi_achi} can be satisfied.

\subsubsection{Coding Scheme}
\label{Prove_theorem1_A}
Assume w.l.o.g. that $2^{n R_k^{\text s}}$, $2^{n (R_k^{\text o} + R_k^{\text a})}$, $2^{n R_k^{\text a}}, \forall k \in {\cal K}'$, and $2^{n R_k^{\text o}}, \forall k \in {\overline {{\cal K}'}}$, are integers.

\begin{figure*}[ht]
	\centering
	\includegraphics[scale=1]{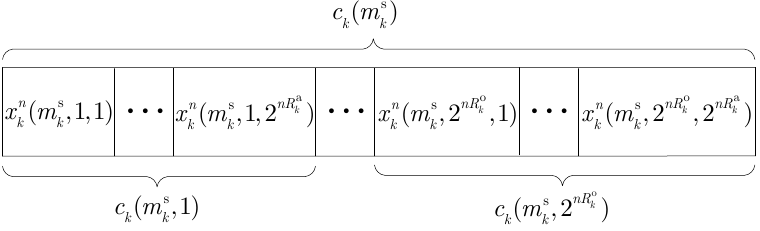}
	\caption{A division of subcodebook ${\mathpzc c}_k (m_k^{\text s})$ of user $k \in {\cal K}'$.}
	\label{Subcodebook}
\end{figure*}
\begin{figure}
	\centering
	\includegraphics[scale=1]{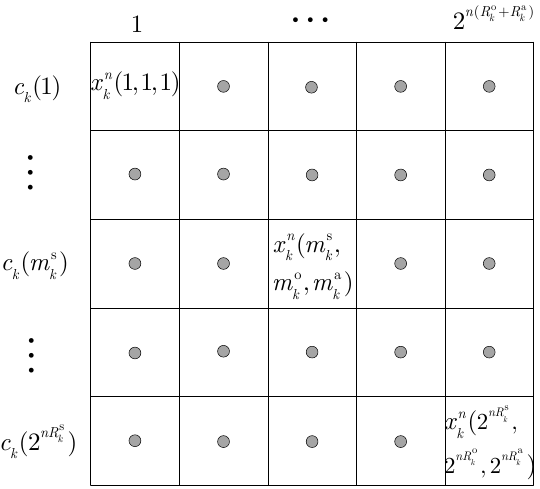}
	\caption{Codebook ${\mathpzc c}_k$ of user $k \in {\cal K}'$.}
	\label{Codebook}
\end{figure}

{\textbf {Codebook generation.}}
For each message pair $(m_k^{\text s}, m_k^{\text o}) \in {\cal M}_k^{\text s} \times {\cal M}_k^{\text o}$ of user $k \in {\cal K}'$, generate a sub-subcodebook ${\mathpzc c}_k (m_k^{\text s}, m_k^{\text o})$ by randomly and independently generating $2^{n R_k^{\text a}}$ sequences $x_k^n(m_k^{\text s}, m_k^{\text o}, m_k^{\text a}), \forall m_k^{\text a} \in {\cal M}_k^{\text a}$, each according to $\prod_{i=1}^n P_{X_k} (x_{ki})$. 
For a given secret message $m_k^{\text s}$, the sub-subcodebooks for all open messages constitute subcodebook ${\mathpzc c}_k (m_k^{\text s})$, i.e., ${\mathpzc c}_k (m_k^{\text s}) = \bigcup_{m_k^{\text o} \in {\cal M}_k^{\text o}} {\mathpzc c}_k (m_k^{\text s}, m_k^{\text o})$.
Fig.~\ref{Subcodebook} gives an example of subcodebook ${\mathpzc c}_k (m_k^{\text s})$.
Then, as shown in Fig. \ref{Codebook}, these subcodebooks constitute the codebook of user $k$, i.e., ${\mathpzc c}_k = \bigcup_{m_k^{\text s} \in {\cal M}_k^{\text s}} {\mathpzc c}_k (m_k^{\text s})$.
For each user $k$ in ${\overline {{\cal K}'}}$, we apply the random coding scheme used in the standard MAC channel with no wiretapping.
In particular, user $k \in {\overline {{\cal K}'}}$ generate its codebook ${\mathpzc c}_k$ by randomly and independently generating $2^{n R_k^{\text o}}$ sequences $x_k^n(m_k^{\text o}), \forall m_k^{\text o} \in {\cal M}_k^{\text o}$, each according to $\prod_{i=1}^n P_{X_k} (x_{ki})$.
The codebooks of all users are then revealed to all transmitters and receivers, including Eve.

{\textbf {Encoding.}} 
To send message pair $(m_k^{\text s}, m_k^{\text o}) \in {\cal M}_k^{\text s} \times {\cal M}_k^{\text o}$, encoder $k \in {\cal K}'$ uniformly chooses a codeword (with index $m_k^{\text a}$) from sub-subcodebook ${\mathpzc c}_k (m_k^{\text s},m_k^{\text o})$ and then transmits $x_k^n (m_k^{\text s}, m_k^{\text o}, m_k^{\text a})$.
To send message $ m_k^{\text o} \in {\cal M}_k^{\text o}$, encoder $k\in {\overline {{\cal K}'}}$ transmits $x_k^n (m_k^{\text o})$.

{\textbf {Decoding.}} 
The decoder at Bob uses joint typicality decoding to find an estimate of the messages and declares that $( {\hat m}_{{\cal K}'}^{\text s}, {\hat m}_{{\cal K}'}^{\text o}, {\hat m}_{\overline {{\cal K}'}}^{\text o} )$ is sent if there exists ${\hat m}_{{\cal K}'}^{\text a} \in \prod_{k \in {\cal K}'} {\cal M}_k^{\text a}$ such that $( {\hat m}_{{\cal K}'}^{\text s}, $ $ {\hat m}_{{\cal K}'}^{\text o}, {\hat m}_{{\cal K}'}^{\text a}, {\hat m}_{\overline {{\cal K}'}}^{\text o} )$ is the unique message-tuple satisfying $( \{ x_k^n ( {\hat m}_k^{\text s}, {\hat m}_k^{\text o}, {\hat m}_k^{\text a} ) \}_{k \in {\cal K}'}, \{ x_k^n ({\hat m}_k^{\text o}) \}_{k \in {\overline {{\cal K}'}}}, y^n )$ $\in {\cal T}_\epsilon^{(n)}( X_{\cal K}, Y)$. 

\subsubsection{Analysis of the Probability of Error}

Since users in ${\cal K}'$ and ${\overline {{\cal K}'}}$ respectively transmit their messages at rate $R_k = R_k^{\text s} + R_k^{\text o} + R_k^{\text a}$ and $R_k = R_k^{\text o}$, and (see (\ref{region_FM3}))
\begin{align}
	\sum\limits_{k \in {\cal S}} (R_k^{\text s} \!+\! R_k^{\text o} \!+\! R_k^{\text a}) \!+\! \sum\limits_{k \in {\cal T}} R_k^{\text o} < I(X_{\cal S}, X_{\cal T}; Y| X_{\overline {\cal S}}, X_{\overline {\cal T}}) \!-\! \epsilon, \forall {\cal S} \subseteq {\cal K}', {\cal T} \subseteq {\overline {{\cal K}'}}, {\cal S} \cup {\cal T} \neq \emptyset,
\end{align}
the rate-tuple $( R_1, \cdots, R_K )$ is inside the capacity region of the MAC channel from all users to Bob.
Then, it can be proven by using the law of large numbers (LLN) and the packing lemma that the average probability of error at Bob vanishes as $n$ goes to infinity, i.e., 
\begin{equation}\label{error0}
	\lim_{n \rightarrow \infty} {\text {Pr}} \left\{ \left( {\hat M}_{\cal K}^{\text s}, {\hat M}_{\cal K}^{\text o}, {\hat M}_{\cal K}^{\text a} \right) \neq \left( M_{\cal K}^{\text s}, M_{\cal K}^{\text o}, M_{\cal K}^{\text a} \right) \right\} = 0,
\end{equation}
Note that for notational convenience, in (\ref{error0}), every user has $M_k^{\text s}$, $M_k^{\text o}$, and $M_k^{\text a}$.
For users in ${\overline {{\cal K}'}}$, which have only open messages, $M_k^{\text s}$ and  $M_k^{\text a}$ are actually constants and do not affect the value of the probability term in (\ref{error0}).
The proof of (\ref{error0}) follows exactly the same steps used in \cite[Subsection 4.5.1]{el2011network} and is omitted here.
Since 
\begin{align}\label{error_relax}
	{\text {Pr}} \left\{ \left( {\hat M}_{\cal K}^{\text s}, {\hat M}_{\cal K}^{\text o}, {\hat M}_{\cal K}^{\text a} \right) \neq \left( M_{\cal K}^{\text s}, M_{\cal K}^{\text o}, M_{\cal K}^{\text a} \right) \right\} & = {\text {Pr}} \left\{ \left( {\hat M}_{\cal K}^{\text s}, {\hat M}_{\cal K}^{\text o} \right) \neq \left( M_{\cal K}^{\text s}, M_{\cal K}^{\text o} \right) ~{\text {or}}~ {\hat M}_{\cal K}^{\text a} \neq M_{\cal K}^{\text a} \right\} \nonumber\\
	& \geq {\text {Pr}} \left\{ \left( {\hat M}_{\cal K}^{\text s}, {\hat M}_{\cal K}^{\text o} \right) \neq \left( M_{\cal K}^{\text s}, M_{\cal K}^{\text o} \right) \right\},
\end{align}
we know that (\ref{Pe}) is true.

\subsubsection{Analysis of the Information Leakage}
\label{A-C}

Define the following total variational distance 
\begin{align}\label{vari_dist2}
	d ( {\cal C}_{\cal K}, m_{{\cal K}'}^{\text s}, m_{\overline {{\cal K}'}}^{\text o} ) \!=\! \left\| P_{Z^n} \!\left( \cdot| {\cal C}_{{\cal K}'}, \left\{ {\cal C}_k (m_k^{\text o}) \right\}_{k \in {\overline {{\cal K}'}}} \right) \!-\! P_{Z^n} \!\left( \cdot| \left\{ {\cal C}_k (m_k^{\text s}) \right\}_{k \in {\cal K}'}, \left\{ {\cal C}_k (m_k^{\text o}) \right\}_{k \in {\overline {{\cal K}'}}} \right) \right\|_1\!.\!\!\!
\end{align}
As shown below, we can get an upper bound on the expectation of $d ( {\cal C}_{\cal K}, m_{{\cal K}'}^{\text s}, m_{\overline {{\cal K}'}}^{\text o} )$
\begin{align}\label{vari_dist_ub2}
	& {\mathbb E} \left[ d ( {\cal C}_{\cal K}, m_{{\cal K}'}^{\text s}, m_{\overline {{\cal K}'}}^{\text o} ) \right] \leq {\mathbb E} \left\| P_{Z^n} \left( \cdot| {\cal C}_{{\cal K}'}, \left\{ {\cal C}_k (m_k^{\text o}) \right\}_{k \in {\overline {{\cal K}'}}} \right) - P_{Z^n} \left( \cdot| \left\{ {\cal C}_k (m_k^{\text o}) \right\}_{k \in {\overline {{\cal K}'}}} \right) \right\|_1 \nonumber\\
	& \quad\quad\quad\quad\quad\quad\quad\quad\quad + {\mathbb E} \left\| P_{Z^n} \left( \cdot| \left\{ {\cal C}_k (m_k^{\text s}) \right\}_{k \in {\cal K}'}, \left\{ {\cal C}_k (m_k^{\text o}) \right\}_{k \in {\overline {{\cal K}'}}} \right) \!-\! P_{Z^n} \left( \cdot| \left\{ {\cal C}_k (m_k^{\text o}) \right\}_{k \in {\overline {{\cal K}'}}} \right) \right\|_1 \nonumber\\
	& \leq 2^{- n \sum_{k \in {\cal K}'} R_k^{\text s} } \sum_{m_{{\cal K}'}^{\text s} \in \prod_{k \in {\cal K}'} {\cal M}_k^{\text s}}\!\!\! {\mathbb E} \left\| P_{Z^n} \left( \cdot| \left\{ {\cal C}_k (m_k^{\text s}) \right\}_{k \in {\cal K}'}, \left\{ {\cal C}_k (m_k^{\text o}) \right\}_{k \in {\overline {{\cal K}'}}} \right) \!-\! P_{Z^n} \left( \cdot| \left\{ {\cal C}_k (m_k^{\text o}) \right\}_{k \in {\overline {{\cal K}'}}} \right) \right\|_1 \nonumber\\
	& + {\mathbb E} \left\| P_{Z^n} \left( \cdot| \left\{ {\cal C}_k (m_k^{\text s}) \right\}_{k \in {\cal K}'}, \left\{ {\cal C}_k (m_k^{\text o}) \right\}_{k \in {\overline {{\cal K}'}}} \right) - P_{Z^n} \left( \cdot| \left\{ {\cal C}_k (m_k^{\text o}) \right\}_{k \in {\overline {{\cal K}'}}} \right) \right\|_1,
\end{align}
where the expectations are taken over the random codebooks, the first step follows from first introducing $P_{Z^n} \left( \cdot| \left\{ {\cal C}_k (m_k^{\text o}) \right\}_{k \in {\overline {{\cal K}'}}} \right)$ and then applying the triangular inequality, and the second step holds by computing $P_{Z^n} \left( \cdot| {\cal C}_{{\cal K}'}, \left\{ {\cal C}_k (m_k^{\text o}) \right\}_{k \in {\overline {{\cal K}'}}} \right)$ over all possible $m_{{\cal K}'}^{\text s}$ and also applying the triangular inequality. 
In addition, it is known from the coding scheme provided above that there are respectively $2^{n (R_k^{\text a} + R_k^{\text o})}$ codewords in ${\cal C}_k (m_k^{\text s}), \forall k \in {\cal K}'$ and one codeword in ${\cal C}_k (m_k^{\text o}), \forall k \in {\overline {{\cal K}'}}$.
Letting $Q_k = \log \left| {\cal C}_k (m_k^{\text o}) \right| = 0, \forall k \in {\overline {{\cal K}'}}$ and $Q_k = \log \left| {\cal C}_k (m_k^{\text s}) \right| = R_k^{\text o} + R_k^{\text a}, \forall k \in {\cal K}'$, since $\sum_{k \in {\cal S}} (R_k^{\text o} + R_k^{\text a}) > I(X_{\cal S}; Z| X_{\overline {{\cal K}'}}), \forall {\cal S} \subseteq {\cal K}', {\cal S} \neq \emptyset$ (see (\ref{region_FM3})), we know that (\ref{condi}) can be satisfied.
Theorem~\ref{theo_varia_dist} can then be applied and yields
\begin{align}\label{ineq7}
	& {\mathbb E} \left\| P_{Z^n} \left( \cdot| \left\{ {\cal C}_k (m_k^{\text s}) \right\}_{k \in {\cal K}'}, \left\{ {\cal C}_k (m_k^{\text o}) \right\}_{k \in {\overline {{\cal K}'}}} \right) - P_{Z^n} \left( \cdot| \left\{ {\cal C}_k (m_k^{\text o}) \right\}_{k \in {\overline {{\cal K}'}}} \right) \right\|_1 \nonumber\\
	\leq & e^{-n \varepsilon}, \forall m_{{\cal K}'}^{\text s} \in \prod_{k \in {\cal K}'} {\cal M}_k^{\text s}.
\end{align}
Based on (\ref{ineq7}), (\ref{vari_dist_ub2}) can be further upper bounded as follows
\begin{align}\label{vari_dist_ub3}
	{\mathbb E} \left[ d ( {\cal C}_{\cal K}, m_{{\cal K}'}^{\text s}, m_{\overline {{\cal K}'}}^{\text o} ) \right] \leq 2 e^{-n \varepsilon} \rightarrow 0.
\end{align}
Now we evaluate the information leakage over all codebooks as follows
\begin{align}\label{inf_leak_C2}
	& I \left( M_{{\cal K}'}^{\text s}; Z^n| M_{\overline {{\cal K}'}}^{\text o}, {\cal C}_{\cal K} \right) = H \left( Z^n| M_{\overline {{\cal K}'}}^{\text o}, {\cal C}_{\cal K} \right) - H \left( Z^n| M_{{\cal K}'}^{\text s}, M_{\overline {{\cal K}'}}^{\text o}, {\cal C}_{\cal K} \right) \nonumber\\
	& = H \left( Z^n| {\cal C}_{{\cal K}'}, \left\{ {\cal C}_k (M_k^{\text o}) \right\}_{k \in {\overline {{\cal K}'}}} \right) - H \left( Z^n| \left\{ {\cal C}_k (M_k^{\text s}) \right\}_{k \in {\cal K}'}, \left\{ {\cal C}_k (M_k^{\text o}) \right\}_{k \in {\overline {{\cal K}'}}} \right) \nonumber\\
	& \overset{\text {(a)}}{\leq} 2^{- n ( \sum_{k \in {\cal K}'} R_k^{\text s} + \sum_{k \in {\overline {{\cal K}'}}} R_k^{\text o} )}
	\sum_{\substack{ m_{{\cal K}'}^{\text s} \in \prod_{k \in {\cal K}'} {\cal M}_k^{\text s},\\ 
			m_{\overline {{\cal K}'}}^{\text o} \in \prod_{k \in {\overline {{\cal K}'}}} {\cal M}_k^{\text o} }}
	\left| H \left( Z^n| {\cal C}_{{\cal K}'}, \left\{ {\cal C}_k (m_k^{\text o}) \right\}_{k \in {\overline {{\cal K}'}}} \right) \right. \nonumber\\
	& \quad\quad\quad\quad\quad\quad\quad\quad\quad\quad\quad\quad\quad\quad\quad \left. -  H \left( Z^n| \left\{ {\cal C}_k (m_k^{\text s}) \right\}_{k \in {\cal K}'}, \left\{ {\cal C}_k (m_k^{\text o}) \right\}_{k \in {\overline {{\cal K}'}}} \right) \right| \nonumber\\
	& \overset{\text {(b)}}{\leq} 2^{- n ( \sum_{k \in {\cal K}'} R_k^{\text s} + \sum_{k \in {\overline {{\cal K}'}}} R_k^{\text o} )}
	\sum_{\substack{ m_{{\cal K}'}^{\text s} \in \prod_{k \in {\cal K}'} {\cal M}_k^{\text s},\\ 
			m_{\overline {{\cal K}'}}^{\text o} \in \prod_{k \in {\overline {{\cal K}'}}} {\cal M}_k^{\text o} }}
	{\mathbb E} \Bigg[ d ( {\cal C}_{\cal K}, m_{{\cal K}'}^{\text s}, m_{\overline {{\cal K}'}}^{\text o} ) \ln \frac{|{\cal Z}|^n}{ d ( {\cal C}_{\cal K}, m_{{\cal K}'}^{\text s}, m_{\overline {{\cal K}'}}^{\text o} ) } \Bigg] \nonumber\\
	& \overset{\text {(c)}}{\leq} 2^{- n ( \sum_{k \in {\cal K}'} R_k^{\text s} + \sum_{k \in {\overline {{\cal K}'}}} R_k^{\text o} )}
	\sum_{\substack{ m_{{\cal K}'}^{\text s} \in \prod_{k \in {\cal K}'} {\cal M}_k^{\text s},\\ 
			m_{\overline {{\cal K}'}}^{\text o} \in \prod_{k \in {\overline {{\cal K}'}}} {\cal M}_k^{\text o} }}
	{\mathbb E} \left[ d ( {\cal C}_{\cal K}, m_{{\cal K}'}^{\text s}, m_{\overline {{\cal K}'}}^{\text o} ) \right] \ln \frac{|{\cal Z}|^n}{{\mathbb E} \left[ d ( {\cal C}_{\cal K}, m_{{\cal K}'}^{\text s}, m_{\overline {{\cal K}'}}^{\text o} ) \right]} \nonumber\\
	& \leq 2 e^{-n \varepsilon} \left( n \ln |{\cal Z}| + n \varepsilon -  \ln 2 \right) \rightarrow 0,
\end{align}
where (\ref{inf_leak_C2}a), (\ref{inf_leak_C2}b), and (\ref{inf_leak_C2}c) follow by respectively applying the triangular inequality, \cite[Lemma~$2.7$]{csiszar2011information}, and Jensen's inequality, and the last step is obtained by using (\ref{vari_dist_ub3}) and the fact that $u \ln \frac{|{\cal Z}|^n}{u}$ is an increasing function of $u$ in $(0, \frac{|{\cal Z}|^n}{e}]$ and $0< 2 e^{-n \varepsilon} < \frac{1}{e} \leq \frac{|{\cal Z}|^n}{e}$ (as $n$ goes to infinity).
Using the chain rule and non-negativity of mutual information,
\begin{align}\label{inf_leak2}
	I \left( M_{{\cal K}'}^{\text s}; Z^n| M_{\overline {{\cal K}'}}^{\text o}, {\cal C}_{\cal K} \right) & = I \left( M_{{\cal K}'}^{\text s}; Z^n, {\cal C}_{\cal K}| M_{\overline {{\cal K}'}}^{\text o} \right) - I \left( M_{{\cal K}'}^{\text s}; {\cal C}_{\cal K}| M_{\overline {{\cal K}'}}^{\text o} \right) \nonumber\\
	& \overset{\text {(a)}}{=} I \left( M_{{\cal K}'}^{\text s}; Z^n, {\cal C}_{\cal K}| M_{\overline {{\cal K}'}}^{\text o} \right) \nonumber\\
	& = I \left( M_{{\cal K}'}^{\text s}; Z^n| M_{\overline {{\cal K}'}}^{\text o} \right) + I \left( M_{{\cal K}'}^{\text s}; {\cal C}_{\cal K}| Z^n, M_{\overline {{\cal K}'}}^{\text o} \right) \nonumber\\
	& \geq I \left( M_{{\cal K}'}^{\text s}; Z^n| M_{\overline {{\cal K}'}}^{\text o} \right),
\end{align}
where (\ref{inf_leak2}a) holds since the choices of random messages at different users are independent of each other and also independent of the choices of random codebooks, resulting in $I \left( M_{{\cal K}'}^{\text s}; {\cal C}_{\cal K}| M_{\overline {{\cal K}'}}^{\text o} \right)$ $= 0$.
It is known from (\ref{inf_leak_C2}) and (\ref{inf_leak2}) that 
\begin{equation}\label{secrecy_metric_ub}
	\lim_{n \rightarrow \infty} I \left( M_{{\cal K}'}^{\text s}; Z^n| M_{\overline {{\cal K}'}}^{\text o} \right) = 0.
\end{equation}
Since $R_k^{\text s} = 0, \forall k \in {\overline {{\cal K}'}}$ and the messages of different users are independent, we have
\begin{align}
	I \left( M_{{\cal K}'}^{\text s}; Z^n| M_{\overline {{\cal K}'}}^{\text o} \right) & \geq I \left( M_{{\cal K}'}^{\text s}; Z^n \right) \nonumber\\
	& = I \left( M_{\cal K}^{\text s}; Z^n \right).
\end{align}
Then, it is known that (\ref{secrecy_metric}) is true.

\subsection{Achievability Proof When (\ref{assumption_0}) does not Hold}

For a given ${\cal K}' \subseteq {\cal K}$, if (\ref{assumption_0}) cannot be satisfied, then, there exists at least one non-empty set ${\cal S} \subseteq {\cal K}'$ such that
\begin{equation}\label{assumption_6}
	I(X_{\cal S}; Y| X_{\overline {\cal S}}, X_{\overline {{\cal K}'}}) - I(X_{\cal S}; Z| X_{\overline {{\cal K}'}}) \leq 0.
\end{equation}
With (\ref{assumption_6}), there are two possible cases, i.e., 
\begin{equation}\label{assumption_60}
	I(X_{{\cal K}'}; Y| X_{\overline {{\cal K}'}}) - I(X_{{\cal K}'}; Z| X_{\overline {{\cal K}'}}) \leq 0,
\end{equation}
and
\begin{equation}\label{assumption_63}
	I(X_{{\cal K}'}; Y| X_{\overline {{\cal K}'}}) - I(X_{{\cal K}'}; Z| X_{\overline {{\cal K}'}}) > 0.
\end{equation}
In the following, we prove the achievability of  ${\mathscr R} (X_{\cal K}, {\cal K}')$ when either (\ref{assumption_60}) or (\ref{assumption_63}) holds.

In the first case, i.e., when (\ref{assumption_60}) holds, by setting ${\cal S} = {\cal K}'$, ${\cal S}' = {\cal S} = {\cal K}'$ and ${\cal T} = \emptyset$ in (\ref{region_DM_exten}), we get
\begin{align}
	\sum\limits_{k \in {\cal K}'} R_k^{\text s} \leq \left[ I(X_{{\cal K}'}; Y| X_{\overline {{\cal K}'}}) - I(X_{{\cal K}'}; Z| X_{\overline {{\cal K}'}}) \right]^+ = 0.
\end{align}
Hence, $R_k^{\text s} = 0, \forall k \in {\cal K}'$.
Considering that $R_k^{\text s} = 0, \forall k \in {\overline {{\cal K}'}}$ in (\ref{region_DM_exten}), we have $R_k^{\text s} = 0, \forall k \in {\cal K}$.
Then, in this case, (\ref{region_DM_exten}) becomes
\begin{equation}\label{region_case1}
	\left\{
	\begin{array}{ll}
		R_k^{\text s} = 0, \forall k \in {\cal K}, \\
		\sum\limits_{k \in {\cal S} \setminus {\cal S}'} R_k^{\text o} + \sum\limits_{k \in {\cal T}} R_k^{\text o}	& \leq \left[ I(X_{\cal S}, X_{\cal T}; Y| X_{\overline {\cal S}}, X_{\overline {\cal T}}) - I(X_{{\cal S}'}; Z| X_{\overline {{\cal K}'}}) \right]^+, \\
		& \forall {\cal S} \subseteq {\cal K}', {\cal S}' \subseteq {\cal S}, {\cal T} \subseteq {\overline {{\cal K}'}},
	\end{array} \right.
\end{equation}
which is included in
\begin{equation}\label{region_case1_2}
	\left\{
	\begin{array}{ll}
		R_k^{\text s} = 0, \forall k \in {\cal K}, \\
		\sum\limits_{k \in {\cal S}} R_k^{\text o} + \sum\limits_{k \in {\cal T}} R_k^{\text o}\leq I(X_{\cal S}, X_{\cal T}; Y| X_{\overline {\cal S}}, X_{\overline {\cal T}}),  \forall {\cal S} \subseteq {\cal K}', {\cal T} \subseteq {\overline {{\cal K}'}},
	\end{array} \right.
\end{equation}
since (\ref{region_case1_2}) consists of only partial inequalities in (\ref{region_case1}) (those with ${\cal S}' = \emptyset$).
Note that (\ref{region_case1_2}) can be seen as the capacity region of a standard DM-MAC channel $P_{Y| X_{\cal K}}$ with no wiretapping and $K$ users, each transmitting at rate $R_k^{\text o}$.
Hence, the achievability of (\ref{region_case1}) is obvious.

In the second case, i.e., when (\ref{assumption_63}) holds, due to (\ref{assumption_6}), there must exist at least one subset ${\cal K}_0 \subsetneqq {\cal K}'$ such that
\begin{equation}\label{assumption_61}
	I(X_{{\cal K}_0}; Y| X_{{\cal K}' \setminus {\cal K}_0}, X_{\overline {{\cal K}'}}) - I(X_{{\cal K}_0}; Z| X_{\overline {{\cal K}'}}) \leq 0,
\end{equation}
and 
\begin{align}\label{assumption_62}
	I(X_{{\cal K}_0 \cup {\cal V}}; Y| X_{{\cal K}' \setminus ({\cal K}_0 \cup {\cal V})}, X_{\overline {{\cal K}'}}) - I(X_{{\cal K}_0 \cup {\cal V}}; Z| X_{\overline {{\cal K}'}}) > 0, \forall {\cal V} \subseteq {\cal K}' \setminus {\cal K}_0, {\cal V} \neq \emptyset.
\end{align}
The inequalities (\ref{assumption_61}) and (\ref{assumption_62}) indicate that ${\cal K}_0$ is the largest set in ${\cal K}'$ which includes all users in ${\cal K}_0$ and ensures (\ref{assumption_61}).
Adding any other users in ${\cal K}' \setminus {\cal K}_0$ to ${\cal K}_0$ results in (\ref{assumption_62}).
Note that if there are multiple subsets in ${\cal K}'$ making (\ref{assumption_61}) and (\ref{assumption_62}) hold, we let ${\cal K}_0$ be any of them.
Let
\begin{align}\label{K_prime}
	{\cal K}'' & = {\cal K}' \setminus {\cal K}_0 \nonumber\\
	& = {\cal K} \setminus ({\overline {{\cal K}'}} \cup {\cal K}_0), \nonumber\\
	{\overline {{\cal K}''}} & = {\cal K} \setminus {\cal K}'' \nonumber\\
	& = {\overline {{\cal K}'}} \cup {\cal K}_0.
\end{align}
Then, we give the following theorem.
\begin{theorem}\label{theo_polytope}
	With ${\cal K}_0$, ${\cal K}''$, and ${\overline {{\cal K}''}}$ defined above, we have
	\begin{equation}\label{cond4}
		I(X_{\cal V}; Y| X_{\overline {\cal V}}, X_{\overline {{\cal K}''}}) - I(X_{\cal V}; Z| X_{\overline {{\cal K}''}}) > 0, \forall {\cal V} \subseteq {\cal K}'', {\cal V} \neq \emptyset,
	\end{equation}
	where ${\overline {\cal V}} = {\cal K}'' \setminus {\cal V}$.
	In addition, if a rate-tuple $(R_1^{\text s}, R_1^{\text o},\cdots, R_K^{\text s}, R_K^{\text o})$ is in region ${\mathscr R} (X_{\cal K}, {\cal K}')$ defined by Theorem~\ref{lemma_DM_exten} and has (\ref{assumption_61}) as well as (\ref{assumption_62}) met, then, it is also in region ${\mathscr R} (X_{\cal K}, {\cal K}'')$, i.e., it satisfies
	\begin{equation}\label{region_DM1}
		\left\{
		\begin{array}{ll}
			R_k^{\text s} = 0, \forall k \in {\overline {{\cal K}''}}, \\
			\sum\limits_{k \in \cal V} R_k^{\text s} + \sum\limits_{k \in {\cal V} \setminus {\cal V}'} R_k^{\text o} + \sum\limits_{k \in {\cal W}} R_k^{\text o} & \leq I(X_{\cal V}, X_{\cal W}; Y| X_{\overline {\cal V}}, X_{\overline {\cal W}}) - I(X_{{\cal V}'}; Z| X_{\overline {{\cal K}''}}), \\
			& \forall {\cal V} \subseteq {\cal K}'', {\cal V}' \subseteq {\cal V}, {\cal W} \subseteq {\overline {{\cal K}''}},
		\end{array} \right.
	\end{equation}
	where ${\overline {\cal W}} = {\overline {{\cal K}''}} \setminus {\cal W}$.
\end{theorem}
\itshape \textbf{Proof:} \upshape
See Appendix \ref{prove_theo_polytope}.
\hfill $\Box$

It can be similarly proven as (\ref{cond2}) that (\ref{cond4}) is equivalent to
\begin{align}\label{assumption_4}
	I(X_{\cal V}, X_{\cal W}; Y| X_{\overline {\cal V}}, X_{\overline {\cal W}}) - I(X_{{\cal V}'}; Z| X_{\overline {{\cal K}''}}) > 0, \forall {\cal V} \subseteq {\cal K}'', {\cal V}' \subseteq {\cal V}, {\cal W} \subseteq {\overline {{\cal K}''}}, {\cal V} \cup {\cal W} \neq \emptyset.
\end{align}
$\left[ \cdot \right]^+$ in (\ref{region_DM1}) can thus be omitted.
Note that to make it easier to read and explain, instead of ${\cal S}$ and ${\cal T}$, we use new notations ${\cal V}$ and ${\cal W}$ in Theorem~\ref{theo_polytope} and Appendix~\ref{prove_theo_polytope} to respectively denote subsets in ${\cal K}''$ and ${\overline {{\cal K}''}}$.

Theorem~\ref{theo_polytope} shows that if a rate-tuple $(R_1^{\text s}, R_1^{\text o},\cdots, R_K^{\text s}, R_K^{\text o})$ is in region ${\mathscr R} (X_{\cal K}, {\cal K}')$ and has (\ref{assumption_61}) as well as (\ref{assumption_62}) met, it is also in region ${\mathscr R} (X_{\cal K}, {\cal K}'')$ and satisfies (\ref{assumption_4}).
Then, its achievability is immediately clear if we could prove that with (\ref{assumption_4}), any rate-tuple in ${\mathscr R} (X_{\cal K}, {\cal K}'')$ is achievable.
Interestingly, this can be realized by using similar techniques provided in the previous subsection.


\section{Two-User Binary-Input Real Adder Channel}
\label{BAC}

We consider a two-user binary-input real adder channel and show that the achievable region proposed in Theorem~$3$ can strictly improve that provided in \cite[Theorem~$1$]{xu2022achievable}.
	
	\subsection{Two-User DM MAC-WT Channel}
	When $K=2$, by respectively letting ${\cal K}'$ in Theorem~$3$ be $\{1, 2\}$, $\{1\}$, $\{2\}$, and $\emptyset$, we obtain the following four regions
	\setcounter{equation}{50}
	\begin{equation}\label{region_12}
		{\mathscr R} (X_{\cal K}, \{1, 2\}) \left\{\!\!\!
		\begin{array}{ll}
			& R_1^{\text s} + R_1^{\text o} \leq I(X_1; Y| X_2) \\
			& R_2^{\text s} + R_2^{\text o} \leq I(X_2; Y| X_1) \\
			& R_1^{\text s} + R_1^{\text o} + R_2^{\text s} + R_2^{\text o} \leq I(X_1, X_2; Y) \\
			& R_1^{\text s} \leq \left[ I(X_1; Y| X_2) - I(X_1; Z) \right]^+ \\
			& R_2^{\text s} \leq \left[ I(X_2; Y| X_1) - I(X_2; Z) \right]^+ \\
			& R_1^{\text s} + R_2^{\text s} \leq \left[ I(X_1, X_2; Y) - I(X_1, X_2; Z) \right]^+ \\
			& R_1^{\text s} + R_1^{\text o} + R_2^{\text s} \leq \left[ I(X_1, X_2; Y) - I(X_2; Z) \right]^+ \\
			& R_1^{\text s} + R_2^{\text s} + R_2^{\text o} \leq \left[ I(X_1, X_2; Y) - I(X_1; Z) \right]^+
		\end{array} \right.,
	\end{equation}
	\begin{equation}\label{region_1}
		{\mathscr R} (X_{\cal K}, \{1\}) \left\{\!\!\!
		\begin{array}{ll}
			& R_2^{\text s} = 0 \\
			& R_1^{\text s} + R_1^{\text o} \leq I(X_1; Y| X_2) \\
			& R_2^{\text o} \leq I(X_2; Y| X_1) \\
			& R_1^{\text s} + R_1^{\text o} + R_2^{\text o} \leq I(X_1, X_2; Y) \\
			& R_1^{\text s} \leq \left[ I(X_1; Y| X_2) - I(X_1; Z| X_2) \right]^+ \\
			& R_1^{\text s} + R_2^{\text o} \leq \left[ I(X_1, X_2; Y) - I(X_1; Z| X_2) \right]^+
		\end{array} \right.,
	\end{equation}
	\begin{equation}\label{region_2}
		{\mathscr R} (X_{\cal K}, \{2\}) \left\{\!\!\!
		\begin{array}{ll}
			& R_1^{\text s} = 0 \\
			& R_1^{\text o} \leq I(X_1; Y| X_2) \\
			& R_2^{\text s} + R_2^{\text o} \leq I(X_2; Y| X_1) \\
			& R_1^{\text o} + R_2^{\text s} + R_2^{\text o} \leq I(X_1, X_2; Y) \\
			& R_2^{\text s} \leq \left[ I(X_2; Y| X_1) - I(X_2; Z| X_1) \right]^+ \\
			& R_1^{\text o} + R_2^{\text s} \leq \left[ I(X_1, X_2; Y) - I(X_2; Z| X_1) \right]^+
		\end{array} \right.,
	\end{equation}
	\begin{equation}\label{region_phi}
		{\mathscr R} (X_{\cal K}, \emptyset) \left\{\!\!\!
		\begin{array}{ll}
			& R_1^{\text s} = 0 \\
			& R_2^{\text s} = 0 \\
			& R_1^{\text o} \leq I(X_1; Y| X_2) \\
			& R_2^{\text o} \leq I(X_2; Y| X_1) \\
			& R_1^{\text o} + R_2^{\text o} \leq I(X_1, X_2; Y)
		\end{array} \right..
	\end{equation}
	For convenience, we denote the following convex hulls
	\begin{align}
		{\mathscr R}_{\text {new}} & = {\text {Conv}} \bigcup_{P_{X_1}P_{X_2}} {\mathscr R} (X_{\cal K}, \{1, 2\}) \cup {\mathscr R} (X_{\cal K}, \{1\}) \cup {\mathscr R} (X_{\cal K}, \{2\}) \cup {\mathscr R} (X_{\cal K}, \emptyset), \nonumber\\
		{\mathscr R}_{\text {old}} & = {\mathscr R}_{\text {new}}^{\{1, 2\}} = {\text {Conv}} \bigcup_{P_{X_1}P_{X_2}} {\mathscr R} (X_{\cal K}, \{1, 2\}), \nonumber\\
		{\mathscr R}_{\text {new}}^{\{1\}} & = {\text {Conv}} \bigcup_{P_{X_1}P_{X_2}} {\mathscr R} (X_{\cal K}, \{1\}).
	\end{align}
	It has been proven in \cite[Theorem~$1$]{xu2022achievable} that ${\mathscr R}_{\text {old}}$ is an achievable rate region of the two-user DM MAC-WT channel under the weak secrecy metric.
	In Theorem~$3$ of this paper, we further propose a new achievable region ${\mathscr R}_{\text {new}}$ under the strong secrecy metric.
	In the following we prove that ${\mathscr R}_{\text {new}}$ can strictly improve ${\mathscr R}_{\text {old}}$.
	Noticing that ${\mathscr R}_{\text {old}}$ is contained in ${\mathscr R}_{\text {new}}$, it is sufficient to prove the improvement by showing that ${\mathscr R}_{\text {new}}$ is not contained in ${\mathscr R}_{\text {old}}$.
	Since ${\mathscr R}_{\text {new}}^{\{1\}}$ is also contained in ${\mathscr R}_{\text {new}}$, it is sufficient to show that there exist points in ${\mathscr R}_{\text {new}}^{\{1\}}$ that are not contained in ${\mathscr R}_{\text {old}}$.  
	This is done in the following through a specific example.
	
	Note that (\ref{region_12}) and (\ref{region_1}) respectively define a four-dimensional space for $(R_1^{\text s}, R_1^{\text o}, R_2^{\text s}, R_2^{\text o})$ and a three-dimensional space for $(R_1^{\text s}, R_1^{\text o}, R_2^{\text o})$.
	To make the following results easier to plot and visually understandable, we let $R_1^{\text o} = R_2^{\text s} = 0$ in (\ref{region_12}) and $R_1^{\text o} = 0$ in (\ref{region_1}), and obtain two two-dimensional spaces for $(R_1^{\text s}, R_2^{\text o})$.
	In particular, if $R_1^{\text o} = R_2^{\text s} = 0$, the 1st, 3rd, 5th, and 7th inequalities in (\ref{region_12}) are redundant, (\ref{region_12}) reduces to
	\begin{equation}\label{region0_12}
		{\hat {\mathscr R}} (X_{\cal K}, \{1, 2\}) \!\left\{\!\!\!\!\!\!\!
		\begin{array}{ll}
			& R_1^{\text s} \leq \min \{\left[ I(X_1; Y| X_2) \!-\! I(X_1; Z) \right]^+, \left[ I(X_1, X_2; Y) \!-\! I(X_1, X_2; Z) \right]^+ \} \\
			& R_2^{\text o} \leq I(X_2; Y| X_1) \\
			& R_1^{\text s} + R_2^{\text o} \leq \left[ I(X_1, X_2; Y) - I(X_1; Z) \right]^+
		\end{array} \right.\!\!\!\!\!,\!\!\!
	\end{equation}
	and (\ref{region_1}) reduces to
	\begin{equation}\label{region0_1}
		{\hat {\mathscr R}} (X_{\cal K}, \{1\}) \left\{\!\!\!
		\begin{array}{ll}
			& R_1^{\text s} \leq \left[ I(X_1; Y| X_2) - I(X_1; Z| X_2) \right]^+ \\
			& R_2^{\text o} \leq I(X_2; Y| X_1) \\
			& R_1^{\text s} + R_2^{\text o} \leq \left[ I(X_1, X_2; Y) - I(X_1; Z| X_2) \right]^+
		\end{array} \right..
	\end{equation}
	Denote the convex hulls of the union of ${\hat {\mathscr R}} (X_{\cal K}, \{1, 2\})$ and ${\hat {\mathscr R}} (X_{\cal K}, \{1\})$ over all $P_{X_1}P_{X_2}$ by
	\begin{align}\label{R_old_new1}
		{\hat {\mathscr R}}_{\text {old}} & = {\hat {\mathscr R}}_{\text {new}}^{\{1, 2\}} = {\text {Conv}} \bigcup_{P_{X_1}P_{X_2}} {\hat {\mathscr R}} (X_{\cal K}, \{1, 2\}), \nonumber\\
		{\hat {\mathscr R}}_{\text {new}}^{\{1\}} & = {\text {Conv}} \bigcup_{P_{X_1}P_{X_2}} {\hat {\mathscr R}} (X_{\cal K}, \{1\}).
	\end{align}

Note that ${\hat {\mathscr R}} (X_{\cal K}, \{1, 2\})$ is the orthogonal projection of ${\mathscr R} (X_{\cal K}, \{1, 2\})$ onto the hyperplane $R_1^{\text o} = R_2^{\text s} = 0$.
The convex hulls of them over all $P_{X_1}P_{X_2}$, i.e., ${\hat {\mathscr R}}_{\text {old}}$ and ${\mathscr R}_{\text {old}}$, thus contain the same achievable set about $(R_1^{\text s}, R_2^{\text o})$.
This ensures that if a rate tuple $(R_1^{\text s}, R_1^{\text o}, R_2^{\text s}, R_2^{\text o})$ is in ${\mathscr R}_{\text {old}}$, then, $(R_1^{\text s}, R_2^{\text o})$ is in ${\hat {\mathscr R}}_{\text {old}}$.
Reversely, if $(R_1^{\text s}, R_2^{\text o})$ is in ${\hat {\mathscr R}}_{\text {old}}$, then, there must exist $(R_1^{\text o}, R_2^{\text s})$ such that $(R_1^{\text s}, R_1^{\text o}, R_2^{\text s}, R_2^{\text o})$ is in ${\mathscr R}_{\text {old}}$.
Similarly, we know that ${\hat {\mathscr R}}_{\text {new}}^{\{1\}}$ and ${\mathscr R}_{\text {new}}^{\{1\}}$ contain the same achievable set about $(R_1^{\text s}, R_2^{\text o})$.
It is thus sufficient to prove that ${\mathscr R}_{\text {new}}$ can strictly improve ${\mathscr R}_{\text {old}}$ by showing that ${\hat {\mathscr R}}_{\text {new}}^{\{1\}}$ contains achievable points $(R_1^{\text s}, R_2^{\text o})$ that are not contained in ${\hat {\mathscr R}}_{\text {old}}$.
	
\subsection{Two-User Binary-Input Real Adder Channel}
	
	\begin{figure*}[ht]
		\centering
		\includegraphics[scale=0.5]{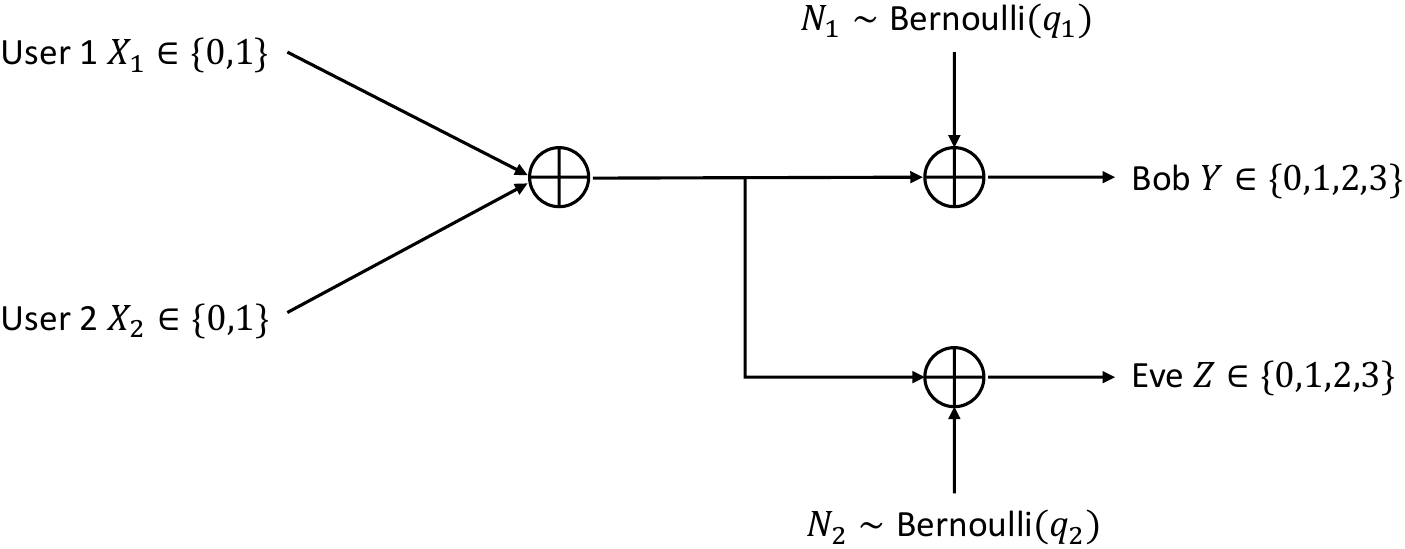}
		\caption{Illustration of a two-user binary-input real adder channel.}
		\label{Binary_adder_channel}
	\end{figure*}
	
We compute the regions (\ref{region0_12}) and (\ref{region0_1}) for the two-user binary-input real adder channels shown in Fig.~\ref{Binary_adder_channel}.
The channel has binary inputs $X_1 \in \{0,1\}$ and $X_2 \in \{0,1\}$, and binary additive noise $N_1 \sim {\text {Bernoulli}} (q_1)$ and $N_2 \sim {\text {Bernoulli}} (q_2)$.
All possible input distributions are parameterized by the probabilities 
$P_{X_1}(1) = 1 - P_{X_1}(0) = \alpha \in [0,1]$ and $P_{X_2}(1) = 1 - P_{X_2}(0) = \beta \in [0,1]$. 
The received signals at Bob and Eve are given by
	\begin{align}\label{binary_adder}
		Y & = X_1 + X_2 + N_1, \nonumber\\
		Z & = X_1 + X_2 + N_2,
	\end{align}
where addition is over the reals, and $Y$ and $Z$ are both quaternary outputs that take values in the alphabet $\{0,1,2,3\}$.
	
	Now we compute the rate upper bounds in (\ref{region0_12}) and (\ref{region0_1}).
	First, based on (\ref{binary_adder}), the mutual information terms in (\ref{region0_12}) can be expressed as
	\begin{align}\label{ubs_region1}
		I(X_1; Y| X_2) - I(X_1; Z) & = H(Y| X_2) - H(Y| X_1, X_2) - H(Z) + H(Z| X_1) \nonumber\\
		& = H(Y| X_2) - H(N_1) - H(Z) + H(Z| X_1), \nonumber\\
		I(X_1, X_2; Y) - I(X_1, X_2; Z) & = H(Y) - H(Y| X_1, X_2) - H(Z) + H(Z| X_1, X_2) \nonumber\\
		& = H(Y) - H(N_1) - H(Z) + H(N_2), \nonumber\\
		I(X_2; Y| X_1) & = H(Y| X_1) - H(Y| X_1, X_2) \nonumber\\
		& = H(Y| X_1) - H(N_1), \nonumber\\
		I(X_1, X_2; Y) - I(X_1; Z) & = H(Y) - H(Y| X_1, X_2) - H(Z) + H(Z| X_1) \nonumber\\
		& = H(Y) - H(N_1) - H(Z) + H(Z| X_1),
	\end{align}
	and the mutual information differences in (\ref{region0_1}) can be expressed as
	\begin{align}\label{ubs_region2}
		I(X_1; Y| X_2) - I(X_1; Z| X_2) & = H(Y| X_2) - H(Y| X_1, X_2) - H(Z| X_2) + H(Z| X_1, X_2) \nonumber\\
		& = H(Y| X_2) - H(N_1) - H(Z| X_2) + H(N_2), \nonumber\\
		I(X_1, X_2; Y) - I(X_1; Z| X_2) & = H(Y) - H(Y| X_1, X_2) - H(Z| X_2) + H(Z| X_1, X_2) \nonumber\\
		& = H(Y) - H(N_1) - H(Z| X_2) + H(N_2). 
	\end{align}
Next, we compute the entropies $H(N_1)$, $H(N_2)$, $H(Y)$, $H(Y| X_1)$, $H(Y| X_2)$, $H(Z)$, $H(Z| X_1)$, and $H(Z| X_2)$.
	
	\begin{table}[h]
		\centering
		\caption{$P_{X_1, X_2, N_1} (x_1, x_2, n_1)$ and $Y$ over different realizations of $(X_1, X_2, N_1)$.}
		\begin{tabular}{|cc|c|c|}
			\hline
			\multicolumn{2}{|l|}{}                   & $P_{X_1, X_2, N_1} (x_1, x_2, n_1)$ & $Y$  \\ \hline
			\multicolumn{1}{|l|}{\multirow{8}{*}{$(X_1, X_2, N_1)$}} & $(0, 0, 0)$ & $(1-\alpha)(1-\beta)(1-q_1)$ & $0$ \\ \cline{2-4} 
			\multicolumn{1}{|l|}{}                  & $(0, 0, 1)$ & $(1-\alpha)(1-\beta)q_1$ & $1$ \\ \cline{2-4} 
			\multicolumn{1}{|l|}{}                  & $(0, 1, 0)$ & $(1-\alpha)\beta(1-q_1)$ & $1$ \\ \cline{2-4} 
			\multicolumn{1}{|l|}{}                  & $(0, 1, 1)$ & $(1-\alpha)\beta q_1$ & $2$ \\ \cline{2-4} 
			\multicolumn{1}{|l|}{}                  & $(1, 0, 0)$ & $\alpha(1-\beta)(1-q_1)$ & $1$ \\ \cline{2-4} 
			\multicolumn{1}{|l|}{}                  & $(1, 0, 1)$ & $\alpha(1-\beta)q_1$ & $2$ \\ \cline{2-4} 
			\multicolumn{1}{|l|}{}                  & $(1, 1, 0)$ & $\alpha \beta (1-q_1)$ & $2$ \\ \cline{2-4} 
			\multicolumn{1}{|l|}{}                  & $(1, 1, 1)$ & $\alpha \beta q_1$ & $3$ \\ \hline
		\end{tabular}
		\label{table1}
	\end{table}
	
	Since both $N_1$ and $N_2$ follow Bernoulli distribution with parameters $q_1$ and $q_2$, we have
	\begin{align}\label{entropy_N1N2}
		H(N_1) & = - q_1 \log q_1 - (1 - q_1) \log (1 - q_1), \nonumber\\
		H(N_2) & = - q_2 \log q_2 - (1 - q_2) \log (1 - q_2).
	\end{align}
	Based on the distributions of $X_1$, $X_2$, and $N_1$, we list the values of $P_{X_1, X_2, N_1} (x_1, x_2, n_1)$ and $Y$ over different realizations of $(X_1, X_2, N_1)$ in Table~\ref{table1}, from which we have
	\begin{align}\label{PY}
		P_Y (0) & = (1-\alpha)(1-\beta)(1-q_1), \nonumber\\
		P_Y (1) & = (1-\alpha)(1-\beta)q_1 + (1-\alpha)\beta(1-q_1) + \alpha(1-\beta)(1-q_1), \nonumber\\
		P_Y (2) & = (1-\alpha)\beta q_1 + \alpha(1-\beta)q_1 + \alpha \beta (1-q_1), \nonumber\\
		P_Y (3) & = \alpha \beta q_1.
	\end{align}
	Accordingly, the entropy of $Y$ can be computed as
	\begin{align}\label{entropy_Y}
		H(Y) = & - \sum_{y \in \{0, 1, 2, 3\}} P_Y (y) \log P_Y (y) \nonumber\\
		= & - (1-\alpha)(1-\beta)(1-q_1) \log [(1-\alpha)(1-\beta)(1-q_1)] \nonumber\\
		& - [(1-\alpha)(1-\beta)q_1 + (1-\alpha)\beta(1-q_1) + \alpha(1-\beta)(1-q_1)] \nonumber\\
		& \times \log [(1-\alpha)(1-\beta)q_1 + (1-\alpha)\beta(1-q_1) + \alpha(1-\beta)(1-q_1)] \nonumber\\
		& - [(1-\alpha)\beta q_1 + \alpha(1-\beta)q_1 + \alpha \beta (1-q_1)] \log [(1-\alpha)\beta q_1 + \alpha(1-\beta)q_1 + \alpha \beta (1-q_1)] \nonumber\\
		& - \alpha \beta q_1 \log (\alpha \beta q_1).
	\end{align}
	
	In Table~\ref{table2}, we list the values of $P_{X_1, Y} (x_1, y)$ and $P_{Y| X_1} (y| x_1)$ over different realizations of $(X_1, Y)$, from which the conditional entropy $H(Y| X_1)$ can be computed as follows
	\begin{align}\label{entropy_Y_X1}
		H(Y| X_1) = & - \sum_{x_1 \in \{0, 1\}} \sum_{y \in \{0, 1, 2, 3\}} P_{X_1, Y} (x_1, y) \log P_{Y| X_1} (y| x_1) \nonumber\\
		= & - (1-\beta)(1-q_1) \log [(1-\beta)(1-q_1)] \nonumber\\
		& - [(1-\beta)q_1 + \beta(1-q_1)] \log [(1-\beta)q_1 + \beta(1-q_1)] - \beta q_1 \log (\beta q_1).
	\end{align}
	
	\begin{table}[h]
		\centering
		\caption{$P_{X_1, Y} (x_1, y)$ and $P_{Y| X_1} (y| x_1)$ over different realizations of $(X_1, Y)$.}
		\begin{tabular}{|cc|c|c|}
			\hline
			\multicolumn{2}{|l|}{}                   & $P_{X_1, Y} (x_1, y)$ & $P_{Y| X_1} (y| x_1)$  \\ \hline
			\multicolumn{1}{|l|}{\multirow{8}{*}{$(X_1, Y)$}} & $(0, 0)$ & $(1-\alpha)(1-\beta)(1-q_1)$ & $(1-\beta)(1-q_1)$ \\ \cline{2-4} 
			\multicolumn{1}{|l|}{}                  & $(0, 1)$ & $(1-\alpha)(1-\beta)q_1 + (1-\alpha)\beta(1-q_1)$ & $(1-\beta)q_1 + \beta(1-q_1)$ \\ \cline{2-4} 
			\multicolumn{1}{|l|}{}                  & $(0, 2)$ & $(1-\alpha)\beta q_1$ & $\beta q_1$ \\ \cline{2-4} 
			\multicolumn{1}{|l|}{}                  & $(0, 3)$ & $0$ & $0$ \\ \cline{2-4} 
			\multicolumn{1}{|l|}{}                  & $(1, 0)$ & $0$ & $0$ \\ \cline{2-4} 
			\multicolumn{1}{|l|}{}                  & $(1, 1)$ & $\alpha(1-\beta)(1-q_1)$ & $(1-\beta)(1-q_1)$ \\ \cline{2-4} 
			\multicolumn{1}{|l|}{}                  & $(1, 2)$ & $\alpha(1-\beta) q_1 + \alpha \beta (1-q_1)$ & $(1-\beta) q_1 + \beta (1-q_1)$ \\ \cline{2-4} 
			\multicolumn{1}{|l|}{}                  & $(1, 3)$ & $\alpha \beta q_1$ & $\beta q_1$ \\ \hline
		\end{tabular}
		\label{table2}
	\end{table}
	
	\begin{table}[h]
		\centering
		\caption{$P_{X_2, Y} (x_2, y)$ and $P_{Y| X_2} (y| x_2)$ over different realizations of $(X_2, Y)$.}
		\begin{tabular}{|cc|c|c|}
			\hline
			\multicolumn{2}{|l|}{}                   & $P_{X_2, Y} (x_2, y)$ & $P_{Y| X_2} (y| x_2)$  \\ \hline
			\multicolumn{1}{|l|}{\multirow{8}{*}{$(X_2, Y)$}} & $(0, 0)$ & $(1-\alpha)(1-\beta)(1-q_1)$ & $(1-\alpha)(1-q_1)$ \\ \cline{2-4} 
			\multicolumn{1}{|l|}{}                  & $(0, 1)$ & $(1-\alpha)(1-\beta)q_1 + \alpha(1-\beta)(1-q_1)$ & $(1-\alpha)q_1 + \alpha(1-q_1)$ \\ \cline{2-4} 
			\multicolumn{1}{|l|}{}                  & $(0, 2)$ & $\alpha(1-\beta)q_1$ & $\alpha q_1$ \\ \cline{2-4} 
			\multicolumn{1}{|l|}{}                  & $(0, 3)$ & $0$ & $0$ \\ \cline{2-4} 
			\multicolumn{1}{|l|}{}                  & $(1, 0)$ & $0$ & $0$ \\ \cline{2-4} 
			\multicolumn{1}{|l|}{}                  & $(1, 1)$ & $(1-\alpha)\beta(1-q_1)$ & $(1-\alpha)(1-q_1)$ \\ \cline{2-4} 
			\multicolumn{1}{|l|}{}                  & $(1, 2)$ & $(1-\alpha)\beta q_1 + \alpha \beta (1-q_1)$ & $(1-\alpha) q_1 + \alpha (1-q_1)$ \\ \cline{2-4} 
			\multicolumn{1}{|l|}{}                  & $(1, 3)$ & $\alpha \beta q_1$ & $\alpha q_1$ \\ \hline
		\end{tabular}
		\label{table3}
	\end{table}
	
	Similarly, $H(Y| X_2)$ can be calculated based on Table~\ref{table3} as
	\begin{align}\label{entropy_Y_X2}
		H(Y| X_2) = & - \sum_{x_2 \in \{0, 1\}} \sum_{y \in \{0, 1, 2, 3\}} P_{X_2, Y} (x_2, y) \log P_{Y| X_2} (y| x_2) \nonumber\\
		= & - (1-\alpha)(1-q_1) \log [(1-\alpha)(1-q_1)] \nonumber\\
		& - [(1-\alpha)q_1 + \alpha(1-q_1)] \log [(1-\alpha)q_1 + \alpha(1-q_1)] - \alpha q_1 \log (\alpha q_1).
	\end{align}
	
	Analogously, the entropies $H(Z)$, $H(Z| X_1)$, and $H(Z| X_2)$ can be computed as follows
	\begin{align}\label{entropy_Z}
		H(Z) = & - \sum_{z \in \{0, 1, 2, 3\}} P_Z (z) \log P_Z (z) \nonumber\\
		= & - (1-\alpha)(1-\beta)(1-q_2) \log [(1-\alpha)(1-\beta)(1-q_2)] \nonumber\\
		& - [(1-\alpha)(1-\beta)q_2 + (1-\alpha)\beta(1-q_2) + \alpha(1-\beta)(1-q_2)] \nonumber\\
		& \times \log [(1-\alpha)(1-\beta)q_2 + (1-\alpha)\beta(1-q_2) + \alpha(1-\beta)(1-q_2)] \nonumber\\
		& - [(1-\alpha)\beta q_2 + \alpha(1-\beta)q_2 + \alpha \beta (1-q_2)] \log [(1-\alpha)\beta q_2 + \alpha(1-\beta)q_2 + \alpha \beta (1-q_2)] \nonumber\\
		& - \alpha \beta q_2 \log (\alpha \beta q_2),
	\end{align}
	\begin{align}\label{entropy_Z_X1}
		H(Z| X_1) = & - \sum_{x_1 \in \{0, 1\}} \sum_{z \in \{0, 1, 2, 3\}} P_{X_1, Z} (x_1, z) \log P_{Z| X_1} (z| x_1) \nonumber\\
		= & - (1-\beta)(1-q_2) \log [(1-\beta)(1-q_2)] \nonumber\\
		& - [(1-\beta)q_2 + \beta(1-q_2)] \log [(1-\beta)q_2 + \beta(1-q_2)] - \beta q_2 \log (\beta q_2),
	\end{align}
	\begin{align}\label{entropy_Z_X2}
		H(Z| X_2) = & - \sum_{x_2 \in \{0, 1\}} \sum_{z \in \{0, 1, 2, 3\}} P_{X_2, Z} (x_2, z) \log P_{Z| X_2} (z| x_2) \nonumber\\
		= & - (1-\alpha)(1-q_2) \log [(1-\alpha)(1-q_2)] \nonumber\\
		& - [(1-\alpha)q_2 + \alpha(1-q_2)] \log [(1-\alpha)q_2 + \alpha(1-q_2)] - \alpha q_2 \log (\alpha q_2).
	\end{align}
	By substituting (\ref{entropy_N1N2}) and (\ref{entropy_Y}) $\sim$ (\ref{entropy_Z_X2}) to (\ref{ubs_region1}) and (\ref{ubs_region2}), the upper bounds in (\ref{region0_12}) and (\ref{region0_1}) can be computed.
	
	\subsection{Numerical Results}
	
	In this subsection, we depict the convex hulls ${\hat {\mathscr R}}_{\text {old}}$ and ${\hat {\mathscr R}}_{\text {new}}^{\{1\}}$ in (\ref{R_old_new1}) for the considered two-user binary-input real adder channel with $(q_1, q_2) = (0.5, 0.75)$, and show that ${\hat {\mathscr R}}_{\text {new}}^{\{1\}}$ contains points that are not in ${\hat {\mathscr R}}_{\text {old}}$.
	The Matlab code for the numerical results has been shared on GitHub for reproducibility at the link:\\ {\code{https://github.com/DrHaoxu/Achievable_region_binary_adder_channel}}.
	
	We first depict ${\hat {\mathscr R}}_{\text {old}}$.
	For convenience, denote
	\begin{align}\label{abc1}
		& a_1 = \min \{\left[ I(X_1; Y| X_2) - I(X_1; Z) \right]^+, \left[ I(X_1, X_2; Y) - I(X_1, X_2; Z) \right]^+ \}, \nonumber\\
		& b = I(X_2; Y| X_1), \nonumber\\
		& c_1 = \left[ I(X_1, X_2; Y) - I(X_1; Z) \right]^+,
	\end{align}
	based on which ${\hat {\mathscr R}} (X_{\cal K}, \{1, 2\})$ in (\ref{region0_12}) can be rewritten as
	\begin{equation}\label{region12}
		{\hat {\mathscr R}} (X_{\cal K}, \{1, 2\}) \left\{\!\!\!
		\begin{array}{ll}
			& R_1^{\text s} \leq a_1 \\
			& R_2^{\text o} \leq b \\
			& R_1^{\text s} + R_2^{\text o} \leq c_1
		\end{array} \right..
	\end{equation}
	Note that in the two-user binary-input real adder channel, the distributions of $X_1$ and $X_2$ are respectively determined by $\alpha \in [0,1]$ and $\beta \in [0,1]$.
	Then, ${\hat {\mathscr R}}_{\text {old}}$ in (\ref{R_old_new1}) can be rewritten as
	\begin{align}\label{R_old}
		{\hat {\mathscr R}}_{\text {old}} = {\text {Conv}} \bigcup_{\alpha \in [0,1], \beta \in [0,1]} {\hat {\mathscr R}} (X_{\cal K}, \{1, 2\}).
	\end{align}
	Using the definition and non-negativity of mutual information, we have
	\begin{align}\label{ineq_mutual1}
		I(X_1, X_2; Y) - I(X_1; Z) & = I(X_1; Y| X_2) + I(X_2; Y) - I(X_1; Z) \nonumber\\
		& \geq I(X_1; Y| X_2) - I(X_1; Z), \nonumber\\
		I(X_1, X_2; Y) - I(X_1; Z) & \geq I(X_1, X_2; Y) - I(X_1; Z) - I(X_2; Z| X_1) \nonumber\\
		& = I(X_1, X_2; Y) - I(X_1, X_2; Z).
	\end{align}
	From (\ref{abc1}) and (\ref{ineq_mutual1}), we know that for any $\alpha$ and $\beta$, $c_1 \geq a_1$ always holds.
	Since $(q_1, q_2) = (0.5, 0.75)$, if $(\alpha, \beta)$ is known, the values of $a_1$, $b$, and $c_1$ can be computed based on the analysis given in the previous subsection.
	As discussed below, ${\hat {\mathscr R}} (X_{\cal K}, \{1, 2\})$ has five possible shapes, depending on the values of $a_1$, $b$, and $c_1$.
	\begin{figure*}[t]
		\centering
		\includegraphics[scale=0.5]{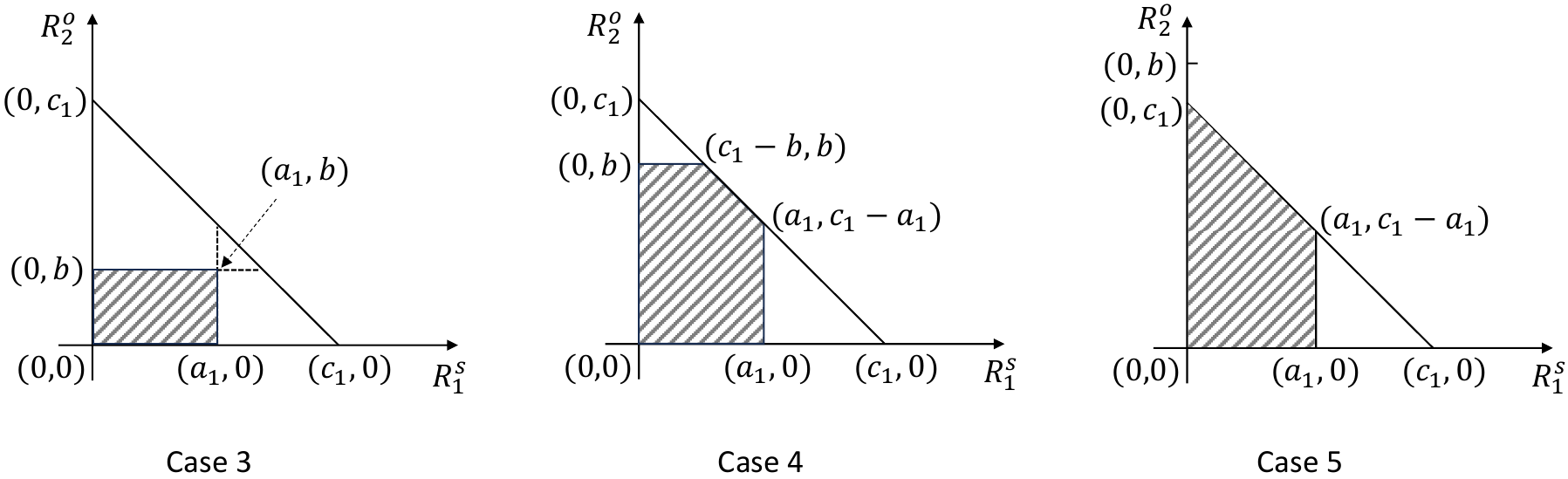}
		\caption{Achievable region ${\hat {\mathscr R}} (X_{\cal K}, \{1, 2\})$ in different cases.}
		\label{Case345}
	\end{figure*}
	\begin{itemize}
		\item Case~1: $c_1 = 0$.
		
		In this case, both $R_1^{\text s}$ and $R_2^{\text o}$ are $0$. ${\hat {\mathscr R}} (X_{\cal K}, \{1, 2\})$ thus contains only one point $(0,0)$.
		
		\item Case~2: $a_1 = 0$.
		
		In this case, $R_1^{\text s}=0$ and ${\hat {\mathscr R}} (X_{\cal K}, \{1, 2\})$ is a line segment between $(0,0)$ and $(0,\min\{b, c_1\})$.
		
		\item Case~3: $a_1 > 0$ and $b \leq c_1 - a_1$.
		
		As shown in Fig.~\ref{Case345}, ${\hat {\mathscr R}} (X_{\cal K}, \{1, 2\})$ in this case is a rectangular region with four vertices $(0,0)$, $(a_1, 0)$, $(a_1, b)$, and $(0,b)$.
		
		\item Case~4: $a_1 > 0$ and $c_1 - a_1 < b \leq c_1$.
		
		As shown in Fig.~\ref{Case345}, ${\hat {\mathscr R}} (X_{\cal K}, \{1, 2\})$ in this case is a pentagon region with five vertices $(0,0)$, $(a_1, 0)$, $(a_1, c_1 - a_1)$, $(c_1 - b, b)$, and $(0,b)$.
		
		\item Case~5: $a_1 > 0$ and $b > c_1$.
		
		As shown in Fig.~\ref{Case345}, ${\hat {\mathscr R}} (X_{\cal K}, \{1, 2\})$ in this case is a trapezoidal region with four vertices $(0,0)$, $(a_1, 0)$, $(a_1, c_1 - a_1)$, and $(0,c_1)$.
	\end{itemize}
	To depict ${\hat {\mathscr R}}_{\text {old}}$, we sample $\alpha$ and $\beta$ in the range [0,1] with step $\delta = 0.01$, record the corner points of ${\hat {\mathscr R}} (X_{\cal K}, \{1, 2\})$ for each given $(\alpha, \beta)$, and then depict the convex hull consisting of all these corner points using Matlab.
	In Fig.~\ref{convex_hull}, the area encircled by the red line is ${\hat {\mathscr R}}_{\text {old}}$.
	
	Now we depict ${\hat {\mathscr R}}_{\text {new}}^{\{1\}}$.
	For convenience, denote
	\begin{align}\label{ac2}
		& a_2 = \left[ I(X_1; Y| X_2) - I(X_1; Z| X_2) \right]^+, \nonumber\\
		& c_2 = \left[ I(X_1, X_2; Y) - I(X_1; Z| X_2) \right]^+,
	\end{align}
	based on which ${\hat {\mathscr R}} (X_{\cal K}, \{1\})$ in (\ref{region0_1}) and ${\hat {\mathscr R}}_{\text {old}}$ in (\ref{R_old_new1}) can be respectively rewritten as
	\begin{equation}\label{region1}
		{\hat {\mathscr R}} (X_{\cal K}, \{1\}) \left\{\!\!\!
		\begin{array}{ll}
			& R_1^{\text s} \leq a_2 \\
			& R_2^{\text o} \leq b \\
			& R_1^{\text s} + R_2^{\text o} \leq c_2
		\end{array} \right.,
	\end{equation}
	\begin{align}\label{R_new}
		{\hat {\mathscr R}}_{\text {new}}^{\{1\}} & = {\text {Conv}} \bigcup_{\alpha \in [0,1], \beta \in [0,1]} {\hat {\mathscr R}} (X_{\cal K}, \{1\}).
	\end{align}
	Since 
	\begin{align}
		I(X_1, X_2; Y) - I(X_1; Z| X_2) & = I(X_1; Y| X_2) + I(X_2; Y) - I(X_1; Z| X_2) \nonumber\\
		& \geq I(X_1; Y| X_2) - I(X_1; Z| X_2),
	\end{align}
	we know that for any $\alpha$ and $\beta$, $c_2 \geq a_2$ always holds.
	Then, ${\hat {\mathscr R}}_{\text {new}}^{\{1\}}$ can be depicted similarly as ${\hat {\mathscr R}}_{\text {old}}$.
	We omit the details for brevity.
	In Fig.~\ref{convex_hull}, the area encircled by the blue line is ${\hat {\mathscr R}}_{\text {new}}^{\{1\}}$.
	
	\begin{figure}
		\begin{minipage}[t]{0.5\linewidth}
			\centering
			\includegraphics[width=3in]{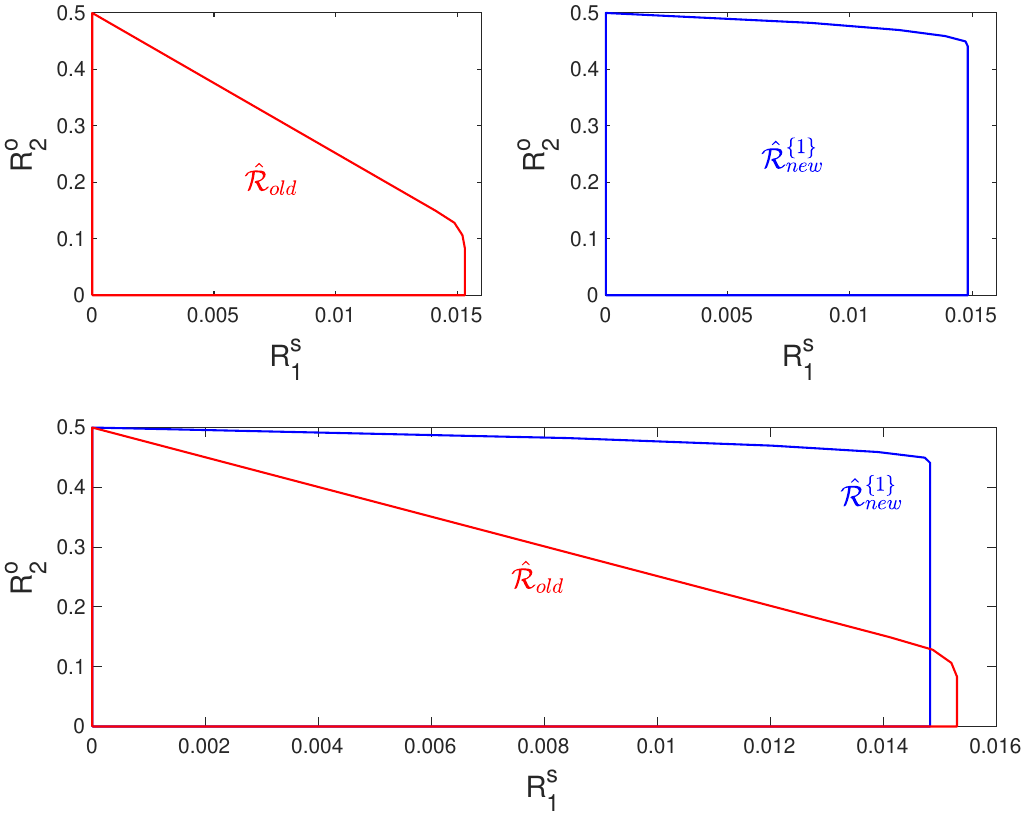}
			\caption{Convex hulls ${\hat {\mathscr R}}_{\text {old}}$ and ${\hat {\mathscr R}}_{\text {new}}^{\{1\}}$ with $(q_1, q_2) = (0.5, 0.75)$.}
			\label{convex_hull}
		\end{minipage}
		\hskip 1ex
		\begin{minipage}[t]{0.5\linewidth}
			\centering
			\includegraphics[width=3in]{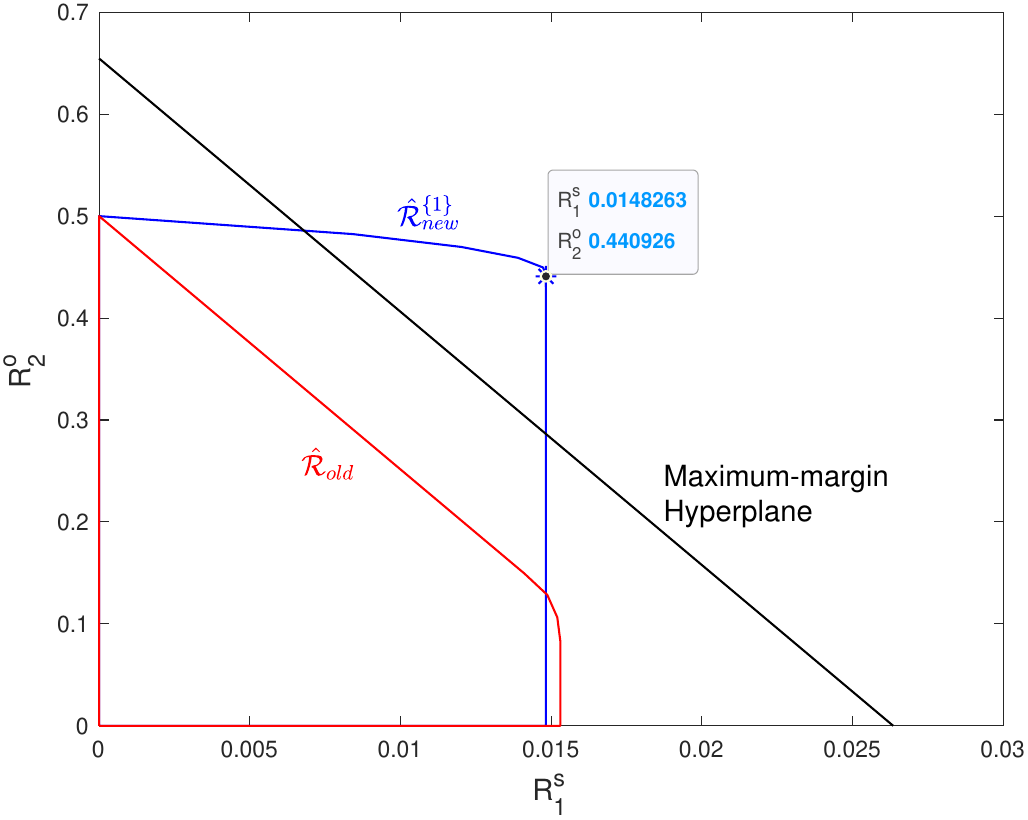}
			\caption{SVM-based separation of an extreme point in ${\hat {\mathscr R}}_{\text {new}}^{\{1\}}$ and ${\hat {\mathscr R}}_{\text {old}}$.}
			\label{SVM}
		\end{minipage}
	\end{figure}
	
	From Fig.~\ref{convex_hull} we see that ${\hat {\mathscr R}}_{\text {old}}$ and ${\hat {\mathscr R}}_{\text {new}}^{\{1\}}$ have non-overlapping areas.
	In the following, we choose a point in ${\hat {\mathscr R}}_{\text {new}}^{\{1\}}$, and show that it is linearly separable from  all the points contained in ${\hat {\mathscr R}}_{\text {old}}$ using a support vector machine (SVM).
	This point is thus not included in ${\hat {\mathscr R}}_{\text {old}}$.
	In particular, let ${\cal J}$ denote the convex set of all corner points of ${\hat {\mathscr R}} (X_{\cal K}, \{1, 2\})$ collected in depicting ${\hat {\mathscr R}}_{\text {old}}$ for all $(\alpha, \beta)$, and ${\cal J}_0$ denote the set of all extreme points of ${\cal J}$.
	Note that an extreme point of a convex set is a point in the set that does not lie on any open line segment between any other two points of the same set.
	Therefore, ${\cal J}_0$ is the minimal convex subset in ${\cal J}$ that has the same convex hull as ${\cal J}$.
	Denote the $j$-th extreme point in ${\cal J}_0$ by $\bm v_j = [R_1^{\text s}(j), R_2^{\text o}(j)]^T$.
	Similarly, we let ${\cal L}$ denote the convex set of all corner points of ${\hat {\mathscr R}} (X_{\cal K}, \{1\})$ collected in depicting ${\hat {\mathscr R}}_{\text {new}}^{\{1\}}$ for all $(\alpha, \beta)$, and ${\cal L}_0$ denote the set of all extreme points of ${\cal L}$.
	We choose an extreme point $\bm v_0 = [0.0148263, 0.440926]^T$ in ${\cal L}_0$.
	To show that $\bm v_0$ and ${\cal J}_0$ are linearly separable, we find a separating hyperplane by considering the following
	support vector machine (SVM) optimization problem:
	\begin{align}\label{opt_SVM}
		\mathop {\min }\limits_{ \bm w, t } \quad & || \bm w ||_2^2  \nonumber\\
		\text{s.t.} \quad\; & \bm w^T \bm v_0 - t \geq 1, \nonumber\\
		&  \bm w^T \bm v_j - t \leq -1, \forall j = 1, \cdots, |{\cal J}_0|,
	\end{align}
	where $\bm w \in {\mathbb R}^{2 \times 1}$ and $t \in {\mathbb R}$ are the optimization variables.
	The hyperplane $\bm w^T \bm v - t = 0$ is known as the maximum-margin hyperplane.
	(\ref{opt_SVM}) is a convex problem and can be easily numerically solved (e.g., using CVX in Matlab). 
	Fig.~\ref{SVM} shows ${\hat {\mathscr R}}_{\text {old}}$, ${\hat {\mathscr R}}_{\text {new}}^{\{1\}}$, $\bm v_0$, and the maximum-margin hyperplane. The existence of the maximum margin solution to (\ref{opt_SVM}) in turns yields that 
$\bm v_0 \in {\hat {\mathscr R}}_{\text {new}}^{\{1\}}$ is not in ${\hat {\mathscr R}}_{\text {old}}$.
Hence, the achievable region ${\mathscr R}_{\text {new}}$ obtained in this paper can strictly improve ${\mathscr R}_{\text {old}}$ 
given in \cite[Theorem~$1$]{xu2022achievable}.
	
	\begin{figure}
		\begin{minipage}[t]{0.5\linewidth}
			\centering
			\includegraphics[width=3in]{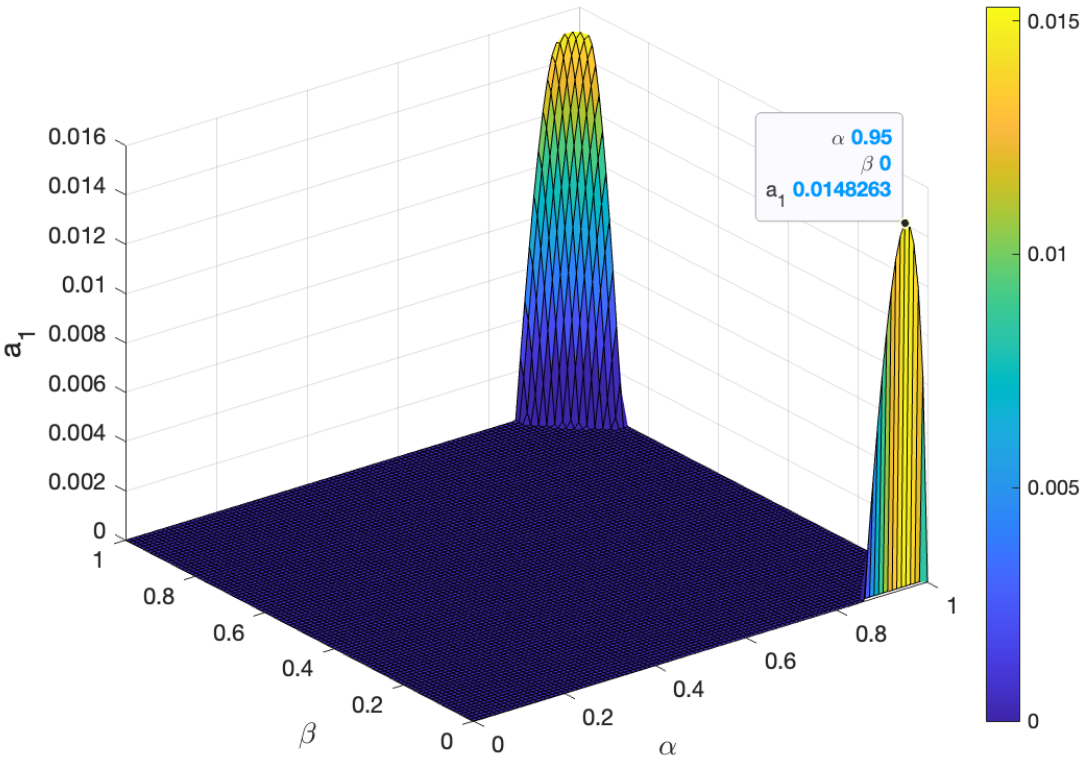}
			\caption{$a_1$ versus $\alpha$ and $\beta$ with ${\hat {\mathscr R}}_{\text {new}}^{\{1\}}$ with $(q_1, q_2) = (0.5, 0.75)$ and sampling size $\delta = 0.01$.}
			\label{a1_VS_alpha_beta}
		\end{minipage}
		\hskip 1ex
		\begin{minipage}[t]{0.5\linewidth}
			\centering
			\includegraphics[width=3in]{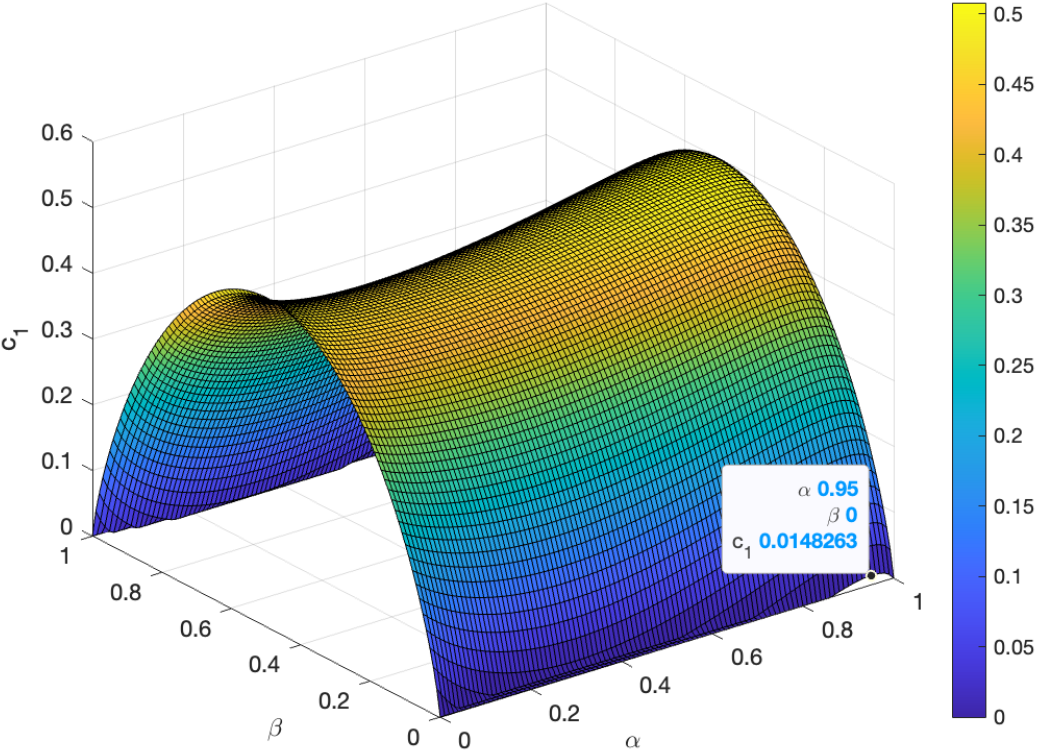}
			\caption{$c_1$ versus $\alpha$ and $\beta$ with ${\hat {\mathscr R}}_{\text {new}}^{\{1\}}$ with $(q_1, q_2) = (0.5, 0.75)$ and sampling size $\delta = 0.01$.}
			\label{c1_VS_alpha_beta}
		\end{minipage}
	\end{figure}
	
	We conclude with a qualitative and insightful interpretation of why, for the considered example, 
	${\hat {\mathscr R}}_{\text {old}}$ and ${\hat {\mathscr R}}_{\text {new}}^{\{1\}}$ have non-overlapping areas. 
	In Fig.~\ref{a1_VS_alpha_beta} and Fig.~\ref{c1_VS_alpha_beta}, we show $a_1$ and $c_1$ versus $\alpha$ and $\beta$ with sampling size $\delta = 0.01$.
	It can be seen from Fig.~\ref{a1_VS_alpha_beta} that for most values of $\alpha$ and $\beta$, $a_1 = 0$. 
	Since $a_1$ is the upper bound to $R_1^{\text s}$ in ${\hat {\mathscr R}} (X_{\cal K}, \{1, 2\})$, we have  is $R_1^{\text s} = 0$ in these cases.
	Only when $\alpha$ is around $0.95$ and $\beta$ is close to $0$ or $1$, will $R_1^{\text s}$ take on a relatively large value.
	However, in these cases, we see from Fig.~\ref{c1_VS_alpha_beta} that $c_1$ is small.
	Since $c_1$ is the upper bound to $R_1^{\text s} + R_2^{\text o}$ in ${\hat {\mathscr R}} (X_{\cal K}, \{1, 2\})$, it is impossible for $R_2^{\text o}$ to take a large value in these cases.
	This explains why when $R_1^{\text s}$ is large, $R_2^{\text o}$ is small in ${\hat {\mathscr R}}_{\text {old}}$ (see Fig.~\ref{convex_hull}).
	
	\begin{figure}
		\begin{minipage}[t]{0.5\linewidth}
			\centering
			\includegraphics[width=3in]{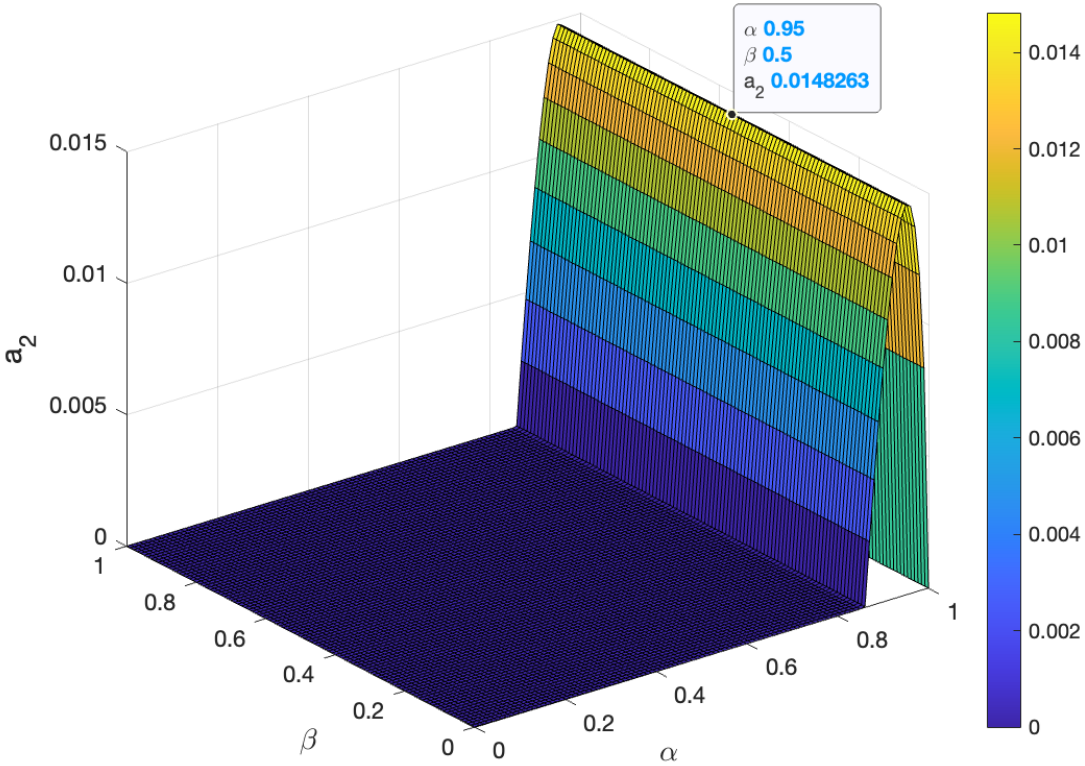}
			\caption{$a_2$ versus $\alpha$ and $\beta$ with ${\hat {\mathscr R}}_{\text {new}}^{\{1\}}$ with $(q_1, q_2) = (0.5, 0.75)$ and sampling size $\delta = 0.01$.}
			\label{a2_VS_alpha_beta}
		\end{minipage}
		\hskip 1ex
		\begin{minipage}[t]{0.5\linewidth}
			\centering
			\includegraphics[width=3in]{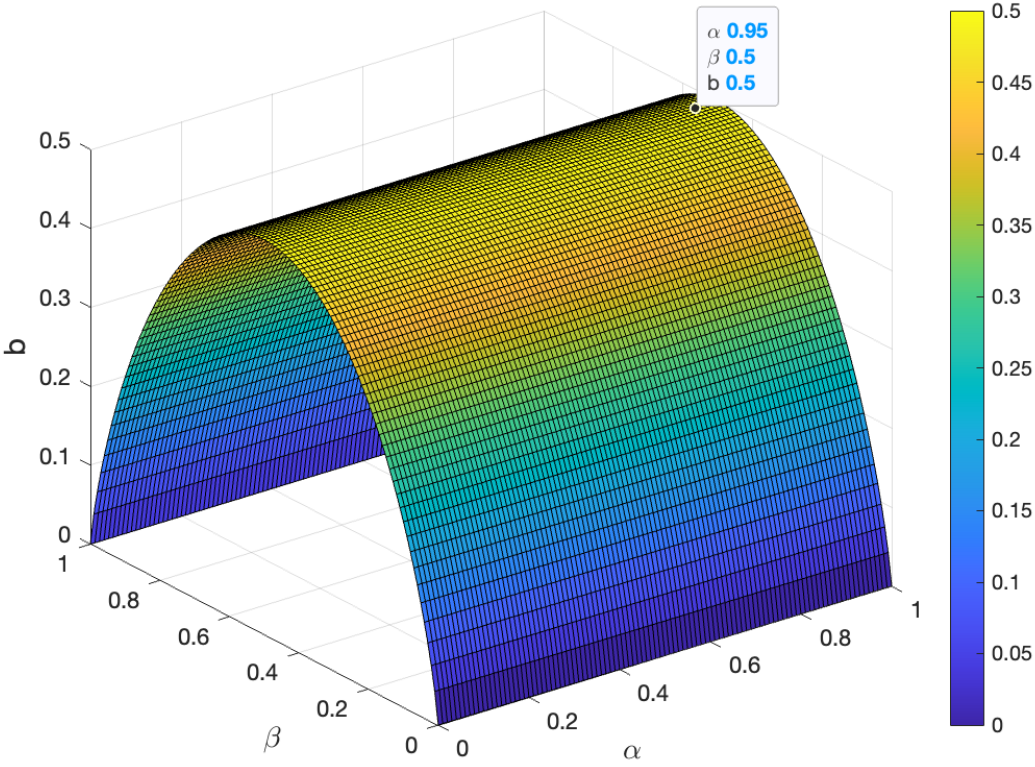}
			\caption{$b$ versus $\alpha$ and $\beta$ with ${\hat {\mathscr R}}_{\text {new}}^{\{1\}}$ with $(q_1, q_2) = (0.5, 0.75)$ and sampling size $\delta = 0.01$.}
			\label{b_VS_alpha_beta}
		\end{minipage}
	\end{figure}
	
	\begin{figure}
		\centering
		\includegraphics[scale=0.5]{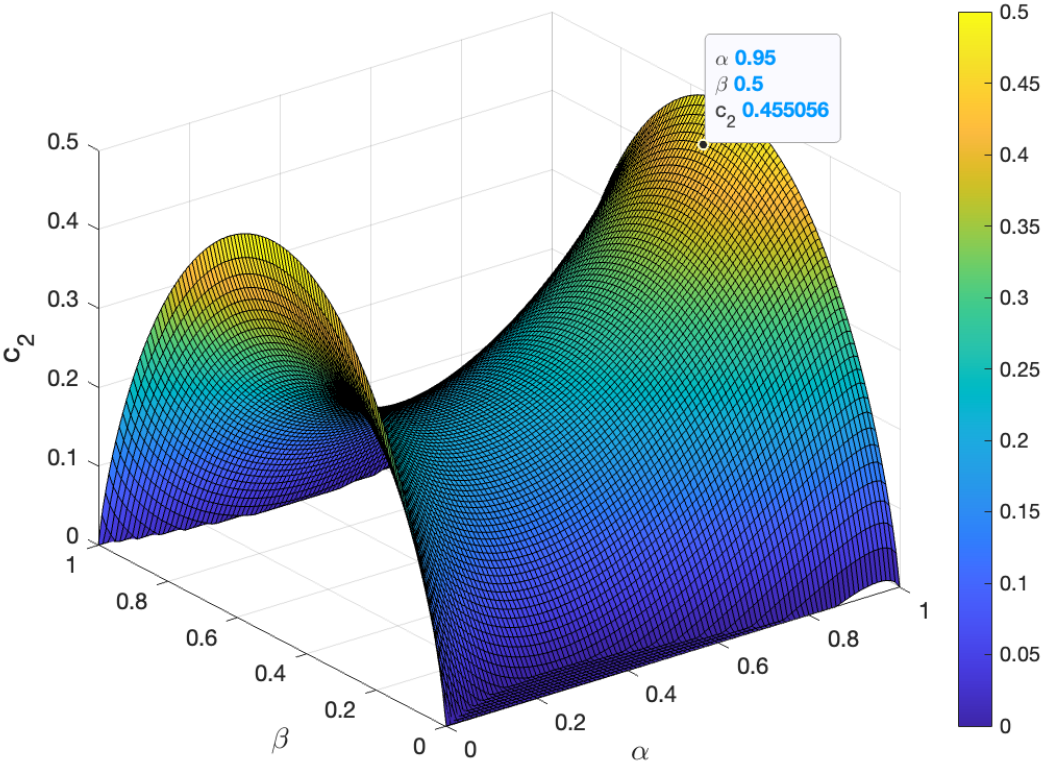}
		\caption{$c_2$ versus $\alpha$ and $\beta$ with ${\hat {\mathscr R}}_{\text {new}}^{\{1\}}$ with $(q_1, q_2) = (0.5, 0.75)$ and sampling size $\delta = 0.01$.}
		\label{c2_VS_alpha_beta}
	\end{figure}
	
	In Fig.~\ref{a2_VS_alpha_beta}, Fig.~\ref{b_VS_alpha_beta}, and Fig.~\ref{c2_VS_alpha_beta}, we plot $a_2$, $b$, and $c_2$ versus $\alpha$ and $\beta$, respectively.
	Different from $a_1$, Fig.~\ref{a2_VS_alpha_beta} shows that whatever the value of $\beta$, as long as $\alpha$ is around $0.95$, $a_2$ takes on a relatively large value.
	When $\alpha$ and $\beta$ are respectively $0.95$ and $0.5$, we see that $a_2$, $b$, and $c_2$ are all relatively large.
	Since $a_2$, $b$, and $c_2$ are respectively upper bounds to $R_1^{\text s}$, $R_2^{\text o}$, and $R_1^{\text s} + R_2^{\text o}$, it is possible for a point $(R_1^{\text s}, R_2^{\text o})$ in the region ${\hat {\mathscr R}} (X_{\cal K}, \{1\})$ to have both large $R_1^{\text s}$ and $R_2^{\text o}$.
	This explains why ${\hat {\mathscr R}}_{\text {new}}^{\{1\}}$ has a non-overlapping area on the upper-right side of ${\hat {\mathscr R}}_{\text {old}}$ in Fig.~\ref{convex_hull}.
	
In the achievability scheme of Section~\ref{achie_proof},  user~$2$ plays different roles in achieving the points in ${\hat {\mathscr R}} (X_{\cal K}, \{1, 2\})$ and ${\hat {\mathscr R}} (X_{\cal K}, \{1\})$, corresponding to different coding schemes, with  different advantages and disadvantages in determining the achievable region (see Remark~\ref{remark_lemma_DM_exten}). 
In particular, to achieve a point in ${\hat {\mathscr R}} (X_{\cal K}, \{1, 2\})$, both the two users have to introduce additional auxiliary messages such that the sum rate of their open and auxiliary messages is beyond Eve's decoding capability, and at the same time, all messages can be decoded by Bob.
In contrast, to achieve a point in ${\hat {\mathscr R}} (X_{\cal K}, \{1\})$, only user~$1$ has to introduce an additional auxiliary messages such that the sum rate of its open and auxiliary messages is beyond Eve's decoding capability.
These conditions are characterized by rate inequalities. 
The numerical results in this section show that for some values of $(\alpha, \beta)$, the conditions of the first coding scheme can be satisfied, but not those of the second one, such that rate point is in ${\hat {\mathscr R}} (X_{\cal K}, \{1, 2\})$ but not in ${\hat {\mathscr R}} (X_{\cal K}, \{1\})$.
For some other values of $(\alpha, \beta)$, the reverse is true.
Therefore, to obtain a larger achievable region, all possibles schemes should be considered.


\section{Conclusions}
\label{conclusion}

In this paper, we studied the information-theoretic secrecy for a $K$-user DM MAC-WT channel, where each user has both secret and open messages for the intended receiver.
Auxiliary messages were introduced to protect the confidential information.
To ensure that the rate of auxiliary messages is large enough for protecting the secret message from Eve and the resulting sum rate (together with the secret and open message rate) does not exceed Bob's decoding capability, we developed an inequality structure involving the rates of all users' secret, open, and auxiliary messages, and also gave the general proof.
We adopted strong secrecy, defined by the mutual information between all confidential messages and the received signal at Eve, as the secrecy metric.
To prove the achievability under this criterion, we analyzed the output statistics in terms of variational distance for the $K$-user standard DM-MAC channel.
In addition, we showed that users with zero secrecy rate may have different options in choosing their coding schemes.
By considering all possible options, we obtained a new achievable region for the considered channel that enlarges previously known results, and the improvement has been verified by a two-user binary-input real adder channel.
Considering the fact that, even for the standard two-user DM MAC-WT channel with only secret messages, a tight converse is still an open problem, in this paper we focused on extending the achievable region with respect to what was previously known, for the $K$-user DM MAC-WT with open and secret messages. The investigation of outer bounds with the aim of eventually proving a matching converse remains as a  fascinating and yet challenging future work.

\appendices

\section{Proof of Lemma~\ref{theorem_FM}}
\label{Prove_theorem_FM}

In this appendix, we prove Lemma~\ref{theorem_FM} by showing that (\ref{rate_region0}) is the projection of (\ref{region_FM1}) onto the hyperplane $\{ R_k^{\text a} = 0, \forall k \in {\cal K}\}$.
Note that due to (\ref{cond3}), the polytope defined by (\ref{rate_region0}) is non-empty.
Since as shown below we can successfully prove that (\ref{rate_region0}) is the projection of (\ref{region_FM1}), we know from \cite[Appendix~D]{el2011network} that the polytope defined by (\ref{region_FM1}) must be non-empty.
As stated in Remark~\ref{remark_FM}, it is impossible to prove Lemma~\ref{theorem_FM} by directly using the Fourier-Motzkin procedure to eliminate all $R_k^{\text a}$ in (\ref{region_FM1}), not only because of its huge complexity but also due to the fact that the elimination strategy works only if $K$ is given.
Hence, we adopt mathematical induction in the following to prove Lemma~\ref{theorem_FM}.

We first consider the base case with $K = 1$.
By eliminating $R_1^{\text a}$ in (\ref{region_FM1}) using the Fourier-Motzkin procedure \cite[Appendix D]{el2011network}, it can be easily proven that (\ref{rate_region0}) is the projection of (\ref{region_FM1}) onto the hyperplane $\{ R_1^{\text a} = 0\}$.
Lemma~\ref{theorem_FM} can thus be proven for this simple case.

Next, we consider the induction step.
Assume that for any given positive integer $K$, (\ref{rate_region0}) is the projection of (\ref{region_FM1}) onto the hyperplane $\{ R_k^{\text a} = 0, \forall k \in {\cal K}\}$.
Then, by this assumption, it is possible to obtain (\ref{rate_region0}) by eliminating the variables $R_k^{\text a}, \forall k \in {\cal K}$ using the Fourier-Motzkin procedure. 
For convenience, in the following we refer to this assumption as the {\em induction assumption}. 
Under the induction assumption, we shall prove that the statement of Lemma 1 holds  for $K+1$ users.
With $K+1$ users, (\ref{rate_region0}) and (\ref{region_FM1}) become
\begin{align}\label{rate_region_K+1}
\sum_{k \in \cal S} R_k^{\text s} + \sum_{k \in {\cal S}\setminus {\cal S}'} R_k^{\text o} \leq I(X_{\cal S}; Y| X_{\overline {\cal S}}) - I(X_{{\cal S}'}; Z), \forall {\cal S} \subseteq {\cal K} \cup \{ K + 1 \}, {\cal S}' \subseteq \cal S,
\end{align}
and
\begin{equation}\label{region_FM_K+1}
\left\{
\begin{array}{ll}
R_k^{\text a} \geq 0, \forall k \in {\cal K} \cup \{ K + 1 \}, \\
\sum\limits_{k \in {\cal S}} (R_k^{\text s} + R_k^{\text o} + R_k^{\text a}) \leq I(X_{\cal S}; Y| X_{\overline {\cal S}}), \forall {\cal S} \subseteq {\cal K} \cup \{ K + 1 \}, \\
\sum\limits_{k \in {\cal S}} (R_k^{\text o} + R_k^{\text a}) \geq I(X_{\cal S}; Z), \forall {\cal S} \subseteq {\cal K} \cup \{ K + 1 \}.
\end{array} \right.
\end{equation}
We need to show that (\ref{rate_region_K+1}) is the projection of (\ref{region_FM_K+1}) onto the hyperplane $\{ R_k^{\text a} = 0, \forall k \in {\cal K} \cup \{ K + 1 \}\}$, i.e., (\ref{rate_region_K+1}) can be obtained by eliminating $R_k^{\text a}, \forall k \in {\cal K}$ as well as $R_{K + 1}^{\text a}$ in (\ref{region_FM_K+1}).
For this purpose, by separating user $K + 1$ from users in set ${\cal K}$, we rewrite (\ref{rate_region_K+1}) equivalently as
\begin{subequations}\label{rate_region_K+1_2}
	\begin{align}
	& \sum_{k \in \cal S} R_k^{\text s} + \sum_{k \in {\cal S}\setminus {\cal S}'} R_k^{\text o} \leq I(X_{\cal S}; Y| X_{\overline {\cal S}}, X_{K + 1}) - I(X_{{\cal S}'}; Z), \forall {\cal S} \subseteq {\cal K}, {\cal S}' \subseteq \cal S,\label{rate_region_K+1_2a}\\
	& \sum_{k \in \cal S} R_k^{\text s} \!+\! R_{K + 1}^{\text s} \!+\! \sum_{k \in {\cal S}\setminus {\cal S}'} R_k^{\text o} \!+\! R_{K + 1}^{\text o} \!\leq\! I(X_{\cal S}, X_{K + 1}; Y| X_{\overline {\cal S}}) \!-\! I(X_{{\cal S}'}; Z), \forall {\cal S} \subseteq {\cal K}, {\cal S}' \subseteq \cal S,\label{rate_region_K+1_2b}\\
	& \sum_{k \in \cal S} R_k^{\text s} \!+\! R_{K + 1}^{\text s} \!+\! \sum_{k \in {\cal S}\setminus {\cal S}'} \!R_k^{\text o} \leq I(X_{\cal S}, X_{K + 1}; Y| X_{\overline {\cal S}}) \!-\! I(X_{{\cal S}'}, X_{K + 1}; Z), \forall {\cal S} \subseteq {\cal K}, {\cal S}' \subseteq \cal S,\label{rate_region_K+1_2c}
	\end{align}
\end{subequations}
and (\ref{region_FM_K+1}) as
\begin{subequations}\label{region_FM_K+1_2}
	\begin{align}
	& R_k^{\text a} \geq 0, \forall k \in {\cal K}, \label{region_FM_K+1_2a}\\
	& \sum\limits_{k \in {\cal S}} (R_k^{\text s} + R_k^{\text o} + R_k^{\text a}) \leq I(X_{\cal S}; Y| X_{\overline {\cal S}}, X_{K + 1}), \forall {\cal S} \subseteq {\cal K}, \label{region_FM_K+1_2b}\\
	& \sum\limits_{k \in {\cal S}} (R_k^{\text o} + R_k^{\text a}) \geq I(X_{\cal S}; Z), \forall {\cal S} \subseteq {\cal K}, \label{region_FM_K+1_2c}\\
	& \sum\limits_{k \in {\cal S}} (R_k^{\text s} \!+\! R_k^{\text o} \!+\! R_k^{\text a}) \leq I(X_{\cal S}, X_{K + 1}; Y| X_{\overline {\cal S}}) \!-\! (R_{K+1}^{\text s} \!+\! R_{K+1}^{\text o} \!+\! R_{K+1}^{\text a}), \forall {\cal S} \subseteq {\cal K}, {\cal S} \neq \emptyset, \label{region_FM_K+1_2d}\\
	& \sum\limits_{k \in {\cal S}} (R_k^{\text o} + R_k^{\text a}) \geq I(X_{\cal S}, X_{K + 1}; Z) - (R_{K+1}^{\text o} + R_{K+1}^{\text a}), \forall {\cal S} \subseteq {\cal K}, {\cal S} \neq \emptyset,\label{region_FM_K+1_2e}\\
	& R_{K+1}^{\text a} \geq 0, \label{region_FM_K+1_2f}\\
	& R_{K+1}^{\text s} + R_{K+1}^{\text o} + R_{K+1}^{\text a} \leq I(X_{K+1}; Y| X_{{\cal K}}), \label{region_FM_K+1_2g}\\
	& R_{K+1}^{\text o} + R_{K+1}^{\text a} \geq I(X_{K+1}; Z). \label{region_FM_K+1_2h}
	\end{align}
\end{subequations}
Note that in (\ref{region_FM_K+1_2d}) and (\ref{region_FM_K+1_2e}) we let ${\cal S} \neq \emptyset$ since otherwise they reduce to (\ref{region_FM_K+1_2g}) and (\ref{region_FM_K+1_2h}), which do not contain $R_k^{\text a}, \forall k \in {\cal K}$.
In the following, we eliminate first $R_k^{\text a}, \forall k \in {\cal K}$ and then $R_{K+1}^{\text a}$.

\subsection{Elimination of $R_k^{\text a}, \forall k \in {\cal K}$}

\begin{figure*}[ht]
	\centering
	\includegraphics[scale=0.6]{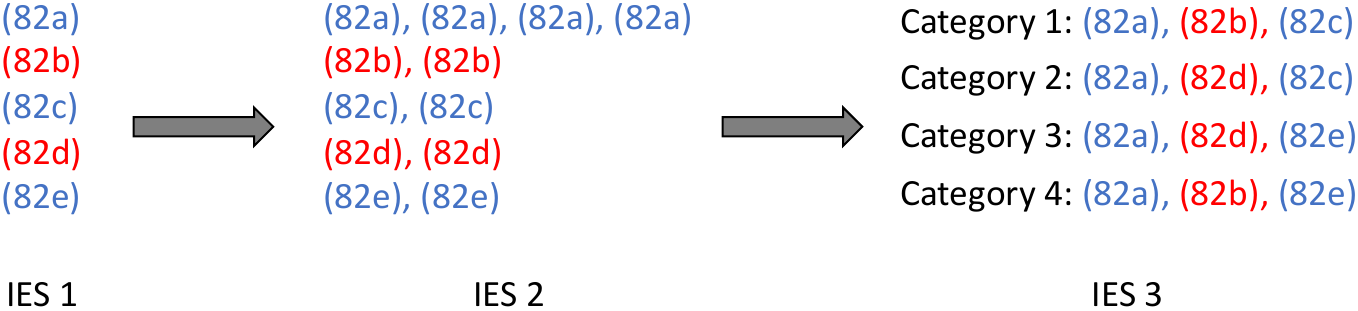}
	\caption{Division of ISE~1.}
	\label{ineq}
\end{figure*}

To eliminate $R_k^{\text a}, \forall k \in {\cal K}$ in (\ref{region_FM_K+1_2}), we focus on (\ref{region_FM_K+1_2a})~$\sim$ (\ref{region_FM_K+1_2e}) since only these inequalities contain $R_k^{\text a}, \forall k \in {\cal K}$ while (\ref{region_FM_K+1_2f})~$\sim$ (\ref{region_FM_K+1_2h}) do not.
Since there are $K$ different $R_k^{\text a}$, as stated above, it is impractical to eliminate $R_k^{\text a}$ one by one.
Hence, instead of eliminating $R_k^{\text a}, \forall k \in {\cal K}$ directly from (\ref{region_FM_K+1_2a})~$\sim$ (\ref{region_FM_K+1_2e}), we divide these inequalities into $4$ categories (see Fig.~\ref{ineq}), which together consider all possible upper and lower bound pairs on $R_k^{\text a}, \forall k \in {\cal K}$, and eliminate $R_k^{\text a}, \forall k \in {\cal K}$ in each category using the induction assumption.

For convenience, we call the inequality system (IES) consisting of (\ref{region_FM_K+1_2a})~$\sim$ (\ref{region_FM_K+1_2e}) IES~1.
As we can see from (\ref{region_FM_K+1_2}), in IES~1, there are three lower bounds on $R_k^{\text a}, \forall k \in {\cal K}$, i.e., (\ref{region_FM_K+1_2a}), (\ref{region_FM_K+1_2c}), and (\ref{region_FM_K+1_2e}), which are marked in blue color in Fig.~\ref{ineq}, and two upper bounds on $R_k^{\text a}, \forall k \in {\cal K}$, i.e., (\ref{region_FM_K+1_2b}) and (\ref{region_FM_K+1_2d}), which are marked in red color.
To eliminate $R_k^{\text a}, \forall k \in {\cal K}$ in IES~1 based on the induction assumption, as shown in Fig.~\ref{ineq}, we first repeat the inequalities in IES~1 and obtain IES~2.
Since the repeated inequalities can neither expand nor shrink the region of IES~1, IES~1 and IES~2 are equivalent.
Then, we change the order of the inequalities in IES~2 and obtain IES~3.
Obviously, IES~3 is equivalent to IES~2, and is thus also equivalent IES~1.
Therefore, eliminating $R_k^{\text a}, \forall k \in {\cal K}$ in IES~1 is equivalent to eliminating those in IES~3, which as we will show below, can be realized by separately eliminating $R_k^{\text a}, \forall k \in {\cal K}$ in each category using the induction assumption.

\subsubsection{Category~$1$}

We include inequalities (\ref{region_FM_K+1_2a}), (\ref{region_FM_K+1_2b}), and (\ref{region_FM_K+1_2c}) in Category~$1$, and rewrite them as follows for clarity
\begin{equation}\label{region_FM_K+1_2_abc}
\left\{
\begin{array}{ll}
R_k^{\text a} \geq 0, \forall k \in {\cal K}, \\
\sum\limits_{k \in {\cal S}} (R_k^{\text s} + R_k^{\text o} + R_k^{\text a}) \leq I(X_{\cal S}; Y| X_{\overline {\cal S}}, X_{K + 1}), \forall {\cal S} \subseteq {\cal K}, \\
\sum\limits_{k \in {\cal S}} (R_k^{\text o} + R_k^{\text a}) \geq I(X_{\cal S}; Z), \forall {\cal S} \subseteq {\cal K}.
\end{array} \right.
\end{equation}
Note that (\ref{region_FM_K+1_2_abc}) has a similar formulation as (\ref{region_FM1}).
Then, from the induction assumption it is known that the projection of (\ref{region_FM_K+1_2_abc}) onto the hyperplane $\{ R_k^{\text a} = 0, \forall k \in {\cal K} \}$ is
\begin{align}\label{R_g_abc}
\sum_{k \in \cal S} R_k^{\text s} + \sum_{k \in {\cal S}\setminus {\cal S}'} R_k^{\text o} \leq I(X_{\cal S}; Y| X_{\overline {\cal S}}, X_{K + 1}) - I(X_{{\cal S}'}; Z), \forall {\cal S} \subseteq {\cal K}, {\cal S}' \subseteq \cal S,
\end{align}
which is the same as (\ref{rate_region_K+1_2a}).

\subsubsection{Category~$2$}

In this category we include (\ref{region_FM_K+1_2a}), (\ref{region_FM_K+1_2d}), and (\ref{region_FM_K+1_2c}), and rewrite them as
\begin{equation}\label{region_FM_K+1_2_adc}
\left\{\!\!\!
\begin{array}{ll}
R_k^{\text a} \geq 0, \forall k \in {\cal K}, \\
\sum\limits_{k \in {\cal S}} (R_k^{\text s} \!+\! R_k^{\text o} \!+\! R_k^{\text a}) \leq I(X_{\cal S}, X_{K + 1}; Y| X_{\overline {\cal S}}) \!-\! (R_{K+1}^{\text s} \!+\! R_{K+1}^{\text o} \!+\! R_{K+1}^{\text a}), \forall {\cal S} \subseteq {\cal K}, {\cal S} \neq \emptyset, \\
\sum\limits_{k \in {\cal S}} (R_k^{\text o} + R_k^{\text a}) \geq I(X_{\cal S}; Z), \forall {\cal S} \subseteq {\cal K}.
\end{array} \right.
\end{equation}
Note that though we let ${\cal S} \neq \emptyset$ in (\ref{region_FM_K+1_2d}), (\ref{region_FM_K+1_2_adc}) still has a similar expression as (\ref{region_FM1}).
It can be checked that if ${\cal S} = \emptyset$, the second inequality of (\ref{region_FM1}) gives $0 \leq 0$, which can be omitted.
Hence, we may also let ${\cal S} \neq \emptyset$ in (\ref{region_FM1}) without changing its formulation.
The induction assumption can thus be used to eliminate $R_k^{\text a}, \forall k \in {\cal K}$ in (\ref{region_FM_K+1_2_adc}) and obtain
\begin{align}\label{R_g_adc}
& \sum_{k \in \cal S} R_k^{\text s} + \sum_{k \in {\cal S}\setminus {\cal S}'} R_k^{\text o} \leq I(X_{\cal S}, X_{K + 1}; Y| X_{\overline {\cal S}}) - (R_{K+1}^{\text s} + R_{K+1}^{\text o} + R_{K+1}^{\text a}) - I(X_{{\cal S}'}; Z),\nonumber\\
& \quad\quad\quad\quad\quad\quad\quad\quad\; \forall {\cal S} \subseteq {\cal K}, {\cal S} \neq \emptyset, {\cal S}' \subseteq \cal S,
\end{align}
which contains $R_{K+1}^{\text a}$ and thus has to be considered in the next step when eliminating $R_{K+1}^{\text a}$.

\subsubsection{Category~$3$}

In Category~$3$ we include (\ref{region_FM_K+1_2a}), (\ref{region_FM_K+1_2d}), and (\ref{region_FM_K+1_2e}), and rewrite them as follows
\begin{equation}\label{region_FM_K+1_2_ade}
\left\{\!\!\!
\begin{array}{ll}
R_k^{\text a} \geq 0, \forall k \in {\cal K}, \\
\sum\limits_{k \in {\cal S}} (R_k^{\text s} \!+\! R_k^{\text o} \!+\! R_k^{\text a}) \leq I(X_{\cal S}, X_{K + 1}; Y| X_{\overline {\cal S}}) \!-\! (R_{K+1}^{\text s} \!+\! R_{K+1}^{\text o} \!+\! R_{K+1}^{\text a}), \forall {\cal S} \subseteq {\cal K}, {\cal S} \neq \emptyset, \\
\sum\limits_{k \in {\cal S}} (R_k^{\text o} + R_k^{\text a}) \geq I(X_{\cal S}, X_{K + 1}; Z) - (R_{K+1}^{\text o} + R_{K+1}^{\text a}), \forall {\cal S} \subseteq {\cal K}, {\cal S} \neq \emptyset.
\end{array} \right.
\end{equation}
Using the induction assumption to eliminate $R_k^{\text a}, \forall k \in {\cal K}$, we have
\begin{align}\label{R_g_ade}
& \sum_{k \in \cal S} R_k^{\text s} + \sum_{k \in {\cal S}\setminus {\cal S}'} R_k^{\text o} \nonumber\\
\leq & I(X_{\cal S}, X_{K + 1}; Y| X_{\overline {\cal S}}) - (R_{K+1}^{\text s} + R_{K+1}^{\text o} + R_{K+1}^{\text a}) - \left[ I(X_{{\cal S}'}, X_{K + 1}; Z) - (R_{K+1}^{\text o} + R_{K+1}^{\text a}) \right],\nonumber\\
= & I(X_{\cal S}, X_{K + 1}; Y| X_{\overline {\cal S}}) - I(X_{{\cal S}'}, X_{K + 1}; Z) - R_{K+1}^{\text s}, \forall {\cal S} \subseteq {\cal K}, {\cal S} \neq \emptyset, {\cal S}' \subseteq {\cal S}, {\cal S}' \neq \emptyset.
\end{align}
By comparing (\ref{R_g_ade}) with (\ref{rate_region_K+1_2c}), it is known that (\ref{R_g_ade}) consists of partial inequalities in (\ref{rate_region_K+1_2c}) with ${\cal S} \subseteq {\cal K}, {\cal S} \neq \emptyset, {\cal S}' \subseteq {\cal S}, {\cal S}' \neq \emptyset$.

\subsubsection{Category~$4$}

We include inequalities (\ref{region_FM_K+1_2a}), (\ref{region_FM_K+1_2b}), and (\ref{region_FM_K+1_2e}) in Category~$4$, and rewrite them as follows
\begin{equation}\label{region_FM_K+1_2_abe}
\left\{
\begin{array}{ll}
R_k^{\text a} \geq 0, \forall k \in {\cal K}, \\
\sum\limits_{k \in {\cal S}} (R_k^{\text s} + R_k^{\text o} + R_k^{\text a}) \leq I(X_{\cal S}; Y| X_{\overline {\cal S}}, X_{K + 1}), \forall {\cal S} \subseteq {\cal K}, \\
\sum\limits_{k \in {\cal S}} (R_k^{\text o} + R_k^{\text a}) \geq I(X_{\cal S}, X_{K + 1}; Z) - (R_{K+1}^{\text o} + R_{K+1}^{\text a}), \forall {\cal S} \subseteq {\cal K}, {\cal S} \neq \emptyset.
\end{array} \right.
\end{equation}
The following projection of (\ref{region_FM_K+1_2_abe}) can then be obtained from the induction assumption
\begin{align}\label{R_g_abe}
& \sum_{k \in \cal S} R_k^{\text s} + \sum_{k \in {\cal S}\setminus {\cal S}'} R_k^{\text o} \leq I(X_{\cal S}; Y| X_{\overline {\cal S}}, X_{K + 1}) - I(X_{{\cal S}'}, X_{K + 1}; Z) + R_{K+1}^{\text o} + R_{K+1}^{\text a},\nonumber\\
& \quad\quad\quad\quad\quad\quad\quad\quad\; \forall {\cal S} \subseteq {\cal K}, {\cal S} \neq \emptyset, {\cal S}' \subseteq {\cal S}, {\cal S}' \neq \emptyset,
\end{align}
which also contains $R_{K+1}^{\text a}$ and has to be considered in the next step when eliminating $R_{K+1}^{\text a}$.

Combining (\ref{region_FM_K+1_2f})~$\sim$ (\ref{region_FM_K+1_2h}), (\ref{R_g_abc}), (\ref{R_g_adc}), (\ref{R_g_ade}), and (\ref{R_g_abe}), we get a projection of (\ref{region_FM_K+1_2}) onto the hyperplane $\{ R_k^{\text a} = 0, \forall k \in {\cal K} \}$.
To further get a projection of (\ref{region_FM_K+1_2}) onto the hyperplane $\{ R_k^{\text a} = 0, \forall k \in {\cal K} \cup \{ K + 1 \}\}$, we have to eliminate $R_{K + 1}^{\text a}$.

\subsection{Elimination of $R_{K+1}^{\text a}$}

In this subsection, we eliminate $R_{K+1}^{\text a}$ in (\ref{region_FM_K+1_2}). Note that in the previous step, we have eliminated $R_k^{\text a}, \forall k \in {\cal K}$ in (\ref{region_FM_K+1_2a})~$\sim$ (\ref{region_FM_K+1_2e}), and obtained (\ref{R_g_abc}), (\ref{R_g_adc}), (\ref{R_g_ade}), and (\ref{R_g_abe}), in which (\ref{R_g_adc}) and (\ref{R_g_abe}) contain $R_{K+1}^{\text a}$. Therefore, when eliminating $R_{K+1}^{\text a}$, we should not consider all the inequalities (\ref{region_FM_K+1_2a})~$\sim$ (\ref{region_FM_K+1_2h}) in (\ref{region_FM_K+1_2}). Instead, we consider part of them, i.e., (\ref{region_FM_K+1_2f})~$\sim$ (\ref{region_FM_K+1_2h}), and also (\ref{R_g_adc}) and (\ref{R_g_abe}).
From (\ref{region_FM_K+1_2g}) and (\ref{R_g_adc}), we get the following upper bounds on $R_{K+1}^{\text a}$
\begin{subequations}\label{R_g_ub}
	\begin{align}
	R_{K+1}^{\text a} & \leq I(X_{K+1}; Y| X_{{\cal K}}) - (R_{K+1}^{\text s} + R_{K+1}^{\text o}), \label{R_g_ub1}\\
	R_{K+1}^{\text a} & \leq I(X_{\cal S}, X_{K + 1}; Y| X_{\overline {\cal S}}) - I(X_{{\cal S}'}; Z) - \Big( \sum_{k \in \cal S} R_k^{\text s} + \sum_{k \in {\cal S}\setminus {\cal S}'} R_k^{\text o} + R_{K+1}^{\text s} + R_{K+1}^{\text o} \Big), \nonumber\\
	& \; \forall {\cal S} \subseteq {\cal K}, {\cal S} \neq \emptyset, {\cal S}' \subseteq \cal S.\label{R_g_ub2}
	\end{align}
\end{subequations}
Moreover, the following lower bounds on $R_{K+1}^{\text a}$ can be obtained from (\ref{region_FM_K+1_2f}), (\ref{region_FM_K+1_2h}), and (\ref{R_g_abe})
\begin{subequations}\label{R_g_lb}
	\begin{align}
	R_{K+1}^{\text a} & \geq 0, \label{R_g_lb1}\\
	R_{K+1}^{\text a} & \geq I(X_{K+1}; Z) - R_{K+1}^{\text o}, \label{R_g_lb2}\\
	R_{K+1}^{\text a} & \geq - I(X_{\cal S}; Y| X_{\overline {\cal S}}, X_{K + 1}) + I(X_{{\cal S}'}, X_{K + 1}; Z) + \sum_{k \in \cal S} R_k^{\text s} + \sum_{k \in {\cal S}\setminus {\cal S}'} R_k^{\text o} - R_{K+1}^{\text o}, \nonumber\\
	& \; \forall {\cal S} \subseteq {\cal K}, {\cal S} \neq \emptyset, {\cal S}' \subseteq {\cal S}, {\cal S}' \neq \emptyset.\label{R_g_lb3}
	\end{align}
\end{subequations}
Comparing these upper and lower bounds, we can eliminate $R_{K+1}^{\text a}$.

Firstly, we compare (\ref{R_g_ub}) with (\ref{R_g_lb1}), and get
\begin{subequations}\label{R_g_ub_lb1}
	\begin{align}
	& R_{K+1}^{\text s} + R_{K+1}^{\text o} \leq I(X_{K+1}; Y| X_{{\cal K}}), \label{R_g_ub1_lb1}\\
	& \sum_{k \in \cal S} R_k^{\text s} + \sum_{k \in {\cal S}\setminus {\cal S}'} R_k^{\text o} + R_{K+1}^{\text s} + R_{K+1}^{\text o} \leq I(X_{\cal S}, X_{K + 1}; Y| X_{\overline {\cal S}}) - I(X_{{\cal S}'}; Z), \nonumber\\
	& \quad\quad\quad\quad\quad\quad\quad\quad\quad\quad\quad\quad\quad\quad\quad\; \forall {\cal S} \subseteq {\cal K}, {\cal S} \neq \emptyset, {\cal S}' \subseteq \cal S.\label{R_g_ub2_lb1}
	\end{align}
\end{subequations}
The inequalities (\ref{R_g_ub1_lb1}) and (\ref{R_g_ub2_lb1}) can be integrated into one formula as follows
\begin{align}\label{R_g_ub1_ub2_lb1}
\sum_{k \in \cal S} R_k^{\text s} \!+\! \sum_{k \in {\cal S}\setminus {\cal S}'} R_k^{\text o} \!+\! R_{K+1}^{\text s} \!+\! R_{K+1}^{\text o} \leq I(X_{\cal S}, X_{K + 1}; Y| X_{\overline {\cal S}}) \!-\! I(X_{{\cal S}'}; Z), \forall {\cal S} \subseteq {\cal K}, {{\cal S}'} \subseteq \cal S,
\end{align}
which is the same as (\ref{rate_region_K+1_2b}).

Secondly, we compare (\ref{R_g_ub}) with (\ref{R_g_lb2}), and get
\begin{subequations}\label{R_g_ub_lb2}
	\begin{align}
	& R_{K+1}^{\text s} \leq I(X_{K+1}; Y| X_{{\cal K}}) - I(X_{K+1}; Z), \label{R_g_ub1_lb2}\\
	& \sum_{k \in \cal S} R_k^{\text s} + \sum_{k \in {\cal S}\setminus {\cal S}'} R_k^{\text o} + R_{K+1}^{\text s} \leq I(X_{\cal S}, X_{K + 1}; Y| X_{\overline {\cal S}}) - I(X_{{\cal S}'}; Z) - I(X_{K+1}; Z), \nonumber\\
	& \quad\quad\quad\quad\quad\quad\quad\quad\quad\quad\quad\quad \forall {\cal S} \subseteq {\cal K}, {\cal S} \neq \emptyset, {\cal S}' \subseteq \cal S.\label{R_g_ub2_lb2}
	\end{align}
\end{subequations}
By separately considering ${\cal S}' = \emptyset$ and ${\cal S}' \neq \emptyset$, we may divide (\ref{R_g_ub2_lb2}) into two formulas as below
\begin{subequations}\label{R_g_ub_lb2_2}
	\begin{align}
	& \sum_{k \in \cal S} (R_k^{\text s} + R_k^{\text o}) + R_{K+1}^{\text s} \leq I(X_{\cal S}, X_{K + 1}; Y| X_{\overline {\cal S}}) - I(X_{K+1}; Z), \forall {\cal S} \subseteq {\cal K}, {\cal S} \neq \emptyset, {\cal S}' = \emptyset, \label{R_g_ub2_lb2_1}\\
	& \sum_{k \in \cal S} R_k^{\text s} + \sum_{k \in {\cal S}\setminus {\cal S}'} R_k^{\text o} + R_{K+1}^{\text s} \leq I(X_{\cal S}, X_{K + 1}; Y| X_{\overline {\cal S}}) - I(X_{{\cal S}'}; Z) - I(X_{K+1}; Z), \nonumber\\
	& \quad\quad\quad\quad\quad\quad\quad\quad\quad\quad\quad\quad \forall {\cal S} \subseteq {\cal K}, {\cal S} \neq \emptyset, {\cal S}' \subseteq {\cal S}, {\cal S}' \neq \emptyset.\label{R_g_ub2_lb2_2}
	\end{align}
\end{subequations}
From (\ref{R_g_ade}), (\ref{R_g_ub1_lb2}), and (\ref{R_g_ub2_lb2_1}), it can be found that these inequalities can be integrated into one formula as follows
\begin{align}\label{rate_region_K+1_2c2}
\sum_{k \in \cal S} R_k^{\text s} \!+\! \sum_{k \in {\cal S}\setminus {\cal S}'} R_k^{\text o} \!+\! R_{K + 1}^{\text s} \!\leq\! I(X_{\cal S}, X_{K + 1}; Y| X_{\overline {\cal S}}) \!-\! I(X_{{\cal S}'}, X_{K + 1}; Z), \forall {\cal S} \subseteq {\cal K}, {\cal S}' \subseteq \cal S,
\end{align}
which is the same as (\ref{rate_region_K+1_2c}).
Combining (\ref{R_g_abc}), (\ref{R_g_ub1_ub2_lb1}), and (\ref{rate_region_K+1_2c2}), it is known that (\ref{rate_region_K+1_2}) or (\ref{rate_region_K+1}) has already been obtained.
All the other inequalities resulted from the elimination procedure should be redundant if Lemma~\ref{theorem_FM} is true.
Hence, we have to prove that (\ref{R_g_ub2_lb2_2}) is redundant.
Since $X_{K + 1}$ is independent of $X_{{\cal S}'}$, (\ref{R_g_ade}) can be rewritten and relaxed as follows
\begin{align}\label{R_g_ade_relax}
& \sum_{k \in \cal S} R_k^{\text s} + \sum_{k \in {\cal S}\setminus {\cal S}'} R_k^{\text o} + R_{K + 1}^{\text s} \nonumber\\
\leq & I(X_{\cal S}, X_{K + 1}; Y| X_{\overline {\cal S}}) - I(X_{{\cal S}'}, X_{K + 1}; Z) \nonumber\\
\leq & I(X_{\cal S}, X_{K + 1}; Y| X_{\overline {\cal S}}) - I(X_{{\cal S}'}; Z) - I(X_{K+1}; Z), \forall {\cal S} \subseteq {\cal K}, {\cal S} \neq \emptyset, {\cal S}' \subseteq {\cal S}, {\cal S}' \neq \emptyset,
\end{align}
where the second inequality is the upper bound in (\ref{R_g_ub2_lb2_2}).
This indicates that (\ref{R_g_ade}) sets a tighter upper bond on $\sum_{k \in \cal S} R_k^{\text s} + \sum_{k \in {\cal S}\setminus {\cal S}'} R_k^{\text o} + R_{K + 1}^{\text s}$ than (\ref{R_g_ub2_lb2_2}).
Since (\ref{R_g_ade}) is included in (\ref{rate_region_K+1_2c2}) or (\ref{rate_region_K+1_2c}), (\ref{R_g_ub2_lb2_2}) is thus redundant.

Finally, we compare (\ref{R_g_ub}) with (\ref{R_g_lb3}), which results in
\begin{subequations}\label{R_g_ub_lb3}
	\begin{align}
	& \sum_{k \in \cal S} R_k^{\text s} + \sum_{k \in {\cal S}\setminus {\cal S}'} R_k^{\text o} + R_{K+1}^{\text s} \leq I(X_{\cal S}; Y| X_{\overline {\cal S}}, X_{K + 1}) + I(X_{K+1}; Y| X_{{\cal K}}) - I(X_{{\cal S}'}, X_{K+1}; Z), \nonumber\\ 
	& \quad\quad\quad\quad\quad\quad\quad\quad\quad\quad\quad\quad \forall {\cal S} \subseteq {\cal K}, {\cal S} \neq \emptyset, {\cal S}' \subseteq {\cal S}, {\cal S}' \neq \emptyset,\vspace{0.5em} \label{R_g_ub1_lb3}\\
	& \sum_{k \in \cal S} R_k^{\text s} \!+\! \sum_{k \in {\cal S}\setminus {\cal S}'}\! R_k^{\text o} \!+\! \sum_{k \in {\cal S}_1} R_k^{\text s} \!+\! \sum_{k \in {\cal S}_1\setminus {\cal S}_1'}\! R_k^{\text o} \!+\! R_{K+1}^{\text s} \leq I(X_{\cal S}; Y| X_{\overline {\cal S}}, X_{K + 1}) - I(X_{{\cal S}'}, X_{K + 1}; Z) \nonumber\\
	& + I(X_{{\cal S}_1}, X_{K + 1}; Y| X_{\overline {{\cal S}_1}}) - I(X_{{\cal S}_1'}; Z), \forall {\cal S}, {\cal S}_1 \subseteq {\cal K}, {\cal S}, {\cal S}_1 \neq \emptyset, {\cal S}' \subseteq {\cal S}, {\cal S}' \neq \emptyset, {\cal S}_1' \subseteq {\cal S}_1.\label{R_g_ub2_lb3}
	\end{align}
\end{subequations}
Note that when comparing (\ref{R_g_ub2}) with (\ref{R_g_lb3}), which gives (\ref{R_g_ub2_lb3}), we replace notations ${\cal S}$ and ${\cal S}'$ in (\ref{R_g_ub2}) with ${\cal S}_1$ and ${\cal S}_1'$, respectively, to avoid ambiguity.
As stated after after (\ref{rate_region_K+1_2c2}), (\ref{R_g_ub1_lb3}) and (\ref{R_g_ub2_lb3}) should be redundant if Lemma~\ref{theorem_FM} is true.
We prove the redundancy in the following.

We first prove that (\ref{R_g_ub1_lb3}) is redundant.
Since $X_k, \forall k \in {\cal K} \cup \{ K + 1 \}$ are independent of each other and ${\overline {\cal S}} \subseteq {\cal K}$, we have
\begin{align}\label{R_g_ade_relax2}
& I(X_{\cal S}, X_{K + 1}; Y| X_{\overline {\cal S}}) - I(X_{{\cal S}'}, X_{K + 1}; Z) \nonumber\\
= & I(X_{\cal S}; Y| X_{\overline {\cal S}}) + I(X_{K+1}; Y| X_{{\cal K}}) - I(X_{{\cal S}'}, X_{K+1}; Z) \nonumber\\
\leq & I(X_{\cal S}; Y| X_{\overline {\cal S}}, X_{K \!+\! 1}) \!+\! I(X_{K \!+\! 1}; Y| X_{{\cal K}}) \!-\! I(X_{{\cal S}'}, X_{K \!+\! 1}; Z), \forall {\cal S} \!\subseteq\! {\cal K}, {\cal S} \!\neq\! \emptyset, {{\cal S}'} \!\subseteq\! {\cal S}, {\cal S}' \!\neq\! \emptyset.
\end{align}
By replacing the corresponding terms in (\ref{R_g_ade_relax}) with (\ref{R_g_ade_relax2}), the redundancy of (\ref{R_g_ub1_lb3}) can be similarly proven as that of (\ref{R_g_ub2_lb2_2}).
Since the redundancy proof of (\ref{R_g_ub2_lb3}) is much more complicated than that of (\ref{R_g_ub2_lb2_2}) and (\ref{R_g_ub1_lb3}), for the sake of clarity, we give the proof in the following separate subsection.

\subsection{Redundancy Proof of (\ref{R_g_ub2_lb3})}

Note that the redundancy of an inequality can be proven by showing that this inequality or a tighter bound can be obtained by linearly combining other inequalities.
The most important step in the proof is to know how to divide the sum rate, which in our case is the left-hand-side term of (\ref{R_g_ub2_lb3}), into several terms such that combining upper bounds on these terms, we can get either (\ref{R_g_ub2_lb3}) or a tighter bound.
Now we show how to do this.
We first rewrite the left-hand-side term of (\ref{R_g_ub2_lb3}) as follows
\begin{equation}\label{term_ab}
\underbrace { \sum_{k \in \cal S} R_k^{\text s} + \sum_{k \in {\cal S}\setminus {\cal S}'} R_k^{\text o} }_{{\text {Term}}~a} + \underbrace { \sum_{k \in {\cal S}_1} R_k^{\text s} + \sum_{k \in {\cal S}_1\setminus {\cal S}_1'} R_k^{\text o} + R_{K+1}^{\text s} }_{{\text {Term}}~b},
\end{equation}
which contains Term~$a$ and Term~$b$.
Note that adding the upper bounds on Term~$a$ and Term~$b$ cannot prove the redundancy.
Therefore, we have to divide (\ref{term_ab}) into several other terms.
We show that it is enough for the proof to divide (\ref{term_ab}) into two terms.
In particular, we exchange the message rates $R_k^{\text s}$ and $R_k^{\text o}$ in Term~$a$ and Term~$b$, and obtain two new terms, i.e., Term~$a'$ and Term~$b'$, whose upper bounds then help prove the redundancy.
We set the following criteria for the exchange process.
\begin{itemize}
	\item Criterion~$1$. Term~$b'$ contains as many secret message rates $R_k^{\text s}$ as possible;
	\item Criterion~$2$. Term~$a'$ contains as many open message rates $R_k^{\text o}$ as possible;
	\item Criterion~$3$. Criterion~$1$ has a higher priority than Criterion~$2$;
	\item Criterion~$4$. For any $k$, there could be only one $R_k^{\text s}$ or $R_k^{\text o}$ in Term~$a'$ or Term~$b'$;
	\item Criterion~$5$. For any $k$, if $R_k^{\text s}$ is not in Term~$a'$ or Term~$b'$, $R_k^{\text o}$ cannot be in this term.
\end{itemize}

\begin{figure}
	\begin{minipage}[t]{0.49\linewidth}
		\centering
		\includegraphics[width=3in]{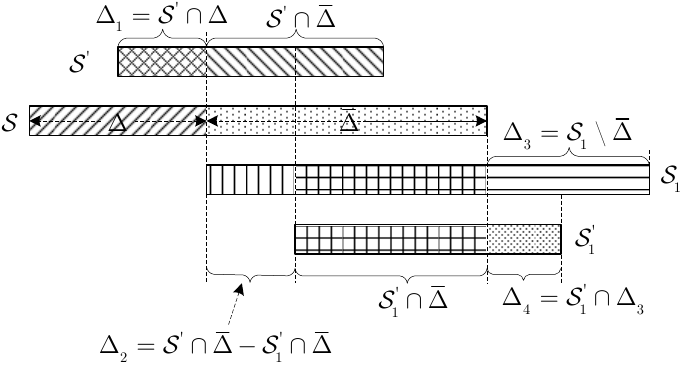}
		\caption{Sets ${\cal S}$, ${\cal S}'$, ${\cal S}_1$, ${\cal S}_1'$, and their divisions.}
		\label{Sets_S}
	\end{minipage}
	\hskip 1ex
	\begin{minipage}[t]{0.49\linewidth}
		\centering
		\includegraphics[width=3in]{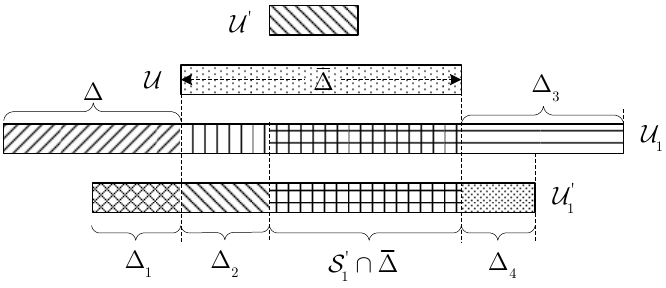}
		\caption{Sets ${\cal U}$, ${\cal U}'$, ${\cal U}_1$, ${\cal U}_1'$ obtained from ${\cal S}$, ${\cal S}'$, ${\cal S}_1$, ${\cal S}_1'$ based on Criteria $1 \sim 5$.}
		\label{Sets_U}
	\end{minipage}
\vspace{-1.0em}
\end{figure}
For ease of understanding, we give an example system with ${\cal K} = \{ 1, 2, 3, 4\}$, ${\cal S} = \{ 1, 3, 4\}$, ${\cal S}' = \{ 4\}$, ${\cal S}_1 = \{ 1, 2, 4\}$, and ${\cal S}_1' = \emptyset$.
In this case, (\ref{term_ab}) takes on form
\begin{equation}\label{term_ab_eg}
\underbrace { R_1^{\text s} + R_1^{\text o} + R_3^{\text s} + R_3^{\text o} + R_4^{\text s} }_{{\text {Term}}~a} + \underbrace { R_1^{\text s} + R_1^{\text o} + R_2^{\text s} + R_2^{\text o} + R_4^{\text s} + R_4^{\text o} + R_5^{\text s} }_{{\text {Term}}~b}.
\end{equation}
Since $R_3^{\text s}$ is not included in Term~$b$, according to Criterion~$1$ and Criterion~$3$, $R_3^{\text s}$ should be moved into Term~$b$.
Note that $R_3^{\text o}$ has also to be moved into Term~$b$ since otherwise Criterion~$5$ is violated.
Since both Term~$a$ and Term~$b$ contain $R_4^{\text s}$, while only Term~$b$ contains $R_4^{\text o}$, according to Criterion~$2$, $R_4^{\text o}$ should be moved into Term~$a$.
Note that we may not move $R_1^{\text o}$ in Term~$b$ into Term~$a$ since otherwise Criterion~$4$ is violated, and may not move $R_2^{\text o}$ into Term~$a$ since otherwise Criterion~$5$ is violated.
With these operations, (\ref{term_ab_eg}) becomes
\begin{equation}\label{term_ab_eg2}
\underbrace { R_1^{\text s} + R_1^{\text o} + R_4^{\text s} + R_4^{\text o} }_{{\text {Term}}~a'} + \underbrace { R_1^{\text s} + R_1^{\text o} + R_2^{\text s} + R_2^{\text o} + R_3^{\text s} + R_3^{\text o} + R_4^{\text s} + R_5^{\text s} }_{{\text {Term}}~b'}.
\end{equation}

In the following, we first describe these operations mathematically and then prove the redundancy of (\ref{R_g_ub2_lb3}).
For convenience, we give an example of sets ${\cal S}$, ${\cal S}'$, ${\cal S}_1$, ${\cal S}_1'$, and their divisions in Fig.~\ref{Sets_S}.
Let ${\overline \Delta}$ denote the set of user indexes which are in both $\cal S$ and ${\cal S}_1$, and $\Delta$ denote the set of indexes which are in $\cal S$ but not in ${\cal S}_1$, i.e.,
\begin{align}\label{Delta}
{\overline \Delta} & = {\cal S} \cap {\cal S}_1, \nonumber\\
\Delta & = {\cal S} - {\cal S}_1.
\end{align}
Let $\Delta_1$ denote the intersection of ${\cal S}'$ and $\Delta$, and $\Delta_2$ denote the set of user indexes which are in ${\cal S}' \cap {\overline \Delta}$ but not in ${\cal S}_1' \cap {\overline \Delta}$, i.e.,
\begin{align}\label{Delta12}
\Delta_1 & = {\cal S}' \cap \Delta, \nonumber\\
\Delta_2 & = {\cal S}' \cap {\overline \Delta} - {\cal S}_1' \cap {\overline \Delta}. 
\end{align}
With the operations described above, let ${\cal U}$, ${\cal U}'$, ${\cal U}_1$, and ${\cal U}_1'$, which respectively correspond to ${\cal S}$, ${\cal S}'$, ${\cal S}_1$, and ${\cal S}_1'$ in (\ref{term_ab}), denote the user indexes in Term~$a'$ and Term~$b'$.
Then, according to Criterion~$1$,
\begin{align}\label{U_U1}
{\cal U} & = {\cal S} \setminus \Delta, \nonumber\\
{\cal U}_1 & = {\cal S}_1 \cup \Delta. 
\end{align}
The complementary sets of ${\cal U}$ and ${\cal U}_1$ are 
\begin{align} \label{U_U1_complement}
{\overline {\cal U}} & = {\cal K} \setminus ({\cal S} \setminus \Delta) \nonumber\\
& = {\overline {\cal S}} \cup \Delta, \nonumber\\
{\overline {{\cal U}_1}} & = {\cal K} \setminus ({\cal S}_1 \cup \Delta) \nonumber\\ 
& = {\overline {{\cal S}_1}} \setminus \Delta,
\end{align}
where ${\overline {\cal S}} = {\cal K} \setminus {\cal S}$ and ${\overline {{\cal S}_1}} = {\cal K} \setminus {\cal S}_1$.
Moreover, according to Criterion~$2$, Criterion~$4$, and Criterion~$5$, we have
\begin{align}\label{U_U1_prime}
{\cal U}' & = {\cal S}' \setminus (\Delta_1 \cup \Delta_2), \nonumber\\
{\cal U}_1' & = {\cal S}_1' \cup (\Delta_1 \cup \Delta_2).
\end{align}
Fig.~\ref{Sets_U} depicts the sets ${\cal U}$, ${\cal U}'$, ${\cal U}_1$, ${\cal U}_1'$ obtained from ${\cal S}$, ${\cal S}'$, ${\cal S}_1$, ${\cal S}_1'$ in Fig.~\ref{Sets_S}.
We thus have 
\begin{align}\label{R_g_ub2_lb3_oper}
& \underbrace { \sum_{k \in \cal S} R_k^{\text s} + \sum_{k \in {\cal S}\setminus {\cal S}'} R_k^{\text o} }_{{\text {Term}}~a} + \underbrace { \sum_{k \in {\cal S}_1} R_k^{\text s} + \sum_{k \in {\cal S}_1\setminus {\cal S}_1'} R_k^{\text o} + R_{K+1}^{\text s} }_{{\text {Term}}~b} \nonumber\\
= & \underbrace { \sum_{k \in \cal U} R_k^{\text s} + \sum_{k \in {\cal U}\setminus {\cal U}'} R_k^{\text o} }_{{\text {Term}}~a'} + \underbrace { \sum_{k \in {\cal U}_1} R_k^{\text s} + \sum_{k \in {\cal U}_1\setminus {\cal U}_1'} R_k^{\text o} + R_{K+1}^{\text s} }_{{\text {Term}}~b'} \nonumber\\
\leq & I(X_{\cal U}; Y| X_{\overline {\cal U}}, X_{K + 1}) - I(X_{{\cal U}'}; Z) + I(X_{{\cal U}_1}, X_{K + 1}; Y| X_{\overline {{\cal U}_1}}) - I(X_{{\cal U}_1'}, X_{K + 1}; Z) \nonumber\\
= & I(X_{\cal S} \setminus X_\Delta; Y| X_{\overline {\cal S}}, X_\Delta, X_{K + 1}) - I(X_{{\cal S}'} \setminus X_{\Delta_1 \cup \Delta_2}; Z) \nonumber\\
+ & I(X_{{\cal S}_1}, X_\Delta, X_{K + 1}; Y| X_{\overline {{\cal S}_1}} \setminus X_\Delta) - I(X_{{\cal S}_1'}, X_{\Delta_1}, X_{\Delta_2}, X_{K + 1}; Z),\nonumber\\
& \forall {\cal S}, {\cal S}_1 \subseteq {\cal K}, {\cal S}, {\cal S}_1 \neq \emptyset, {\cal S}' \subseteq {\cal S}, {\cal S}' \neq \emptyset, {\cal S}_1' \subseteq {\cal S}_1,
\end{align}
where the inequality results from (\ref{R_g_abc}) and (\ref{rate_region_K+1_2c2}).
On the other hand, based on the definitions of $\Delta$, $\Delta_1$, and $\Delta_2$, and the chain rule of mutual information, (\ref{R_g_ub2_lb3}) can be rewritten as follows
\begin{align}\label{R_g_ub2_lb3_re}
& \sum_{k \in \cal S} R_k^{\text s} + \sum_{k \in {\cal S}\setminus {\cal S}'} R_k^{\text o} + \sum_{k \in {\cal S}_1} R_k^{\text s} + \sum_{k \in {\cal S}_1\setminus {\cal S}_1'} R_k^{\text o} + R_{K+1}^{\text s} \nonumber\\
\leq & I(X_{\cal S}; Y| X_{\overline {\cal S}}, X_{K + 1}) - I(X_{{\cal S}'}, X_{K + 1}; Z) + I(X_{{\cal S}_1}, X_{K + 1}; Y| X_{\overline {{\cal S}_1}}) - I(X_{{\cal S}_1'}; Z) \nonumber\\
= & I(X_{\cal S} \setminus X_\Delta, X_\Delta; Y| X_{\overline {\cal S}}, X_{K + 1}) - I(X_{{\cal S}'} \setminus X_{\Delta_1 \cup \Delta_2}, X_{\Delta_1 \cup \Delta_2}, X_{K + 1}; Z) \nonumber\\
+ & I(X_{{\cal S}_1}, X_{K + 1}; Y| X_{\overline {{\cal S}_1}}) - I(X_{{\cal S}_1'}; Z) \nonumber\\
= & I(X_{\cal S} \setminus X_\Delta; Y| X_{\overline {\cal S}}, X_\Delta, X_{K + 1}) + I(X_\Delta; Y| X_{\overline {\cal S}}, X_{K + 1}) - I(X_{{\cal S}'} \setminus X_{\Delta_1 \cup \Delta_2}; Z)\nonumber\\
- & I( X_{\Delta_1}, X_{\Delta_2}, X_{K + 1}; Z| X_{{\cal S}'} \setminus X_{\Delta_1 \cup \Delta_2}) + I(X_{{\cal S}_1}, X_{K + 1}; Y| X_{\overline {{\cal S}_1}}) - I(X_{{\cal S}_1'}; Z), \nonumber\\
& \forall {\cal S}, {\cal S}_1 \subseteq {\cal K}, {\cal S}, {\cal S}_1 \neq \emptyset, {\cal S}' \subseteq {\cal S}, {\cal S}' \neq \emptyset, {\cal S}_1' \subseteq {\cal S}_1.
\end{align}
In the following, we show that the upper bound in (\ref{R_g_ub2_lb3_oper}) is no larger and is thus tighter than that in (\ref{R_g_ub2_lb3_re}).
Then, (\ref{R_g_ub2_lb3_re}) is redundant.
Neglecting the common terms $I(X_{\cal S} \setminus X_\Delta; Y| X_{\overline {\cal S}}, X_\Delta, X_{K + 1})$ and $I(X_{{\cal S}'} \setminus X_{\Delta_1 \cup \Delta_2}; Z)$ in (\ref{R_g_ub2_lb3_oper}) and (\ref{R_g_ub2_lb3_re}), we prove
\begin{align}\label{relation1}
I(X_{{\cal S}_1}, X_\Delta, X_{K + 1}; Y| X_{\overline {{\cal S}_1}} \setminus X_\Delta) \leq I(X_\Delta; Y| X_{\overline {\cal S}}, X_{K + 1}) + I(X_{{\cal S}_1}, X_{K + 1}; Y| X_{\overline {{\cal S}_1}}),
\end{align}
and
\begin{align}\label{relation2}
I(X_{{\cal S}_1'}, X_{\Delta_1}, X_{\Delta_2}, X_{K + 1}; Z) \geq I( X_{\Delta_1}, X_{\Delta_2}, X_{K + 1}; Z| X_{{\cal S}'} \setminus X_{\Delta_1 \cup \Delta_2}) + I(X_{{\cal S}_1'}; Z).
\end{align}

From the definitions of $\overline \Delta$ and $\Delta$ in (\ref{Delta}), it is known that $\Delta \cap {\overline \Delta} = \emptyset$ and ${\cal S} = \Delta \cup {\overline \Delta}$.
Hence,
\begin{align}\label{S_comple}
{\overline {\cal S}} & = {\cal K} \setminus {\cal S} \nonumber\\
& = ({\cal K} \setminus {\overline \Delta}) \setminus \Delta.
\end{align}
Moreover, since ${\overline \Delta} \subseteq {\cal S}_1$,
\begin{align}\label{S_prime_comple}
{\overline {{\cal S}_1}} & = {\cal K} \setminus {\cal S}_1 \nonumber\\
& \subseteq {\cal K} \setminus {\overline \Delta}.
\end{align}
Based on (\ref{S_comple}) and (\ref{S_prime_comple}), we have
\begin{align}\label{S_prime_S_comple}
{\overline {{\cal S}_1}} \setminus \Delta & \subseteq ({\cal K} \setminus {\overline \Delta}) \setminus \Delta \nonumber\\
& = {\overline {\cal S}}.
\end{align}
Using the chain rule of mutual information, (\ref{S_prime_S_comple}), and the fact that $X_k, \forall k \in {\cal K} \cup \{ K + 1 \}$ are independent of each other, we have
\begin{align}\label{relation1_2}
I(X_{{\cal S}_1}, X_\Delta, X_{K + 1}; Y| X_{\overline {{\cal S}_1}} \setminus X_\Delta) & = I(X_\Delta; Y| X_{\overline {{\cal S}_1}} \setminus X_\Delta) + I(X_{{\cal S}_1}, X_{K + 1}; Y| X_{\overline {{\cal S}_1}}) \nonumber\\
& \leq I(X_\Delta; Y| X_{\overline {\cal S}}, X_{K + 1}) + I(X_{{\cal S}_1}, X_{K + 1}; Y| X_{\overline {{\cal S}_1}}).
\end{align}
The inequation (\ref{relation1}) is thus true.
On the other hand, since ${\cal S} = \Delta \cup {\overline \Delta}$, $\Delta \cap {\overline \Delta} = \emptyset$, and ${\cal S}' \subseteq {\cal S}$, as shown in Fig.~\ref{Sets_S}, ${\cal S}'$ can be divided into two disjoint parts as follows
\begin{align}\label{S_1_disjoint}
{\cal S}' & = ({\cal S}' \cap \Delta) \cup ({\cal S}' \cap {\overline \Delta}) \nonumber\\
& = \Delta_1 \cup ({\cal S}' \cap {\overline \Delta}),
\end{align}
where we used the definition of $\Delta_1$ in (\ref{Delta12}).
Hence,
\begin{equation}\label{S_1_Delta}
{\cal S}' \cap {\overline \Delta} = {\cal S}' \setminus \Delta_1.
\end{equation}
Let $\Delta_3$ denote the set of user indexes which are in ${\cal S}_1$ but not in $\cal S$, and $\Delta_4$ denote the intersection of ${\cal S}_1'$ and $\Delta_3$, i.e., 
\begin{align}\label{Delta34}
\Delta_3 & = {\cal S}_1 - {\cal S} \nonumber\\
& = {\cal S}_1 \setminus {\overline \Delta}, \nonumber\\
\Delta_4 & = {\cal S}_1' \cap \Delta_3. 
\end{align}
It can then be similarly proven as (\ref{S_1_Delta}) that
\begin{equation}\label{S_1_Delta_prime}
{\cal S}_1' \cap {\overline \Delta} = {\cal S}_1' \setminus \Delta_4,
\end{equation}
which can also be found from Fig.~\ref{Sets_S}.
From (\ref{S_1_Delta}), (\ref{S_1_Delta_prime}), and the definition of $\Delta_2$ in (\ref{Delta12}), ${\cal S}'$ in (\ref{S_1_disjoint}) can be further divided into three disjoint parts as follows
\begin{align}\label{S_1_disjoint_2}
{\cal S}' & = \Delta_1 \cup ({\cal S}' \cap {\overline \Delta}) \nonumber\\
& = \Delta_1 \cup \Delta_2 \cup \left[ ({\cal S}' \cap {\overline \Delta}) \cap ({\cal S}_1' \cap {\overline \Delta}) \right] \nonumber\\
& = \Delta_1 \cup \Delta_2 \cup \left[ ({\cal S}' \setminus \Delta_1) \cap ({\cal S}_1' \setminus \Delta_4) \right] \nonumber\\
& = \Delta_1 \cup \Delta_2 \cup ({\cal S}' \cap {\cal S}_1'), 
\end{align}
where the last step holds since $\Delta_1 \cap \Delta_4 = \emptyset$.
Accordingly, we have
\begin{align}\label{S1_S1_prime}
{\cal S}' \setminus (\Delta_1 \cup \Delta_2) & = {\cal S}' \cap {\cal S}_1' \nonumber\\
& \subseteq {\cal S}_1'.
\end{align}
Then, using the chain rule of mutual information, (\ref{S1_S1_prime}), and the fact that $X_k, \forall k \in {\cal K} \cup \{ K + 1 \}$ are independent of each other, we have
\begin{align}\label{relation2_2}
I(X_{{\cal S}_1'}, X_{\Delta_1}, X_{\Delta_2}, X_{K + 1}; Z) & = I( X_{\Delta_1}, X_{\Delta_2}, X_{K + 1}; Z| X_{{\cal S}_1'}) + I(X_{{\cal S}_1'}; Z)\nonumber\\
& \geq I( X_{\Delta_1}, X_{\Delta_2}, X_{K + 1}; Z| X_{{\cal S}'} \setminus X_{\Delta_1 \cup \Delta_2}) + I(X_{{\cal S}_1'}; Z),
\end{align}
i.e., (\ref{relation2}) is true.
Combining (\ref{R_g_ub2_lb3_oper}), (\ref{R_g_ub2_lb3_re}), (\ref{relation1_2}), and (\ref{relation2_2}), it is known that (\ref{R_g_ub2_lb3}) is redundant.

So far we have shown that (\ref{rate_region_K+1}) (or (\ref{rate_region_K+1_2})) can be obtained by eliminating $R_k^{\text a}, \forall k \in {\cal K} \cup \{ K + 1 \}$ in (\ref{region_FM_K+1}) (or (\ref{region_FM_K+1_2})), and all the other inequalities resulted from the elimination procedure, i.e., (\ref{R_g_ub2_lb2_2}), (\ref{R_g_ub1_lb3}), and (\ref{R_g_ub2_lb3}), are redundant.
As a result, (\ref{rate_region_K+1}) is the projection of (\ref{region_FM_K+1}) onto the hyperplane $\{ R_k^{\text a} = 0, \forall k \in {\cal K} \cup \{ K + 1 \}\}$.
Lemma~\ref{theorem_FM} is thus proven.

\section{Proof of Theorem~\ref{lemma_FM_gene_K2}}
\label{Prove_lemma_FM_gene_K2}

Since ${\overline {{\cal K}'}}$ has $2^{\left| {\overline {{\cal K}'}} \right| }$ subsets, we may divide the inequality system (\ref{region_FM2}) into $2^{\left| {\overline {{\cal K}'}} \right|}$ subsystems with each one corresponding to a subset ${\cal T} \subseteq {\overline {{\cal K}'}}$.
For any ${\cal T} \subseteq {\overline {{\cal K}'}}$, the inequality subsystem is 
\begin{equation}\label{region_FM2_sub}
\left\{
\begin{array}{ll}
R_k^{\text a} \geq 0, \forall k \in {\cal K}', \\
\sum\limits_{k \in {\cal S}} (R_k^{\text s} + R_k^{\text o} + R_k^{\text a}) + \sum\limits_{k \in {\cal T}} R_k^{\text o} \leq I(X_{\cal S}, X_{\cal T}; Y| X_{\overline {\cal S}}, X_{\overline {\cal T}}), \forall {\cal S} \subseteq {\cal K}', \\
\sum\limits_{k \in {\cal S}} (R_k^{\text o} + R_k^{\text a}) \geq I(X_{\cal S}; Z| X_{\overline {{\cal K}'}}), \forall {\cal S} \subseteq {\cal K}'.
\end{array} \right.
\end{equation}
It is obvious that eliminating $R_k^{\text a}, \forall k \in {\cal K}'$ in (\ref{region_FM2}) is equivalent to eliminating $R_k^{\text a}, \forall k \in {\cal K}'$ in (\ref{region_FM2_sub}) for all ${\cal T} \subseteq {\overline {{\cal K}'}}$.
Due to the assumption $I(X_{\cal S}; Y| X_{\overline {\cal S}}, X_{\overline {{\cal K}'}}) \geq I(X_{\cal S}; Z| X_{\overline {{\cal K}'}}), \forall {\cal S} \subseteq {\cal K}'$ made in Theorem~\ref{lemma_FM_gene_K2}, for a given ${\cal T} \subseteq {\overline {{\cal K}'}}$, we have
\begin{align}
I(X_{\cal S}, X_{\cal T}; Y| X_{\overline {\cal S}}, X_{\overline {\cal T}}) & \geq I(X_{\cal S}; Y| X_{\overline {\cal S}}, X_{\overline {{\cal K}'}}) \nonumber\\
& \geq I(X_{\cal S}; Z| X_{\overline {{\cal K}'}}), \forall {\cal S} \subseteq {\cal K}'.
\end{align}
Then, by replacing $I(X_{\cal S}; Y| X_{\overline {\cal S}})$ in (\ref{region_FM1}) with $I(X_{\cal S}, X_{\cal T}; Y| X_{\overline {\cal S}}, X_{\overline {\cal T}}) - \sum_{k \in {\cal T}} R_k^{\text o}$ and $I(X_{\cal S}; Z)$ with $I(X_{\cal S}; Z| X_{\overline {{\cal K}'}})$, we can eliminate $R_k^{\text a}, \forall k \in {\cal K}'$ in (\ref{region_FM2_sub}) based on Lemma~\ref{theorem_FM} and get
\begin{align}\label{rate_region0_sub}
\sum_{k \in \cal S} R_k^{\text s} \!+\! \sum_{k \in {\cal S} \setminus {\cal S}'} R_k^{\text o} \!\leq\! I(X_{\cal S}, X_{\cal T}; Y| X_{\overline {\cal S}}, X_{\overline {\cal T}}) \!-\! \sum_{k \in {\cal T}} R_k^{\text o} \!-\! I(X_{{\cal S}'}; Z| X_{\overline {{\cal K}'}}), \forall {\cal S} \subseteq {\cal K}', {\cal S}' \subseteq {\cal S}.
\end{align}
Combining the inequalities (\ref{rate_region0_sub}) for all ${\cal T} \subseteq {\overline {{\cal K}'}}$, (\ref{region_DM0}) can be obtained, and Theorem~\ref{lemma_FM_gene_K2} is thus proven.

\section{Proof of Theorem~\ref{theo_varia_dist}}
\label{prove_theo_varia_dist}

For notational convenience, in this appendix, we use $p(\cdot)$ to denote the probability of a variable's realization.
For example, in the following formulas (\ref{eq5}) and (\ref{eq6}),
\begin{align}
	p(x_{\overline {{\cal K}'}}^n) & = {\text {Pr}} \left\{ X_{\overline {{\cal K}'}}^n = x_{\overline {{\cal K}'}}^n \right\}, \nonumber\\
	p({\mathpzc c}_{{\cal K}'}) & = {\text {Pr}} \left\{ {\cal C}_{{\cal K}'} = {\mathpzc c}_{{\cal K}'} \right\}.
\end{align}
In addition, if not specified, the realization of a variable in a $\sum$ or $\prod$ operator takes values from the set consisting of all possible values.
For example, in (\ref{eq5}c),
\begin{align}
	x_{\overline {{\cal K}'}}^n \in \prod_{k \in {\overline {{\cal K}'}}} {\cal X}_k^n~ {\text {and}}~ z^n \in {\cal Z}^n.
\end{align}
Now we prove Theorem~\ref{theo_varia_dist}.
First, the left-hand-side term of (\ref{exp_vari_dist}) can be rewritten as
\begin{align}\label{eq5}
	{\mathbb E} \left\| P_{Z^n} (\cdot| {\cal C}_{\cal K}) - P_{Z^n} (\cdot| {\cal C}_{\overline {{\cal K}'}}) \right\|_1 & \overset{\text {(a)}}{=} {\mathbb E} \left\| P_{Z^n} (\cdot| {\cal C}_{{\cal K}'}, X_{\overline {{\cal K}'}}^n) - P_{Z^n} (\cdot| X_{\overline {{\cal K}'}}^n) \right\|_1 \nonumber\\
	& = \sum_{x_{\overline {{\cal K}'}}^n} p(x_{\overline {{\cal K}'}}^n) {\mathbb E} \left\| P_{Z^n} (\cdot| {\cal C}_{{\cal K}'}, x_{\overline {{\cal K}'}}^n) - P_{Z^n} (\cdot| x_{\overline {{\cal K}'}}^n) \right\|_1 \nonumber\\
	& \overset{\text {(b)}}{=} \sum_{x_{\overline {{\cal K}'}}^n} p(x_{\overline {{\cal K}'}}^n) {\mathbb E} \left\| P_{Z^n} (\cdot| {\cal C}_{{\cal K}'}, x_{\overline {{\cal K}'}}^n) - {\mathbb E} \left[ P_{Z^n} (\cdot| {\cal C}_{{\cal K}'}, x_{\overline {{\cal K}'}}^n) \right] \right\|_1 \nonumber\\
	& \overset{\text {(c)}}{=} \sum_{x_{\overline {{\cal K}'}}^n, z^n} p(x_{\overline {{\cal K}'}}^n)  {\mathbb E} \left| P_{Z^n} (z^n| {\cal C}_{{\cal K}'}, x_{\overline {{\cal K}'}}^n) - {\mathbb E} \left[ P_{Z^n} (z^n| {\cal C}_{{\cal K}'}, x_{\overline {{\cal K}'}}^n) \right] \right|,\!\!\!
\end{align}
where (\ref{eq5}a) holds since $Q_k = 0, \forall k \in {\overline {{\cal K}'}}$ and each codebook ${\cal C}_k, \forall k \in {\overline {{\cal K}'}}$ thus has only one codeword $X_k^n$, (\ref{eq5}b) holds since
\begin{align}\label{eq6}
	{\mathbb E} \left[ P_{Z^n} (\cdot| {\cal C}_{{\cal K}'}, x_{\overline {{\cal K}'}}^n) \right] & = \sum_{{\mathpzc c}_{{\cal K}'}} p({\mathpzc c}_{{\cal K}'}) P_{Z^n} (\cdot| {\mathpzc c}_{{\cal K}'}, x_{\overline {{\cal K}'}}^n) \nonumber\\
	& = \sum_{{\mathpzc c}_{{\cal K}'}} P_{Z^n} (\cdot, {\mathpzc c}_{{\cal K}'}| x_{\overline {{\cal K}'}}^n) \nonumber\\
	& = P_{Z^n} (\cdot| x_{\overline {{\cal K}'}}^n),
\end{align}
and (\ref{eq5}c) uses the definition of total variational distance.
Since (see the definition in (\ref{p1}))
\begin{align}\label{eq7}
	P_{Z^n} (z^n| {\cal C}_{{\cal K}'}, x_{\overline {{\cal K}'}}^n) = 2^{- n \sum_{k \in {\cal K}'} Q_k} \sum_{m_{{\cal K}'} \in \prod_{k \in {\cal K}'} {\cal M}_k} P_{Z^n} (z^n| \{X_k^n (m_k) \}_{k \in {\cal K}'}, x_{\overline {{\cal K}'}}^n),
\end{align}
we divide it into two parts as follows based on whether $\{X_k^n (m_k) \}_{k \in {\cal K}'}$, $x_{\overline {{\cal K}'}}^n$, and $z^n$ are jointly typical or not
\begin{align}\label{eq8}
	& {\hat P}_{Z^n} (z^n| {\cal C}_{{\cal K}'}, x_{\overline {{\cal K}'}}^n) =  2^{- n \sum_{k \in {\cal K}'} Q_k} \sum_{m_{{\cal K}'} \in \prod_{k \in {\cal K}'} {\cal M}_k} \Big\{ P_{Z^n} (z^n| \{X_k^n (m_k) \}_{k \in {\cal K}'}, x_{\overline {{\cal K}'}}^n) \nonumber\\
	& \quad\quad\quad\quad\quad\quad\quad\quad\quad\quad\quad\quad\quad\quad\quad\quad\quad\quad \times O (\{X_k^n (m_k) \}_{k \in {\cal K}'}, x_{\overline {{\cal K}'}}^n, z^n) \Big\}, \nonumber\\
	& {\tilde P}_{Z^n} (z^n| {\cal C}_{{\cal K}'}, x_{\overline {{\cal K}'}}^n) = 2^{- n \sum_{k \in {\cal K}'} Q_k} \sum_{m_{{\cal K}'} \in \prod_{k \in {\cal K}'} {\cal M}_k} \Big\{ P_{Z^n} (z^n| \{X_k^n (m_k) \}_{k \in {\cal K}'}, x_{\overline {{\cal K}'}}^n) \nonumber\\
	& \quad\quad\quad\quad\quad\quad\quad\quad\quad\quad\quad\quad\quad\quad\quad\quad\quad\quad \times {\tilde O} (\{X_k^n (m_k) \}_{k \in {\cal K}'}, x_{\overline {{\cal K}'}}^n, z^n) \Big\},
\end{align}
where
\begin{align}
	& O (\{X_k^n (m_k) \}_{k \in {\cal K}'}, x_{\overline {{\cal K}'}}^n, z^n) = \left\{
	\begin{array}{ll}
		1, & {\text {if}}~ (\{X_k^n (m_k) \}_{k \in {\cal K}'}, x_{\overline {{\cal K}'}}^n, z^n) \in {\cal T}_\epsilon^{(n)}, \\
		0, & {\text {otherwise}},
	\end{array} \right.\label{eq8.0}\\
	& {\tilde O} (\{X_k^n (m_k) \}_{k \in {\cal K}'}, x_{\overline {{\cal K}'}}^n, z^n) = \left\{
	\begin{array}{ll}
		1, & {\text {if}}~ (\{X_k^n (m_k) \}_{k \in {\cal K}'}, x_{\overline {{\cal K}'}}^n, z^n) \notin {\cal T}_\epsilon^{(n)}, \\
		0, & {\text {otherwise}}.
	\end{array} \right. \label{eq8.1}
\end{align}
Using (\ref{eq8}) and the triangular inequality, (\ref{eq5}) can be upper bounded as follows
\begin{align}\label{ineq8}
	{\mathbb E}\! \left\|\! P_{Z^n} (\cdot| {\cal C}_{\cal K}) \!-\! P_{Z^n} (\cdot| {\cal C}_{\overline {{\cal K}'}}) \!\right\|_1 & \leq \sum_{x_{\overline {{\cal K}'}}^n, z^n} p(x_{\overline {{\cal K}'}}^n) {\mathbb E} \left| {\tilde P}_{Z^n} (z^n| {\cal C}_{{\cal K}'}, x_{\overline {{\cal K}'}}^n) \!-\! {\mathbb E} \left[ {\tilde P}_{Z^n} (z^n| {\cal C}_{{\cal K}'}, x_{\overline {{\cal K}'}}^n) \right] \right| \nonumber\\
	& + \sum_{(x_{\overline {{\cal K}'}}^n, z^n) \in {\cal T}_\epsilon^{(n)}}\!\!\! p(x_{\overline {{\cal K}'}}^n) {\mathbb E} \!\left|\! {\hat P}_{Z^n}\! (z^n| {\cal C}_{{\cal K}'}, x_{\overline {{\cal K}'}}^n) \!-\! {\mathbb E} \!\left[\! {\hat P}_{Z^n}\! (z^n| {\cal C}_{{\cal K}'}, x_{\overline {{\cal K}'}}^n) \!\right] \!\right|\!.
\end{align}
Now we further upper bound (\ref{ineq8}) by separately evaluating its two summation terms.
First,
\begin{align}\label{ineq9}
	& \sum_{x_{\overline {{\cal K}'}}^n, z^n} p(x_{\overline {{\cal K}'}}^n) {\mathbb E} \left| {\tilde P}_{Z^n} (z^n| {\cal C}_{{\cal K}'}, x_{\overline {{\cal K}'}}^n) - {\mathbb E} \left[ {\tilde P}_{Z^n} (z^n| {\cal C}_{{\cal K}'}, x_{\overline {{\cal K}'}}^n) \right] \right| \nonumber\\
	\leq & 2 \sum_{x_{\overline {{\cal K}'}}^n, z^n} p(x_{\overline {{\cal K}'}}^n) {\mathbb E} \left[ {\tilde P}_{Z^n} (z^n| {\cal C}_{{\cal K}'}, x_{\overline {{\cal K}'}}^n) \right] \nonumber\\
	= & 2 \sum_{x_{\overline {{\cal K}'}}^n, z^n} p(x_{\overline {{\cal K}'}}^n) 2^{- n \sum_{k \in {\cal K}'} Q_k} \!\!\!\sum_{m_{{\cal K}'} \in \prod_{k \in {\cal K}'} {\cal M}_k}\!\!\!\! {\mathbb E}\! \left[\! P_{Z^n}\! (z^n| \{\! X_k^n (m_k) \!\}_{k \in {\cal K}'}, x_{\overline {{\cal K}'}}^n) {\tilde O} (\{\! X_k^n (m_k) \!\}_{k \in {\cal K}'}, x_{\overline {{\cal K}'}}^n, z^n) \!\right]  \nonumber\\
	\overset{\text {(a)}}{=} & 2 \sum_{x_{\overline {{\cal K}'}}^n, z^n} p(x_{\overline {{\cal K}'}}^n) {\mathbb E} \left[ P_{Z^n} (z^n| X_{{\cal K}'}^n, x_{\overline {{\cal K}'}}^n) {\tilde O} (X_{{\cal K}'}^n, x_{\overline {{\cal K}'}}^n, z^n) \right] \nonumber\\
	\overset{\text {(b)}}{=} & 2 \sum_{(x_{{\cal K}'}^n, x_{\overline {{\cal K}'}}^n, z^n) \notin {\cal T}_\epsilon^{(n)}} p(x_{{\cal K}'}^n) p(x_{\overline {{\cal K}'}}^n) p(z^n| x_{{\cal K}'}^n, x_{\overline {{\cal K}'}}^n) \nonumber\\
	= & 2 \sum_{(x_{\cal K}^n, z^n) \notin {\cal T}_\epsilon^{(n)}} p(x_{\cal K}^n, z^n) \nonumber\\
	= & 2 (1 - {\text {Pr}} \{ (x_{\cal K}^n, z^n) \in {\cal T}_\epsilon^{(n)} \}) \nonumber\\
	\overset{\text {(c)}}{\leq} & 4 \prod_{k = 1}^K |{\cal X}_k| |{\cal Z}| e^{- n \epsilon^2 \mu},
\end{align}
where (\ref{ineq9}a) follows from the symmetry in codebook generation, (\ref{ineq9}b) holds due to (\ref{eq8.1}), (\ref{ineq9}c) is obtained by using \cite[Theorem~$1.1$]{kramer2008topics}, and $\mu$ is the smallest value of $p(x_{\cal K}, z)$.
Note that in this appendix, we choose a sufficiently small $\epsilon$ such that $0 < \epsilon < \mu$.

Next, we evaluate the second summation term of (\ref{ineq8}).
To this end, we first bound each expectation in the term.
Specifically, for a given $(x_{\overline {{\cal K}'}}^n, z^n) \in {\cal T}_\epsilon^{(n)}$,
\begin{align}\label{ineq10}
	{\mathbb E} \! \left|\! {\hat P}_{Z^n} (z^n| {\cal C}_{{\cal K}'}, x_{\overline {{\cal K}'}}^n) \!-\! {\mathbb E} \!\left[\! {\hat P}_{Z^n} (z^n| {\cal C}_{{\cal K}'}, x_{\overline {{\cal K}'}}^n) \!\right] \!\right| & \leq \sqrt{ {\mathbb E} \!\left[ \left( {\hat P}_{Z^n} (z^n| {\cal C}_{{\cal K}'}, x_{\overline {{\cal K}'}}^n) - {\mathbb E} \!\left[ {\hat P}_{Z^n} (z^n| {\cal C}_{{\cal K}'}, x_{\overline {{\cal K}'}}^n) \right]  \right)^2 \right]} \nonumber\\
	& = \sqrt{{\text {Var}} \left( {\hat P}_{Z^n} (z^n| {\cal C}_{{\cal K}'}, x_{\overline {{\cal K}'}}^n) \right)},
\end{align}
where the first step holds since for a convex function $f(u) = u^2$, using Jensen's inequality, $\left( {\mathbb E} \left[ u \right] \right)^2 \leq {\mathbb E} \left[ u^2 \right]$.
Using the definition of ${\hat P}_{Z^n} (z^n| {\cal C}_{{\cal K}'}, x_{\overline {{\cal K}'}}^n)$, its variance in (\ref{ineq10}) can be computed as (\ref{ineq11}) given below
\begin{align}\label{ineq11}
	& {\text {Var}} \left( {\hat P}_{Z^n} (z^n| {\cal C}_{{\cal K}'}, x_{\overline {{\cal K}'}}^n) \right) \nonumber\\
	= & {\mathbb E} \left[ \left( {\hat P}_{Z^n} (z^n| {\cal C}_{{\cal K}'}, x_{\overline {{\cal K}'}}^n) \right)^2 \right] - \left( {\mathbb E} \left[ {\hat P}_{Z^n} (z^n| {\cal C}_{{\cal K}'}, x_{\overline {{\cal K}'}}^n) \right] \right)^2 \nonumber\\
	= & 2^{- 2 n \sum_{k \in {\cal K}'} Q_k}  \left\{ {\mathbb E} \left[ \left( \sum_{m_{{\cal K}'} \in \prod_{k \in {\cal K}'} {\cal M}_k} P_{Z^n} (z^n| \{X_k^n (m_k) \}_{k \in {\cal K}'}, x_{\overline {{\cal K}'}}^n) O (\{X_k^n (m_k) \}_{k \in {\cal K}'}, x_{\overline {{\cal K}'}}^n, z^n) \right) \right. \right. \nonumber\\
	& \quad\quad\quad\quad\quad\quad \left. \times \left( \sum_{{\hat m}_{{\cal K}'} \in \prod_{k \in {\cal K}'} {\cal M}_k} P_{Z^n} (z^n| \{X_k^n ({\hat m}_k) \}_{k \in {\cal K}'}, x_{\overline {{\cal K}'}}^n) O (\{X_k^n ({\hat m}_k) \}_{k \in {\cal K}'}, x_{\overline {{\cal K}'}}^n, z^n) \right) \right] \nonumber\\
	& \quad\quad\quad\quad\quad\quad - {\mathbb E} \left[ \sum_{m_{{\cal K}'} \in \prod_{k \in {\cal K}'} {\cal M}_k} P_{Z^n} (z^n| \{X_k^n (m_k) \}_{k \in {\cal K}'}, x_{\overline {{\cal K}'}}^n) O (\{X_k^n (m_k) \}_{k \in {\cal K}'}, x_{\overline {{\cal K}'}}^n, z^n) \right] \nonumber\\
	& \quad\quad\quad\quad\quad\quad \left. \times {\mathbb E} \left[ \sum_{{\hat m}_{{\cal K}'} \in \prod_{k \in {\cal K}'} {\cal M}_k} P_{Z^n} (z^n| \{X_k^n ({\hat m}_k) \}_{k \in {\cal K}'}, x_{\overline {{\cal K}'}}^n) O (\{X_k^n ({\hat m}_k) \}_{k \in {\cal K}'}, x_{\overline {{\cal K}'}}^n, z^n) \right] \right\} \nonumber\\
	= & 2^{- 2 n \sum_{k \in {\cal K}'} Q_k} \sum_{m_{{\cal K}'}} \sum_{{\hat m}_{{\cal K}'}} \left\{ {\mathbb E} \left[ P_{Z^n} (z^n| \{X_k^n (m_k) \}_{k \in {\cal K}'}, x_{\overline {{\cal K}'}}^n) O (\{X_k^n (m_k) \}_{k \in {\cal K}'}, x_{\overline {{\cal K}'}}^n, z^n) \right.\right. \nonumber\\
	& \quad\quad\quad\quad\quad\quad\quad\quad\quad \times \left. P_{Z^n} (z^n| \{X_k^n ({\hat m}_k) \}_{k \in {\cal K}'}, x_{\overline {{\cal K}'}}^n) O (\{X_k^n ({\hat m}_k) \}_{k \in {\cal K}'}, x_{\overline {{\cal K}'}}^n, z^n) \right]  \nonumber\\
	& \quad\quad\quad\quad\quad\quad\quad\quad\quad - {\mathbb E} \left[ P_{Z^n} (z^n| \{X_k^n (m_k) \}_{k \in {\cal K}'}, x_{\overline {{\cal K}'}}^n) O (\{X_k^n (m_k) \}_{k \in {\cal K}'}, x_{\overline {{\cal K}'}}^n, z^n) \right] \nonumber\\
	& \quad\quad\quad\quad\quad\quad\quad\quad\quad \left. \times {\mathbb E} \left[ P_{Z^n} (z^n| \{X_k^n ({\hat m}_k) \}_{k \in {\cal K}'}, x_{\overline {{\cal K}'}}^n) O (\{X_k^n ({\hat m}_k) \}_{k \in {\cal K}'}, x_{\overline {{\cal K}'}}^n, z^n) \right] \right\} \nonumber\\
	= & 2^{- n \sum_{k \in {\cal K}'} Q_k} \sum_{{\hat m}_{{\cal K}'}} \left\{ {\mathbb E} \left[ P_{Z^n} (z^n| \{X_k^n (1) \}_{k \in {\cal K}'}, x_{\overline {{\cal K}'}}^n) O (\{X_k^n (1) \}_{k \in {\cal K}'}, x_{\overline {{\cal K}'}}^n, z^n) \right.\right. \nonumber\\
	& \quad\quad\quad\quad\quad\quad\quad \times \left. P_{Z^n} (z^n| \{X_k^n ({\hat m}_k) \}_{k \in {\cal K}'}, x_{\overline {{\cal K}'}}^n) O (\{X_k^n ({\hat m}_k) \}_{k \in {\cal K}'}, x_{\overline {{\cal K}'}}^n, z^n) \right]  \nonumber\\
	& \quad\quad\quad\quad\quad\quad\quad - {\mathbb E} \left[ P_{Z^n} (z^n| \{X_k^n (1) \}_{k \in {\cal K}'}, x_{\overline {{\cal K}'}}^n) O (\{X_k^n (1) \}_{k \in {\cal K}'}, x_{\overline {{\cal K}'}}^n, z^n) \right] \nonumber\\
	& \quad\quad\quad\quad\quad\quad\quad \left. \times {\mathbb E} \left[ P_{Z^n} (z^n| \{X_k^n ({\hat m}_k) \}_{k \in {\cal K}'}, x_{\overline {{\cal K}'}}^n) O (\{X_k^n ({\hat m}_k) \}_{k \in {\cal K}'}, x_{\overline {{\cal K}'}}^n, z^n) \right] \right\},
\end{align}
in which the last step follows from the symmetry in codebook generation.
Note that if none of ${\hat m}_k, \forall k \in {\cal K}'$ in (\ref{ineq11}) is $1$, i.e., 
\begin{equation}\label{noneq1}
	{\hat m}_k \neq 1, \forall k \in {\cal K}',
\end{equation}
$\{X_k^n ({\hat m}_k) \}_{k \in {\cal K}'}$ is independent of $\{X_k^n (1) \}_{k \in {\cal K}'}$.
In this case, the expectation difference in the last step of (\ref{ineq11}) is $0$, i.e.,
\begin{align}\label{eq9}
	& {\mathbb E} \left[ P_{Z^n} (z^n| \{X_k^n (1) \}_{k \in {\cal K}'}, x_{\overline {{\cal K}'}}^n) O (\{X_k^n (1) \}_{k \in {\cal K}'}, x_{\overline {{\cal K}'}}^n, z^n) \right. \nonumber\\
	& \times \left. P_{Z^n} (z^n| \{X_k^n ({\hat m}_k) \}_{k \in {\cal K}'}, x_{\overline {{\cal K}'}}^n) O (\{X_k^n ({\hat m}_k) \}_{k \in {\cal K}'}, x_{\overline {{\cal K}'}}^n, z^n) \right]  \nonumber\\
	- & {\mathbb E} \left[ P_{Z^n} (z^n| \{X_k^n (1) \}_{k \in {\cal K}'}, x_{\overline {{\cal K}'}}^n) O (\{X_k^n (1) \}_{k \in {\cal K}'}, x_{\overline {{\cal K}'}}^n, z^n) \right] \nonumber\\
	& \times {\mathbb E} \left[ P_{Z^n} (z^n| \{X_k^n ({\hat m}_k) \}_{k \in {\cal K}'}, x_{\overline {{\cal K}'}}^n) O (\{X_k^n ({\hat m}_k) \}_{k \in {\cal K}'}, x_{\overline {{\cal K}'}}^n, z^n) \right] = 0.
\end{align}
As a result, to evaluate (\ref{ineq11}), we only need to consider those ${\hat m}_{{\cal K}'} \in \prod_{k \in {\cal K}'} {\cal M}_k$ in which at least one ${\hat m}_k$ is $1$, i.e., at least one user in ${\cal K}'$ transmits the first codeword.
Let ${\cal A} \subsetneqq {\cal K}'$ and ${\overline {\cal A}} = {\cal K}' \setminus {\cal A}$ denote sets of users which respectively satisfy
\begin{align}\label{eq10}
	{\hat m}_k & \neq 1, \forall k \in {\cal A}, \nonumber\\
	{\hat m}_k & = 1, \forall k \in {\overline {\cal A}}.
\end{align}
Using (\ref{eq9}) and (\ref{eq10}), ${\text {Var}} \left( {\hat P}_{Z^n} (z^n| {\cal C}_{{\cal K}'}, x_{\overline {{\cal K}'}}^n) \right)$ in (\ref{ineq11}) can be upper bounded as 
\begin{align}\label{ineq12}
	& {\text {Var}} \left( {\hat P}_{Z^n} (z^n| {\cal C}_{{\cal K}'}, x_{\overline {{\cal K}'}}^n) \right) \nonumber\\
	= & 2^{- n \sum_{k \in {\cal K}'} Q_k} \sum_{{\cal A} \subsetneqq {\cal K}'} \sum_{m_{\cal A} \in \prod_{k \in {\cal A}} {\cal M}_k \setminus \{1\}} {\Bigg \{} {\mathbb E} \left[ P_{Z^n} (z^n| \{X_k^n (1) \}_{k \in {\cal K}'}, x_{\overline {{\cal K}'}}^n) O (\{X_k^n (1) \}_{k \in {\cal K}'}, x_{\overline {{\cal K}'}}^n, z^n) \right. \nonumber\\
	& \times \left. P_{Z^n} (z^n| \{X_k^n ({\hat m}_k) \}_{k \in {\cal A}}, \{X_k^n (1) \}_{k \in \overline {\cal A}}, x_{\overline {{\cal K}'}}^n) O (\{X_k^n ({\hat m}_k) \}_{k \in {\cal A}}, \{X_k^n (1) \}_{k \in \overline {\cal A}}, x_{\overline {{\cal K}'}}^n, z^n) \right]  \nonumber\\
	& - {\mathbb E} \left[ P_{Z^n} (z^n| \{X_k^n (1) \}_{k \in {\cal K}'}, x_{\overline {{\cal K}'}}^n) O (\{X_k^n (1) \}_{k \in {\cal K}'}, x_{\overline {{\cal K}'}}^n, z^n) \right] \nonumber\\
	& \times {\mathbb E} \left[ P_{Z^n} (z^n| \{X_k^n ({\hat m}_k) \}_{k \in {\cal A}}, \{X_k^n (1) \}_{k \in \overline {\cal A}}, x_{\overline {{\cal K}'}}^n) O (\{X_k^n ({\hat m}_k) \}_{k \in {\cal A}}, \{X_k^n (1) \}_{k \in \overline {\cal A}}, x_{\overline {{\cal K}'}}^n, z^n) \right] {\Bigg \}} \nonumber\\
	\leq & 2^{- n \sum_{k \in {\cal K}'} Q_k} \sum_{{\cal A} \subsetneqq {\cal K}'} \sum_{m_{\cal A} \in \prod_{k \in {\cal A}} {\cal M}_k \setminus \{1\}} {\mathbb E} {\Big [} P_{Z^n} (z^n| \{X_k^n (1) \}_{k \in {\cal K}'}, x_{\overline {{\cal K}'}}^n) O (\{X_k^n (1) \}_{k \in {\cal K}'}, x_{\overline {{\cal K}'}}^n, z^n) \nonumber\\
	& \times P_{Z^n} (z^n| \{X_k^n ({\hat m}_k) \}_{k \in {\cal A}}, \{X_k^n (1) \}_{k \in \overline {\cal A}}, x_{\overline {{\cal K}'}}^n) O (\{X_k^n ({\hat m}_k) \}_{k \in {\cal A}}, \{X_k^n (1) \}_{k \in \overline {\cal A}}, x_{\overline {{\cal K}'}}^n, z^n) {\Big ]},
\end{align}
in which the last step holds since all non-negative subtrahends are omitted.
We further bound each expectation term in the last step of (\ref{ineq12}) as follows
\begin{align}\label{ineq13}
	& {\mathbb E} {\Big [} P_{Z^n} (z^n| \{X_k^n (1) \}_{k \in {\cal K}'}, x_{\overline {{\cal K}'}}^n) O (\{X_k^n (1) \}_{k \in {\cal K}'}, x_{\overline {{\cal K}'}}^n, z^n) \nonumber\\
	& \times P_{Z^n} (z^n| \{X_k^n ({\hat m}_k) \}_{k \in {\cal A}}, \{X_k^n (1) \}_{k \in \overline {\cal A}}, x_{\overline {{\cal K}'}}^n) O (\{X_k^n ({\hat m}_k) \}_{k \in {\cal A}}, \{X_k^n (1) \}_{k \in \overline {\cal A}}, x_{\overline {{\cal K}'}}^n, z^n) {\Big ]} \nonumber\\
	= & \sum_{\substack{ 
			\{x_k^n (1) \}_{k \in {\cal K}'}, \{x_k^n ({\hat m}_k) \}_{k \in {\cal A}}:\\
			(\{x_k^n (1) \}_{k \in {\cal K}'}, x_{\overline {{\cal K}'}}^n, z^n) \in {\cal T}_\epsilon^{(n)},\\ 
			(\{x_k^n ({\hat m}_k) \}_{k \in {\cal A}}, \{x_k^n (1) \}_{k \in \overline {\cal A}}, x_{\overline {{\cal K}'}}^n, z^n) \in {\cal T}_\epsilon^{(n)} }
	}
	{\Big [} p (\{x_k^n (1) \}_{k \in {\cal K}'}) p (\{x_k^n ({\hat m}_k) \}_{k \in {\cal A}}) p (z^n| \{x_k^n (1) \}_{k \in {\cal K}'}, x_{\overline {{\cal K}'}}^n)\nonumber\\
	& \quad\quad\quad\quad\quad\quad\quad\quad\quad\quad\quad\quad\quad\quad\quad  \times p (z^n| \{x_k^n ({\hat m}_k) \}_{k \in {\cal A}}, \{x_k^n (1) \}_{k \in \overline {\cal A}}, x_{\overline {{\cal K}'}}^n) {\Big ]}\nonumber\\
	= & \sum_{\substack{ \{x_k^n (1) \}_{k \in {\cal K}'}, \{x_k^n ({\hat m}_k) \}_{k \in {\cal A}}:\\
			(\{x_k^n (1) \}_{k \in {\cal K}'}, x_{\overline {{\cal K}'}}^n, z^n) \in {\cal T}_\epsilon^{(n)},\\ 
			(\{x_k^n ({\hat m}_k) \}_{k \in {\cal A}}, \{x_k^n (1) \}_{k \in \overline {\cal A}}, x_{\overline {{\cal K}'}}^n, z^n) \in {\cal T}_\epsilon^{(n)} }}
	p (\{x_k^n (1) \}_{k \in {\cal K}'}, z^n| x_{\overline {{\cal K}'}}^n) p (\{x_k^n ({\hat m}_k) \}_{k \in {\cal A}}, z^n| \{x_k^n (1) \}_{k \in \overline {\cal A}}, x_{\overline {{\cal K}'}}^n) \nonumber\\
	\overset{\text {(a)}}{\leq} & \sum_{\substack{ \{x_k^n (1) \}_{k \in {\cal K}'}:\\
			(\{x_k^n (1) \}_{k \in {\cal K}'}, x_{\overline {{\cal K}'}}^n, z^n) \in {\cal T}_\epsilon^{(n)}}}
	\!\!\!\!\!\!p (\{x_k^n (1) \}_{k \in {\cal K}'}, z^n| x_{\overline {{\cal K}'}}^n) \!\!\left[ \sum_{\{x_k^n ({\hat m}_k) \}_{k \in {\cal A}}} p (\{x_k^n ({\hat m}_k) \}_{k \in {\cal A}}, z^n| \{x_k^n (1) \}_{k \in \overline {\cal A}}, x_{\overline {{\cal K}'}}^n) \right] \nonumber\\
	= & \sum_{\substack{ \{x_k^n (1) \}_{k \in {\cal K}'}:\\
			(\{x_k^n (1) \}_{k \in {\cal K}'}, x_{\overline {{\cal K}'}}^n, z^n) \in {\cal T}_\epsilon^{(n)} }}
	p (\{x_k^n (1) \}_{k \in {\cal K}'}, z^n| x_{\overline {{\cal K}'}}^n) p ( z^n| \{x_k^n (1) \}_{k \in \overline {\cal A}}, x_{\overline {{\cal K}'}}^n) \nonumber\\
	\overset{\text {(b)}}{\leq} & 2^{- n H(Z| X_{\overline {\cal A}}, X_{\overline {{\cal K}'}}) (1 - \epsilon)} \sum_{\{x_k^n (1) \}_{k \in {\cal K}'}} p (\{x_k^n (1) \}_{k \in {\cal K}'}, z^n| x_{\overline {{\cal K}'}}^n) \nonumber\\
	= & 2^{- n H(Z| X_{\overline {\cal A}}, X_{\overline {{\cal K}'}}) (1 - \epsilon)} p ( z^n| x_{\overline {{\cal K}'}}^n) \nonumber\\
	\leq & 2^{- n \left[ H(Z| X_{\overline {{\cal K}'}}) + H(Z| X_{\overline {\cal A}}, X_{\overline {{\cal K}'}}) \right] (1 - \epsilon)},
\end{align}
where (\ref{ineq13}a) holds since in contrast to the previous step, $\{ x_k^n ({\hat m}_k) \}_{k \in {\cal A}}$ no longer has to satisfy $(\{x_k^n ({\hat m}_k) \}_{k \in {\cal A}}, \{x_k^n (1) \}_{k \in \overline {\cal A}}, x_{\overline {{\cal K}'}}^n, z^n) \in {\cal T}_\epsilon^{(n)}$, and (\ref{ineq13}b) holds since, on one hand, for any $( \{x_k^n (1) \}_{k \in \overline {\cal A}}, x_{\overline {{\cal K}'}}^n, z^n) \in {\cal T}_\epsilon^{(n)}$,
\begin{equation}
	p ( z^n| \{x_k^n (1) \}_{k \in \overline {\cal A}}, x_{\overline {{\cal K}'}}^n) \leq 2^{- n H(Z| X_{\overline {\cal A}}, X_{\overline {{\cal K}'}}) (1 - \epsilon)},
\end{equation}
and on the other hand, $\{x_k^n (1) \}_{k \in {\cal K}'}$ no longer has to satisfy $(\{x_k^n (1) \}_{k \in {\cal K}'}, x_{\overline {{\cal K}'}}^n, z^n) \in {\cal T}_\epsilon^{(n)}$ as in the previous step.
Substituting (\ref{ineq13}) into (\ref{ineq12}), we have
\begin{align}\label{ineq14}
	{\text {Var}} \left( {\hat P}_{Z^n} (z^n| {\cal C}_{{\cal K}'}, x_{\overline {{\cal K}'}}^n) \right) & \leq 2^{- n \sum_{k \in {\cal K}'} Q_k} \sum_{{\cal A} \subsetneqq {\cal K}'} \sum_{m_{\cal A} \in \prod_{k \in {\cal A}} {\cal M}_k \setminus \{1\}} 2^{- n \left[ H(Z| X_{\overline {{\cal K}'}}) + H(Z| X_{\overline {\cal A}}, X_{\overline {{\cal K}'}}) \right] (1 - \epsilon)} \nonumber\\
	& \overset{\text {(a)}}{=} \sum_{{\cal A} \subsetneqq {\cal K}'} \prod_{k \in {\cal A}} \left( 2^{n Q_k} -1 \right) 2^{- n \left[ \sum_{k \in {\cal K}'} Q_k + H(Z| X_{\overline {{\cal K}'}}) + H(Z| X_{\overline {\cal A}}, X_{\overline {{\cal K}'}}) \right] (1 - \epsilon)} \nonumber\\
	& \leq \sum_{{\cal A} \subsetneqq {\cal K}'} 2^{n \sum_{k \in {\cal A}} Q_k} 2^{- n \left[ \sum_{k \in {\cal K}'} Q_k + H(Z| X_{\overline {{\cal K}'}}) + H(Z| X_{\overline {\cal A}}, X_{\overline {{\cal K}'}}) \right] (1 - \epsilon)} \nonumber\\
	& = \sum_{{\cal A} \subsetneqq {\cal K}'} 2^{- n \left[ \sum_{k \in {\overline {\cal A}}} Q_k + H(Z| X_{\overline {{\cal K}'}}) + H(Z| X_{\overline {\cal A}}, X_{\overline {{\cal K}'}}) \right] (1 - \epsilon)},
\end{align}
where (\ref{ineq14}a) holds since for any $k \in {\cal A}$, ${\hat m}_k$ has $2^{n Q_k} -1$ possible values.
Combining (\ref{ineq10}) and (\ref{ineq14}), we have
\begin{align}\label{ineq15}
	{\mathbb E} \!\left|\! {\hat P}_{Z^n} (z^n| {\cal C}_{{\cal K}'}, x_{\overline {{\cal K}'}}^n) \!-\! {\mathbb E} \!\left[\! {\hat P}_{Z^n} (z^n| {\cal C}_{{\cal K}'}, x_{\overline {{\cal K}'}}^n) \!\right] \!\right| & \!\leq\! \sqrt{\sum_{{\cal A} \subsetneqq {\cal K}'}\! 2^{- n \left[ \sum_{k \in {\overline {\cal A}}} Q_k + H(Z| X_{\overline {{\cal K}'}}) + H(Z| X_{\overline {\cal A}}, X_{\overline {{\cal K}'}}) \right] (1 - \epsilon)}} \nonumber\\
	& \leq \sum_{{\cal A} \subsetneqq {\cal K}'}\!\! 2^{- \frac{n}{2} \left[ \sum_{k \in {\overline {\cal A}}} Q_k \!+\! H(Z| X_{\overline {{\cal K}'}}) \!+\! H(Z| X_{\overline {\cal A}}, X_{\overline {{\cal K}'}}) \right] (1 - \epsilon)}.
\end{align}
Accordingly, the second summation term of (\ref{ineq8}) can be upper bounded as follows
\begin{align}\label{ineq16}
	& \sum_{(x_{\overline {{\cal K}'}}^n, z^n) \in {\cal T}_\epsilon^{(n)}} p(x_{\overline {{\cal K}'}}^n) {\mathbb E} \left| {\hat P}_{Z^n} (z^n| {\cal C}_{{\cal K}'}, x_{\overline {{\cal K}'}}^n) - {\mathbb E} \left[ {\hat P}_{Z^n}(z^n| {\cal C}_{{\cal K}'}, x_{\overline {{\cal K}'}}^n) \right] \right| \nonumber\\
	\leq & 2^{n H(X_{\overline {{\cal K}'}}, Z) (1 + \epsilon)} 2^{- n H(X_{\overline {{\cal K}'}}) (1 - \epsilon)} \sum_{{\cal A} \subsetneqq {\cal K}'} 2^{- \frac{n}{2} \left[ \sum_{k \in {\overline {\cal A}}} Q_k + H(Z| X_{\overline {{\cal K}'}}) + H(Z| X_{\overline {\cal A}}, X_{\overline {{\cal K}'}}) \right] (1 - \epsilon)} \nonumber\\
	= & \sum_{{\cal A} \subsetneqq {\cal K}'} 2^{- \frac{n}{2} \left[ \left( \sum_{k \in {\overline {\cal A}}} Q_k - I(X_{\overline {\cal A}}; Z| X_{\overline {{\cal K}'}}) \right) (1 - \epsilon) - 3 \epsilon H(X_{\overline {{\cal K}'}}, Z) \right]}.
\end{align}
For any ${\cal A} \subsetneqq {\cal K}'$, we know that ${\overline {\cal A}} \subseteq {\cal K}'$ and ${\overline {\cal A}} \neq \emptyset$.
Then, it is known from (\ref{condi}) that
\begin{equation}
	\sum_{k \in {\overline {\cal A}}} Q_k - I(X_{\overline {\cal A}}; Z| X_{\overline {{\cal K}'}}) > 0.
\end{equation}
Since $\epsilon$ is an arbitrarily small positive number, we have
\begin{equation}
	\Bigg[ \sum_{k \in {\overline {\cal A}}} Q_k - I(X_{\overline {\cal A}}; Z| X_{\overline {{\cal K}'}}) \Bigg] (1 - \epsilon) - 3 \epsilon H(X_{\overline {{\cal K}'}}, Z) > 0.
\end{equation}
Then, the upper bound (\ref{ineq16}) vanishes exponentially in $n$.
Substituting (\ref{ineq9}) and (\ref{ineq16}) into (\ref{ineq8}), Theorem~\ref{theo_varia_dist} can be proven.

\section{Proof of Theorem~\ref{max_R_s_joint}}
\label{Prove_max_R_s_joint}

We first prove (\ref{R_s_joint_DM}).
As we have shown in Lemma~\ref{achi_secrecy_only2}, if $R_k^{\text o} = 0, \forall k \in {\cal K}$, (\ref{region_DM_exten}) becomes (\ref{region_secrecy_only2}), which can be divided into
\begin{equation}\label{region_DM_exten_3}
	\left\{
	\begin{array}{ll}
		R_k^{\text s} = 0, \forall k \in {\overline {{\cal K}'}}, \\
		\sum\limits_{k \in \cal S} R_k^{\text s} \leq \left[I(X_{\cal S}; Y| X_{\overline {\cal S}}, X_{\overline {{\cal K}'}}) - I(X_{\cal S}; Z| X_{\overline {{\cal K}'}}) \right]^+, \forall {\cal S} \subsetneqq {\cal K}',
	\end{array} \right.
\end{equation}
and
\begin{equation}\label{region_DM_exten_8}
	\sum\limits_{k \in {\cal K}'} R_k^{\text s} \leq \left[I(X_{{\cal K}'}; Y| X_{\overline {{\cal K}'}}) - I(X_{{\cal K}'}; Z| X_{\overline {{\cal K}'}}) \right]^+.
\end{equation}
(\ref{region_DM_exten_8}) shows that for any rate-tuple in ${\mathscr R} (X_{\cal K}, {\cal K}')$, the sum secrecy rate $\sum_{k \in {\cal K}'} R_k^{\text s}$ is no larger than $\left[I(X_{{\cal K}'}; Y| X_{\overline {{\cal K}'}}) - I(X_{{\cal K}'}; Z| X_{\overline {{\cal K}'}}) \right]^+$.
We now show that this upper bound is achievable.
To this end, we only need to prove that the inequality (\ref{region_DM_exten_8}) is not redundant, i.e., any linear combination of inequalities in (\ref{region_DM_exten_3}) does not generate (\ref{region_DM_exten_8}) or a tighter upper bound to $\sum_{k \in {\cal K}'} R_k^{\text s}$.
In \cite[Appendix~E]{xu2022achievable}, we have provided the proof for the case with ${\cal K}' = {\cal K}$.
When ${\cal K}' \subsetneqq {\cal K}$, we could complete the proof by following similar steps.
We omit the details here for brevity.
Note that for a given ${\cal K}' \subseteq {\cal K}$, $R_k^{\text s} = 0, \forall k \in {\overline {{\cal K}'}}$.
Hence, the achievable upper bound on $\sum_{k \in {\cal K}} R_k^{\text s}$ given by (\ref{region_DM_exten_3}) and (\ref{region_DM_exten_8}) is $\left[I(X_{{\cal K}'}; Y| X_{\overline {{\cal K}'}}) - I(X_{{\cal K}'}; Z| X_{\overline {{\cal K}'}}) \right]^+$.
Then, considering all possible ${\cal K}' \subseteq {\cal K}$, the maximum achievable sum secrecy rate $\sum_{k \in {\cal K}} R_k^{\text s}$ is
\begin{equation}\label{R_s_joint_DM2}
	R^{\text s} (X_{\cal K}) = \mathop {\max }\limits_{{\cal K}' \subseteq {\cal K}} \left\{ \left[ I(X_{{\cal K}'}; Y| X_{\overline {{\cal K}'}}) - I(X_{{\cal K}'}; Z| X_{\overline {{\cal K}'}}) \right]^+ \right\}.
\end{equation}
(\ref{R_s_joint_DM}) is thus true.

Let ${\cal K}'^*$ denote the subset in ${\cal K}$ which achieves (\ref{R_s_joint_DM2}) and assume 
\begin{equation}\label{ineq0}
	I(X_{{\cal K}'^*}; Y| X_{\overline {{\cal K}'^*}}) - I(X_{{\cal K}'^*}; Z| X_{\overline {{\cal K}'^*}}) > 0,
\end{equation}
since otherwise we have $R_k^{\text s} = 0, \forall k \in {\cal K}$, i.e., the system reduces to a standard DM-MAC channel with only open messages.
Before proving the second part of Theorem~\ref{max_R_s_joint}, i.e., (\ref{R_o_K1_DM}), we first show that with ${\cal K}'^*$ defined above, we have
\begin{equation}\label{ineq1}
	I(X_{\cal S}; Y| X_{\overline {\cal S}}, X_{\overline {{\cal K}'^*}}) - I(X_{\cal S}; Z| X_{\overline {{\cal K}'^*}}) \geq 0, \forall {\cal S} \subsetneqq {\cal K}'^*.
\end{equation}
(\ref{ineq1}) can be proven by reductio ad absurdum.
If (\ref{ineq1}) is not true, then there exist subsets in ${\cal K}'^*$ such that the corresponding inequalities in (\ref{ineq1}) do not hold.
W.l.o.g., we assume that there exists only one subset ${\cal S}_0$ in ${\cal K}'^*$ such that
\begin{equation}\label{ineq2}
	I(X_{{\cal S}_0}; Y| X_{\overline {{\cal S}_0}}, X_{\overline {{\cal K}'^*}}) - I(X_{{\cal S}_0}; Z| X_{\overline {{\cal K}'^*}}) < 0.
\end{equation}
Using the chain rule of mutual information and the fact that $X_k, \forall k \in {\cal K}$ are independent of each other, the left-hand side term of (\ref{ineq0}) is upper bounded by
\begin{align}\label{ineq0_ub}
	& I(X_{{\cal K}'^*}; Y| X_{\overline {{\cal K}'^*}}) - I(X_{{\cal K}'^*}; Z| X_{\overline {{\cal K}'^*}}) \nonumber\\
	= & I(X_{{\cal K}'^* \setminus {\cal S}_0}, X_{{\cal S}_0}; Y| X_{\overline {{\cal K}'^*}}) - I(X_{{\cal K}'^* \setminus {\cal S}_0}, X_{{\cal S}_0}; Z| X_{\overline {{\cal K}'^*}}) \nonumber\\
	= & I(X_{{\cal K}'^* \setminus {\cal S}_0}; Y| X_{{\cal S}_0}, X_{\overline {{\cal K}'^*}}) - I(X_{{\cal K}'^* \setminus {\cal S}_0}; Z| X_{\overline {{\cal K}'^*}}) + I(X_{{\cal S}_0}; Y| X_{\overline {{\cal K}'^*}}) - I(X_{{\cal S}_0}; Z| X_{{\cal K}'^* \setminus {\cal S}_0}, X_{\overline {{\cal K}'^*}}) \nonumber\\
	\leq & I(X_{{\cal K}'^* \setminus {\cal S}_0}; Y| X_{{\cal S}_0}, X_{\overline {{\cal K}'^*}}) - I(X_{{\cal K}'^* \setminus {\cal S}_0}; Z| X_{\overline {{\cal K}'^*}}) + I(X_{{\cal S}_0}; Y| X_{\overline {{\cal S}_0}}, X_{\overline {{\cal K}'^*}}) - I(X_{{\cal S}_0}; Z| X_{\overline {{\cal K}'^*}}) \nonumber\\
	< & I(X_{{\cal K}'^* \setminus {\cal S}_0}; Y| X_{{\cal S}_0}, X_{\overline {{\cal K}'^*}}) - I(X_{{\cal K}'^* \setminus {\cal S}_0}; Z| X_{\overline {{\cal K}'^*}}),
\end{align}
where the last step holds due to (\ref{ineq2}).
Then, ${\cal K}' = {\cal K}'^* \setminus {\cal S}_0$ and ${\overline {{\cal K}'}} = {\overline {{\cal K}'^*}} \cup {\cal S}_0$ result in a larger value in (\ref{R_s_joint_DM2}) than ${\cal K}'^*$, i.e., $R^{\text s} (X_{\cal K})$ can be increased.
This is contradicted to the assumption that ${\cal K}'^*$ achieves (\ref{R_s_joint_DM2}).
When there are more subsets in ${\cal K}'^*$ such that the inequalities in (\ref{ineq1}) do not hold, we may prove by following similar steps that $R^{\text s} (X_{\cal K})$ can be further increased.
As a result, if ${\cal K}'^* \subseteq {\cal K}$ achieves (\ref{R_s_joint_DM2}) and (\ref{ineq0}) is true, we have (\ref{ineq1}).

Now we show that if users in ${\cal K}'^*$ transmit their confidential messages at sum rate $R^{\text s} (X_{\cal K})$, the maximum achievable sum rate at which users in ${\cal K}$ could send their open messages is given by (\ref{R_o_K1_DM}).
We divide the users in ${\cal K}$ into two classes, i.e., ${\cal K}'^*$ and ${\overline {{\cal K}'^*}}$, and separately consider their maximum sum open message rate.

First, if $( \{R_k^{\text s}, R_k^{\text o}\}_{k \in {\cal K}'^*}, \{R_k^{\text s} = 0, R_k^{\text o}\}_{k \in {\overline {{\cal K}'^*}}} )$ is a rate-tuple in region ${\mathscr R} (X_{\cal K}, {\cal K}'^*)$ defined by Theorem~\ref{lemma_DM_exten} and $\sum_{k \in {\cal K}'^*} R_k^{\text s} = R^{\text s} (X_{\cal K})$, which is assumed to be positive, by setting ${\cal S} = {\cal K}'^*$, ${\cal S}' = \emptyset$, and ${\cal T} = \emptyset$ in (\ref{region_DM_exten}), we get
\begin{align}\label{R_o_DM_max}
	\sum_{k \in {\cal K}'^*} R_k^{\text o} \leq I(X_{{\cal K}'^*}; Y| X_{\overline {{\cal K}'^*}}) - R^{\text s} (X_{\cal K}) 
	= I(X_{{\cal K}'^*}; Z| X_{\overline {{\cal K}'^*}}),
\end{align}
indicating that the sum rate at which users in ${\cal K}'^*$ can encode their open messages is no larger than $I(X_{{\cal K}'^*}; Z| X_{\overline {{\cal K}'^*}})$.
Then, we prove that this rate is achievable.
Since the rate-tuple is in region ${\mathscr R} (X_{\cal K}, {\cal K}'^*)$, with inequalities (\ref{ineq0}) and (\ref{ineq1}), we could use Theorem~\ref{lemma_FM_gene_K2} and find $R_k^{\text a}, \forall k \in {\cal K}'^*$ such that
\begin{equation}\label{Fourier_Motzkin}
	\left\{
	\begin{array}{ll}
		R_k^{\text a} \geq 0, \forall k \in {\cal K}'^*, \\
		\sum\limits_{k \in {\cal S}} (R_k^{\text s} + R_k^{\text o} + R_k^{\text a}) + \sum\limits_{k \in {\cal T}} R_k^{\text o} \leq I(X_{\cal S}, X_{\cal T}; Y| X_{\overline {\cal S}}, X_{\overline {\cal T}}), \forall {\cal S} \subseteq {\cal K}'^*, {\cal T} \subseteq {\overline {{\cal K}'^*}}, \\
		\sum\limits_{k \in {\cal S}} (R_k^{\text o} + R_k^{\text a}) \geq I(X_{\cal S}; Z| X_{\overline {{\cal K}'^*}}), \forall {\cal S} \subseteq {\cal K}'^*.
	\end{array} \right.
\end{equation}
From (\ref{Fourier_Motzkin}) it is known that if $\sum_{k \in {\cal K}'^*} R_k^{\text o} < I(X_{{\cal K}'^*}; Z| X_{\overline {{\cal K}'^*}})$, we can always split partial rate in $R_k^{\text a}$ to $R_k^{\text o}$, and get ${\hat R}_k^{\text a}$ as well as ${\hat R}_k^{\text o}$, such that
\begin{align}\label{R_o_g_hat}
	{\hat R}_k^{\text a} & \geq 0, \forall k \in {\cal K}'^*, \nonumber\\
	{\hat R}_k^{\text o} + {\hat R}_k^{\text a} & = R_k^{\text o} + R_k^{\text a}, \forall k \in {\cal K}'^*,
\end{align}
and
\begin{equation}\label{max_R_o2}
	\sum_{k \in {\cal K}'^*} {\hat R}_k^{\text o} = I(X_{{\cal K}'^*}; Z| X_{\overline {{\cal K}'^*}}).
\end{equation}
With (\ref{R_o_g_hat}), it can be verified that the new rate-tuple $( \{R_k^{\text s}, {\hat R}_k^{\text o}, {\hat R}_k^{\text a}\}_{k \in {\cal K}'^*}, \{R_k^{\text s} = 0, R_k^{\text o}\}_{k \in {\overline {{\cal K}'^*}}} )$ is in the region defined by (\ref{Fourier_Motzkin}), i.e., it satisfies all inequalities in (\ref{Fourier_Motzkin}). 
Since ${\mathscr R} (X_{\cal K}, {\cal K}'^*)$ can be obtained by projecting (\ref{Fourier_Motzkin}) onto hyperplane $\{ R_k^{\text a} = 0, \forall k \in {\cal K}'^* \}$, then, according to the property of Fourier-Motzkin elimination \cite[Appendix D]{el2011network}, we know that rate-tuple $( \{R_k^{\text s}, {\hat R}_k^{\text o}\}_{k \in {\cal K}'^*}, \{R_k^{\text s} = 0, R_k^{\text o}\}_{k \in {\overline {{\cal K}'^*}}} )$ is in region ${\mathscr R} (X_{\cal K}, {\cal K}'^*)$.
Due to (\ref{max_R_o2}), with this rate-tuple, users in ${\cal K}'^*$ could send their open messages at sum rate $I(X_{{\cal K}'^*}; Z| X_{\overline {{\cal K}'^*}})$.

So far we have shown that the maximum achievable sum secrecy rate is $R^{\text s} (X_{\cal K})$, and if $\sum_{k \in {\cal K}'^*} R_k^{\text s} = R^{\text s} (X_{\cal K})$, the maximum achievable sum rate $\sum_{k \in {\cal K}'^*} R_k^{\text o}$ is $I(X_{{\cal K}'^*}; Z| X_{\overline {{\cal K}'^*}})$.
Now we show that if 
\begin{align}\label{two_max}
	\sum_{k \in {\cal K}'^*} R_k^{\text s} & = R^{\text s} (X_{\cal K}), \nonumber\\
	\sum_{k \in {\cal K}'^*} R_k^{\text o} & = I(X_{{\cal K}'^*}; Z| X_{\overline {{\cal K}'^*}}),
\end{align}
the maximum achievable sum rate $\sum_{k \in {\overline {{\cal K}'^*}}} R_k^{\text o}$ is $I (X_{\overline {{\cal K}'^*}}; Y)$.
Setting ${\cal S} = {\cal K}'^*$, ${\cal S}' = \emptyset$, and ${\cal T} = {\overline {{\cal K}'^*}}$ in (\ref{region_DM_exten}), we have
\begin{align}\label{max_sum_rate_o}
	\sum_{k \in {\overline {{\cal K}'^*}}} R_k^{\text o} & \leq I (X_{{\cal K}'^*}, X_{\overline {{\cal K}'^*}}; Y) - \sum_{k \in {\cal K}'^*} \left( R_k^{\text s} + R_k^{\text o} \right) \nonumber\\
	& = I (X_{{\cal K}'^*}, X_{\overline {{\cal K}'^*}}; Y) - R^{\text s} (X_{\cal K}) - I(X_{{\cal K}'^*}; Z| X_{\overline {{\cal K}'^*}}) \nonumber\\
	& = I (X_{\overline {{\cal K}'^*}}; Y).
\end{align}
Note that from the perspective of Bob, we are considering a DM-MAC channel with $K$ users.
Then, it is known from \cite[Chapter~$4$]{el2011network} that (\ref{max_sum_rate_o}) can hold with equality.
This can be realized by letting Bob treat the signal of users in ${\cal K}'^*$ as noise and decode the information of users in ${\overline {{\cal K}'^*}}$.
Hence, the maximum achievable sum rate at which users in ${\overline {{\cal K}'^*}}$ could send their open messages is $I(X_{\overline {{\cal K}'^*}}; Y)$.
Combining (\ref{two_max}) and (\ref{max_sum_rate_o}) (with equality), it is known that (\ref{R_o_K1_DM}) is true.
Theorem~\ref{max_R_s_joint} is thus proven.

\section{Proof of Theorem~\ref{theo_polytope}}
\label{prove_theo_polytope}

We first prove (\ref{cond4}).
Using the chain rule of mutual information, the left-hand-side term of (\ref{assumption_62}) is upper bounded by
\begin{subequations}\label{ineq_63}
\begin{align}
	& I(X_{{\cal K}_0 \cup {\cal V}}; Y| X_{{\cal K}' \setminus ({\cal K}_0 \cup {\cal V})}, X_{\overline {{\cal K}'}}) - I(X_{{\cal K}_0 \cup {\cal V}}; Z| X_{\overline {{\cal K}'}}) \nonumber\\
	= & I(X_{{\cal K}_0}, X_{\cal V}; Y| X_{{\cal K}' \setminus ({\cal K}_0 \cup {\cal V})}, X_{\overline {{\cal K}'}}) - I(X_{{\cal K}_0}, X_{\cal V}; Z| X_{\overline {{\cal K}'}}) \label{ineq_63_a}\\
	= & I \!(X_{{\cal K}_0}; \!Y| X_{{\cal K}' \setminus {\cal K}_0}, \!X_{\overline {{\cal K}'}}) \!-\! I \!(X_{{\cal K}_0}; \!Z| X_{\overline {{\cal K}'}}) \!+\! I \!(X_{\cal V}; \!Y| X_{{\cal K}' \setminus ({\cal K}_0 \cup {\cal V})}, \!X_{\overline {{\cal K}'}}) \!-\! I \!(X_{\cal V}; \!Z| X_{{\overline {{\cal K}'}} \cup {\cal K}_0}) \label{ineq_63_c}\\
	\leq & I(X_{\cal V}; Y| X_{{\cal K}' \setminus ({\cal K}_0 \cup {\cal V})}, X_{\overline {{\cal K}'}}) - I(X_{\cal V}; Z| X_{{\overline {{\cal K}'}} \cup {\cal K}_0}) \label{ineq_63_d}\\
	\leq & I(X_{\cal V}; Y| X_{{\cal K}' \setminus ({\cal K}_0 \cup {\cal V})}, X_{\overline {{\cal K}'} \cup {\cal K}_0}) - I(X_{\cal V}; Z| X_{{\overline {{\cal K}'}} \cup {\cal K}_0}) \label{ineq_63_e}\\
	= & I(X_{\cal V}; Y| X_{\overline {\cal V}}, X_{\overline {{\cal K}''}}) - I(X_{\cal V}; Z| X_{\overline {{\cal K}''}}), \forall {\cal V} \subseteq {\cal K}'', {\cal V} \neq \emptyset, \label{ineq_63_f}
\end{align}
\end{subequations}
where (\ref{ineq_63_d}) follows by using (\ref{assumption_61}), (\ref{ineq_63_e}) holds by adding $X_{{\cal K}_0}$ and using the fact that $X_k, \forall k \in {\cal K}$ are independent of each other, and the last step follows by using the definitions of ${\overline {{\cal K}''}}$ and ${\overline {\cal V}}$ in (\ref{K_prime}) and (\ref{cond4}).
Combining (\ref{assumption_62}) and (\ref{ineq_63}), we know that (\ref{cond4}) is true.

Next, we show that for any rate-tuple in region ${\mathscr R} (X_{\cal K}, {\cal K}')$ defined by Theorem~\ref{lemma_DM_exten}, if (\ref{assumption_61}) and (\ref{assumption_62}) can be satisfied, it is also in region ${\mathscr R} (X_{\cal K}, {\cal K}'')$.
If $(R_1^{\text s}, R_1^{\text o},\cdots, R_K^{\text s}, R_K^{\text o})$ is in region ${\mathscr R} (X_{\cal K}, {\cal K}')$ and satisfies (\ref{assumption_61}), by setting ${\cal S} = {\cal K}_0$, ${\cal S}' = {\cal S} = {\cal K}_0$ and ${\cal T} = \emptyset$ in (\ref{region_DM_exten}), we get
\begin{align}
	\sum\limits_{k \in {\cal K}_0} R_k^{\text s} \leq \left[ I(X_{{\cal K}_0}; Y| X_{{\cal K}' \setminus {\cal K}_0}, X_{\overline {{\cal K}'}}) - I(X_{{\cal K}_0}; Z| X_{\overline {{\cal K}'}}) \right]^+ = 0.
\end{align}
Hence, $R_k^{\text s} = 0, \forall k \in {\cal K}_0$.
Considering that $R_k^{\text s} = 0, \forall k \in {\overline {{\cal K}'}}$ in (\ref{region_DM_exten}), we have
\begin{equation}\label{Rs_zero_K0}
R_k^{\text s} = 0, \forall k \in {\overline {{\cal K}''}},
\end{equation}
which is required by (\ref{region_DM1}).
To prove that $(R_1^{\text s}, R_1^{\text o},\cdots, R_K^{\text s}, R_K^{\text o})$ is also in region ${\mathscr R} (X_{\cal K}, {\cal K}'')$, we need to further verify the upper bounds on $\sum_{k \in \cal V} R_k^{\text s} + \sum_{k \in {\cal V} \setminus {\cal V}'} R_k^{\text o} + \sum_{k \in {\cal W}} R_k^{\text o}$ for all possible set choices given in (\ref{region_DM1}).
To this end, we separately prove
\begin{align}\label{part1}
\sum\limits_{k \in \cal V} R_k^{\text s} + \sum\limits_{k \in {\cal V} \setminus {\cal V}'} R_k^{\text o} + \sum\limits_{k \in {\cal W}} R_k^{\text o} & \leq I(X_{\cal V}, X_{\cal W}; Y| X_{\overline {\cal V}}, X_{{\overline {{\cal K}''}} \setminus {\cal W}}) - I(X_{{\cal V}'}; Z| X_{\overline {{\cal K}''}}), \nonumber\\
& \forall {\cal V} \subseteq {\cal K}'', {\cal V}' \subseteq {\cal V}, {\cal W} \subseteq {\overline {{\cal K}'}},
\end{align}
and
\begin{align}\label{part2}
\sum\limits_{k \in \cal V} R_k^{\text s} + \sum\limits_{k \in {\cal V} \setminus {\cal V}'} R_k^{\text o} + \sum\limits_{k \in {\cal W}} R_k^{\text o} & \leq I(X_{\cal V}, X_{\cal W}; Y| X_{\overline {\cal V}}, X_{{\overline {{\cal K}''}} \setminus {\cal W}}) - I(X_{{\cal V}'}; Z| X_{\overline {{\cal K}''}}), \nonumber\\
& \forall {\cal V} \subseteq {\cal K}'', {\cal V}' \subseteq {\cal V}, {\cal W} \subseteq {\overline {{\cal K}''}}, {\cal W} \cap {\cal K}_0 \neq \emptyset.
\end{align}
Since in (\ref{region_DM1}), ${\cal W} \subseteq {\overline {{\cal K}''}} \triangleq {\overline {{\cal K}'}} \cup {\cal K}_0$, (\ref{part1}) together with (\ref{part2}) takes into account all possible choices of ${\cal W}$ in (\ref{region_DM1}).

We first prove (\ref{part1}).
Since $R_k^{\text s} = 0, \forall k \in {\cal K}_0$,
\begin{subequations}\label{prove_part1}
\begin{align}
& \sum_{k \in \cal V} R_k^{\text s} + \sum_{k \in {\cal V} \setminus {\cal V}'} R_k^{\text o} + \sum_{k \in {\cal W}} R_k^{\text o} \nonumber\\
= & \sum_{k \in {\cal V} \cup {\cal K}_0} R_k^{\text s} + \sum_{k \in {\cal V} \cup {\cal K}_0 \setminus ({\cal V}' \cup {\cal K}_0)} R_k^{\text o} + \sum_{k \in {\cal W}} R_k^{\text o} \label{prove_part1_b}\\
\leq & I(X_{{\cal V} \cup {\cal K}_0}, X_{\cal W}; Y| X_{{\cal K}' \setminus ({\cal V} \cup {\cal K}_0)}, X_{{\overline {{\cal K}'}} \setminus {\cal W}}) - I(X_{{\cal V}' \cup {\cal K}_0}; Z| X_{\overline {{\cal K}'}}) \label{prove_part1_c}\\
= & I(X_{{\cal K}_0}; Y| X_{{\cal K}' \setminus {\cal K}_0}, X_{\overline {{\cal K}'}}) - I(X_{{\cal K}_0}; Z| X_{\overline {{\cal K}'}}) \label{prove_part1_d}\\
+ & I(X_{\cal V}, X_{\cal W}; Y| X_{{\cal K}' \setminus ({\cal V} \cup {\cal K}_0)}, X_{{\overline {{\cal K}'}} \setminus {\cal W}}) - I(X_{{\cal V}'}; Z| X_{{\overline {{\cal K}'}} \cup {\cal K}_0}) \label{prove_part1_e}\\
\leq & I(X_{\cal V}, X_{\cal W}; Y| X_{{\cal K}' \setminus ({\cal V} \cup {\cal K}_0)}, X_{{\overline {{\cal K}'}} \setminus {\cal W}}) - I(X_{{\cal V}'}; Z| X_{{\overline {{\cal K}'}} \cup {\cal K}_0}) \label{prove_part1_f}\\
= & I(X_{\cal V}, X_{\cal W}; Y| X_{\overline {\cal V}}, X_{{\overline {{\cal K}'}} \setminus {\cal W}}) - I(X_{{\cal V}'}; Z| X_{\overline {{\cal K}''}}) \label{prove_part1_g}\\
\leq & I(X_{\cal V}, X_{\cal W}; Y| X_{\overline {\cal V}}, X_{{\overline {{\cal K}'}} \setminus {\cal W}}, X_{{\cal K}_0}) - I(X_{{\cal V}'}; Z| X_{\overline {{\cal K}''}}), \label{prove_part1_h}\\
= & I(X_{\cal V}, X_{\cal W}; Y| X_{\overline {\cal V}}, X_{{\overline {{\cal K}''}} \setminus {\cal W}}) - I(X_{{\cal V}'}; Z| X_{\overline {{\cal K}''}}), \forall {\cal V} \subseteq {\cal K}'', {\cal V}' \subseteq {\cal V}, {\cal W} \subseteq {\overline {{\cal K}'}},
\end{align}
\end{subequations}
where (\ref{prove_part1_c}) is obtained by using the fact that $({\cal V} \cup {\cal K}_0) \subseteq {\cal K}'$, ${\cal W} \subseteq {\overline {{\cal K}'}}$, $(R_1^{\text s}, R_1^{\text o},\cdots, R_K^{\text s}, R_K^{\text o})$ is in region ${\mathscr R} (X_{\cal K}, {\cal K}')$, and setting ${\cal S} = {\cal V} \cup {\cal K}_0$, ${\cal S}' = {\cal V}' \cup {\cal K}_0$, and ${\cal T} = {\cal W}$ in (\ref{region_DM_exten}).
In addition, (\ref{prove_part1_f}) follows from using (\ref{assumption_61}), (\ref{prove_part1_g}) is true since ${\overline {\cal V}} = {\cal K}'' \setminus {\cal V} = {\cal K}' \setminus ({\cal V} \cup {\cal K}_0)$ and ${\overline {{\cal K}''}} = {\overline {{\cal K}'}} \cup {\cal K}_0$, and (\ref{prove_part1_h}) is obtained by adding $X_{{\cal K}_0}$ and using the fact that $X_k, \forall k \in {\cal K}$ are independent of each other.
(\ref{part1}) is thus proven.

Next we prove (\ref{part2}), in which ${\cal W} \subseteq {\overline {{\cal K}''}} \triangleq {\overline {{\cal K}'}} \cup {\cal K}_0$ and ${\cal W} \cap {\cal K}_0 \neq \emptyset$.
Obviously, ${\cal W}$ can be divided into two disjoint subsets, ${\cal W}_1$ and ${\cal W}_2$, with ${\cal W}_1 \subseteq {\overline {{\cal K}'}}$ and ${\cal W}_2 \subseteq {\cal K}_0$.
Then,
\begin{subequations}\label{prove_part2_1}
\begin{align}
\sum_{k \in {\cal W}_2} R_k^{\text o} & = \sum_{k \in {\cal W}_2} R_k^{\text s} + \sum_{k \in {\cal W}_2} R_k^{\text o} \label{prove_part2_1a}\\
& \leq I(X_{{\cal W}_2}; Y| X_{{\cal K}' \setminus {\cal W}_2}, X_{\overline {{\cal K}'}}) \label{prove_part2_1b}\\
& = I(X_{{\cal W}_2}; Y| X_{{\cal K}''}, X_{\overline {{\cal K}'}}, X_{{\cal K}_0 \setminus {\cal W}_2}), \label{prove_part2_1c}
\end{align}
\end{subequations}
where (\ref{prove_part2_1a}) holds since ${\cal W}_2 \subseteq {\cal K}_0$ and $R_k^{\text s} = 0, \forall k \in {\cal K}_0$ (see (\ref{Rs_zero_K0})), and (\ref{prove_part2_1c}) is true since ${\cal K}' = {\cal K}'' \cup {\cal K}_0$.
Note that as defined in (\ref{assumption_61}), ${\cal K}_0 \subsetneqq {\cal K}'$, making ${\cal W}_2 \subseteq {\cal K}'$.
In addition, $(R_1^{\text s}, R_1^{\text o},\cdots, R_K^{\text s}, R_K^{\text o})$ is in region ${\mathscr R} (X_{\cal K}, {\cal K}')$ and thus satisfies (\ref{region_DM_exten}).
Hence, (\ref{prove_part2_1b}) is obtained by setting ${\cal S} = {\cal W}_2$, ${\cal S}' = \emptyset$, and ${\cal T} = \emptyset$ in (\ref{region_DM_exten}).
Then,
\begin{subequations}\label{prove_part2}
\begin{align}
& \sum_{k \in \cal V} R_k^{\text s} + \sum_{k \in {\cal V} \setminus {\cal V}'} R_k^{\text o} + \sum_{k \in {\cal W}} R_k^{\text o} \nonumber\\ 
= & \sum_{k \in \cal V} R_k^{\text s} + \sum_{k \in {\cal V} \setminus {\cal V}'} R_k^{\text o} + \sum_{k \in {\cal W}_1} R_k^{\text o} + \sum_{k \in {\cal W}_2} R_k^{\text o} \label{prove_part2_a}\\
\leq & I(X_{\cal V}, X_{{\cal W}_1}; Y| X_{\overline {\cal V}}, X_{{\overline {{\cal K}'}} \setminus {\cal W}_1}) - I(X_{{\cal V}'}; Z| X_{\overline {{\cal K}''}}) + I(X_{{\cal W}_2}; Y| X_{{\cal K}''}, X_{\overline {{\cal K}'}}, X_{{\cal K}_0 \setminus {\cal W}_2}) \label{prove_part2_b}\\
\leq & I(X_{\cal V}, X_{{\cal W}_1}; Y| X_{\overline {\cal V}}, X_{{\overline {{\cal K}'}} \setminus {\cal W}_1}, X_{{\cal K}_0 \!\setminus\! {\cal W}_2}) \!-\! I(X_{{\cal V}'}; Z| X_{\overline {{\cal K}''}}) \!+\! I(X_{{\cal W}_2}; Y| X_{{\cal K}''}, X_{\overline {{\cal K}'}}, X_{{\cal K}_0 \!\setminus\! {\cal W}_2}) \label{prove_part2_c}\\
= & I(X_{\cal V}, X_{{\cal W}_1}, X_{{\cal W}_2}; Y| X_{\overline {\cal V}}, X_{{\overline {{\cal K}'}} \setminus {\cal W}_1}, X_{{\cal K}_0 \setminus {\cal W}_2}) - I(X_{{\cal V}'}; Z| X_{\overline {{\cal K}''}}) \label{prove_part2_d}\\
= & I(X_{\cal V}, X_{{\cal W}}; Y| X_{\overline {\cal V}}, X_{{\overline {{\cal K}''}} \setminus {\cal W}}) \!-\! I(X_{{\cal V}'}; Z| X_{\overline {{\cal K}''}}), \forall {\cal V} \!\subseteq\! {\cal K}'', {\cal V}' \!\subseteq\! {\cal V}, {\cal W} \!\subseteq\! {\overline {{\cal K}''}}, {\cal W} \!\cap\! {\cal K}_0 \neq \emptyset, \label{prove_part2_e}
\end{align}
\end{subequations}
where (\ref{prove_part2_b}) follows from using (\ref{prove_part1_g}) (since ${\cal W}_1 \subseteq {\overline {{\cal K}'}}$) and (\ref{prove_part2_1}), and (\ref{prove_part2_c}) is obtained by adding $X_{{\cal K}_0 \setminus {\cal W}_2}$ and using the fact that $X_k, \forall k \in {\cal K}$ are independent of each other.
Combining (\ref{Rs_zero_K0}), (\ref{part1}), and (\ref{part2}), it is known that $(R_1^{\text s}, R_1^{\text o},\cdots, R_K^{\text s}, R_K^{\text o})$ satisfies (\ref{region_DM1}) and is thus also in ${\mathscr R} (X_{\cal K}, {\cal K}'')$.
Theorem~\ref{theo_polytope} is then proven.

\bibliographystyle{IEEEtran}
\bibliography{IEEEabrv,Ref}

\end{document}